\documentclass[12pt]{article}
\pdfoutput=1
\usepackage[a4paper]{geometry}
\usepackage{jheppub, amsmath,amssymb,amsfonts,amsxtra, mathrsfs, makeidx,graphics,graphicx,amsthm,epsfig, bm,longtable,float, color,tikz,mathtools,xfrac,footnote,rotating, lscape, makecell, environ,mathtools, empheq}

\usepackage{subfig}
\usepackage{multirow}
\usepackage{adjustbox}

\usepackage{amsthm}
\usepackage{hyperref}

\usepackage{amsmath}
\usepackage{amsfonts}
\usepackage{commath}
\usepackage[english]{babel}
\usepackage{setspace}

\usepackage{epsfig,amssymb} 
\usepackage{amsmath}

\usepackage{amscd,color}
\usepackage{amsmath,graphicx}
\usepackage{verbatim,booktabs}

\usepackage{latexsym}
\usepackage{amsmath,amsfonts,amssymb,amsthm}
\usepackage{amsmath,amsthm}

\DeclareMathOperator{\Tr}{Tr}
\DeclareMathOperator{\rk}{rk}

\title{
\begin{center}
Dualities for three-dimensional \\ $\mathcal{N}=2$ $SU(N_c)$ chiral adjoint SQCD
\end{center}}

\author[a]{Antonio Amariti}
\author[b,c]{and Marco Fazzi} 
\affiliation[a]{INFN, Sezione di Milano, Via Celoria 16, I-20133 Milano, Italy}
\affiliation[b]{INFN, Sezione di Milano--Bicocca, Piazza della Scienza 3, I-20126 Milano, Italy}
\affiliation[c]{Dipartimento di Fisica, Universit\`a di Milano--Bicocca, Piazza della Scienza 3, I-20126 Milano, Italy}

\emailAdd{antonio.amariti@mi.infn.it, marco.fazzi@mib.infn.it}

\abstract{We study dualities for 3d $\mathcal{N}=2$ $SU(N_c)$  SQCD at Chern--Simons level $k$ in presence of an adjoint
with polynomial superpotential. 
The dualities are dubbed \emph{chiral} because there is a  different amount of fundamentals $N _f$ and antifundamentals $N _a$. 
We build a complete classification of such dualities in terms of $|N_f  - N _a|$  and $k$.
The classification is obtained by studying the flow from the non-chiral case, and 
we corroborate our proposals by matching the three-sphere partition functions.
Finally, we revisit the case of $SU(N_c)$ SQCD without the adjoint, comparing our results with previous literature.}

\begin{document}

\maketitle

\section{Introduction}
\label{sec:intro}

The rich web of 3d $\mathcal{N}=2$ infrared dualities represents an active field of research   
both because of its similarities with less or non-supersymmetric cases, and because of its connection
with exact results in mathematical physics through localization.
This latter aspect led to a systematic study of many physical dualities, once the formal structure of the three-sphere partition function of $\mathcal{N}=2$ models was successfully computed via localization in \cite{Kapustin:2009kz,Jafferis:2010un,Hama:2010av,Hama:2011ea}.
Indeed it was soon realized \cite{Willett:2011gp,Benini:2011mf} 
that the identities among the partition functions of the dualities known at the time, namely  
Aharony \cite{Aharony:1997gp} and Giveon--Kutasov  \cite{Giveon:2008zn} dualities, already appeared in the mathematical literature as
integral identities among hyperbolic hypergeometric gamma functions \cite{VanDeBult}.
This result was interesting also because many other integral identities were available in the mathematical literature, suggesting the existence of yet-to-be-discovered dualities. Indeed for $U(N_c)$ SQCD the systematic analysis of \cite{Benini:2011mf} provided a classification scheme for \emph{chiral} dualities. This name originates from the fact that such dualities feature in general a different number of 
fundamental and antifundamental matter fields.

In this paper we study chiral dualities for 3d $\mathcal{N}=2$ SQCD with $SU(N_c)$ gauge group, Chern--Simons (CS) interactions, and adjoint matter,
extending the analysis of \cite{Aharony:2014uya,Hwang:2015wna,Nii:2019qdx,Nii:2020xgd}.
For consistency with the classification of \cite{Benini:2011mf},
we divide the dualities into two main classes denoted $[p,q]$ and $[p,q]^*$ (that we will review in section \ref{sec:review}).
In this classification it is possible to recover non-chiral dualities as well: 
Aharony duality corresponds to the $[0,0]$ case,
and Giveon--Kutasov duality  corresponds to the $[p,p]$ case. 
(Furthermore the $[0,0]$ and the $[p,0]$ cases require further singlets and interactions in the dual phase, 
with respect to the other dualities in the $[p,q]$ family.)
 
Two generalizations of the results of \cite{Benini:2011mf} have been discussed in the literature.
 \begin{itemize}
 \item 
 In \cite{Aharony:2014uya} chiral dualities for $SU(N_c)$ SQCD are tackled.\footnote{Other dualities involving $U(N_c)$ and $SU(N_c)$ chiral SQCD characterized by symmetry enhancements are discussed in \cite{Fazzi:2018rkr,Amariti:2018wht,Benvenuti:2018bav}. See also \cite{Benini:2017dud,Amariti:2018gdc}, where chiral dualities for SQCD with a monopole superpotential are discussed.} The authors identify two classes of dualities that are reminiscent of the $[p,q]$ and $[p,q]^*$ classes of \cite{Benini:2011mf}. Actually the relation between 
\cite{Aharony:2014uya} and \cite{Benini:2011mf} is more involved.

Indeed by applying the standard procedure of 
gauging the topological symmetry of the $U(N_c)$ case of \cite{Benini:2011mf} one does not  
recover the $SU(N_c)$ case of \cite{Aharony:2014uya}.
This is due to the presence of CS interactions: the application
of local mirror symmetry is necessary to obtain the expected duality.
This problem is reminiscent of the one discussed in \cite{Aharony:2013dha} to connect 
Giveon--Kutasov duality for $U(N_c)_k$ to the one for non-chiral
$SU(N_c)_k$ SQCD obtained from circle reduction of 4d Seiberg duality (in addition to a real mass flow).
\item Another generalization of \cite{Benini:2011mf} has been proposed in \cite{Hwang:2015wna}, 
which studies dualities for $U(N_c)$ chiral SQCD with adjoint matter. In this case the authors discuss the extension of the  $[p,q]^*$ case, as well as the $[p,0]$ one, but neglect the general $[p,q]$ case. More recently, some discussion on dualities for $SU(N_c)$ chiral SQCD with adjoint matter appeared in \cite{Nii:2019qdx}. The author focuses on a specific $[p,q]^*$ case with vanishing CS level.
\end{itemize}
Motivated by these results, in this paper we provide a
complete classification of 3d $\mathcal{N}=2$ infrared dualities for 
$SU(N_c)$ chiral adjoint SQCD.
In order to obtain such a classification we first reconsider the case of
$SU(N_c)$ chiral SQCD (section \ref{SUAF}), obtaining a classification that slightly differs
from the one of \cite{Aharony:2014uya},\footnote{Again we expect that the two classification
schemes are related by a mirror transformation.} 
and then discuss the generalization of \cite{Benini:2011mf} to the case of $U(N_c)$ with
adjoint matter (section \ref{sec:adjUN}).
Finally we study the case of $SU(N_c)$ with adjoint matter and provide a classification
of chiral dualities (section \ref{sec:adjSUN}).
As a check of our proposals we match the three-sphere partition functions. Our results are summarized in tables \ref{tab:USU} and \ref{tab:USUadj}. In section \ref{sec:conc} we briefly present our conclusions.

Several appendices complement our analysis. In appendix \ref{appC} we show how to distinguish the $SU(N_c)_{k_1}$ from the $U(1)_{k_2}$ CS contribution in the localized three-sphere partition function of $U(N_c)_{k_1,k_1+N_c k_2}$. In appendix \ref{appAF} we write down the real mass flow (starting from the $SU(N_c)$ version of Giveon--Kutasov duality appearing in \cite{Aharony:2013dha}) producing the dualities of \citep{Aharony:2014uya} at the level of the partition function. In appendix \ref{appB} we write down the real mass flow (again starting from the $SU(N_c)_k$ duality of \cite{Aharony:2013dha}) producing the dualities of section \ref{SUAF}, i.e. a reformulation of the chiral dualities of \cite{Aharony:2014uya}. The interrelation between the sections and appendices of this paper and previous literature is presented in figures \ref{pqfig}, \ref{p0fig}, and \ref{pq*fig}. Finally in appendix \ref{sec:BCCbranes} we discuss the D3-D5-NS5 brane engineering of the dualities of \cite{Benini:2011mf}.

\section{\texorpdfstring{Known dualities for $U(N_c)$ and $SU(N_c)$ chiral SQCD}{Known dualities for U(Nc) and SU(Nc) chiral SQCD}}
\label{sec:review}

The mathematical notation $[p,q]$, $[p,q]^*$ we mentioned in the introduction can be traded for a more physical one in terms of the effective CS level on the Coulomb branch of a $U(N_c)_k$ gauge theory. Defining 
\begin{equation}\label{eq:kpm}
k_\pm \equiv k \pm \frac{1}{2}(N_f-N_a)\ ,
\end{equation}
we have \cite{Benini:2011mf}:
\begin{subequations}\label{eq:pq}
\begin{align}
& [p,q]_a \equiv [-k_+,-k_-]_a\ ,  && [p,q]_b \equiv [k_+,k_-]_b\ , \\
& [p,q]_a^* \equiv [-k_+,k_-]_a^*\ , && [p,q]_b^* \equiv [k_+,-k_-]_b^*\ ,
\end{align}
\end{subequations}
where $\pm k_\pm$ represents the effective CS level on the Coulomb branch for positive and negative values of the real scalar $\sigma$ in the vector multiplet. The ``type'' of the  theory (i.e. the subscript $a$ or $b$) is then determined by requiring $p,q \geq 0$. (In practice, given $\{k,N_f,N_a\}$ one computes $k_\pm$ via \eqref{eq:kpm} and picks the case where both $p$ and $q$ are non-negative. This selects the type of the theory.) Notice that the dual of a type-$a$ theory (be it in the $[p,q]$ or $[p,q]^*$ class) is a type-$b$ theory \cite{Benini:2011mf}.\footnote{\label{foot:type}One can go from type $a$ to type $b$ by acting with a $C$, $P$, or $CP$ transformation (as explained in \cite[Sec. 3.6]{Benini:2011mf}). In this paper the electric phase will always be a type-$b$ theory.} Moreover the cases $[p,0]$ and $[0,q]$ are obtained as limiting cases of $[p,q]$ in the obvious way.

With this notation in place, we are ready to collect the results that have been obtained in the literature so far. Chiral dualities for $U(N_c)$ gauge theories were studied in \cite{Benini:2011mf}. In the $[p,q]$ case we have a duality between:
\begin{itemize}
\item an {\bf electric} model, consisting of $U(N_c)_{k}$ SQCD with $N_f$ 
fundamentals $Q$, $N_a$ antifundamentals $\tilde Q$, $|N_f-N_a| < 2k $,
and vanishing superpotential, $W=0$;
\item a {\bf magnetic} model consisting of $U(\widetilde N_c)_{-k}$ SQCD, where 
$\widetilde N_c \equiv \frac{1}{2}(N_f+N_a) +k-N_c$, with $N_f$ 
fundamentals, $N_a$ antifundamentals, the meson $M = Q \tilde Q$, 
and $W=M q \tilde q$.
\end{itemize}
Further, when $p$ and/or $q$ are vanishing these dualities are modified by the presence 
of extra singlets in the dual, magnetic, phase. These singlets are identified with the monopole
operators of the electric phase.
For example if both $p$ and $q$ are vanishing (the case of Aharony duality) we have:
\begin{itemize}
\item an {\bf electric} model, consisting of $U(N_c)$ SQCD with $N_f$ 
fundamentals $Q$, $N_f$ anti-fundamentals $\tilde Q$,
and $W=0$;
\item a {\bf magnetic} model consisting of $U(\widetilde N_c)$ SQCD, where 
$\widetilde N_c \equiv N_f-N_c$, with $N_f$ 
fundamentals, $N_f$ antifundamentals, the meson $M = Q \tilde Q$,
two new singlets $T_{\pm}$ identified with the monopole and the anti-monopole of the
electric phase, and $W=M q \tilde q + T_+ t_- + T_- t_+$,
where $t_{\pm}$ are the monopole and the antimonopole of the dual phase.
\end{itemize}
If instead only $q=0$ we have:\footnote{The situation for $p=0$ is analogous, and the two are related by a 
parity transformation.}
\begin{itemize}
\item an {\bf electric} model, consisting of $U(N_c)_{(N_f-N_a)/2}$ SQCD with $N_f$ 
fundamentals $Q$, $N_a<N_f$ antifundamentals $\tilde Q$,
and $W=0$;
\item a {\bf magnetic} model consisting of $U(\widetilde N_c)_{(N_a-N_f)/2}$ SQCD, where 
$\widetilde N_c \equiv N_f-N_c$, with $N_f$ 
fundamentals, $N_a$ antifundamentals, the meson $M = Q \tilde Q$,
a new singlet $T_{+}$, and $W=M q \tilde q + T_+ t_- $.
\end{itemize}
In the $[p,q]^*$ case we have a duality between:
\begin{itemize}
\item an {\bf electric} model, consisting of $U(N_c)_{k}$ SQCD with $N_f$ 
fundamentals $Q$, $N_a$ antifundamentals $\tilde Q$, $|N_f-N_a| > 2k $,
and $W=0$;
\item a {\bf magnetic} model consisting of $U(\widetilde N_c)_{-k}$ SQCD, where 
$\widetilde N_c \equiv \max (N_f, N_a) -N_c$, with $N_f$ 
fundamentals, $N_a$ antifundamentals, the meson $M = Q \tilde Q$,
and $W=M q \tilde q$.
\end{itemize}

In the case of $SU(N_c)$ a similar classification was obtained in \cite{Aharony:2014uya}.
The authors obtained the analog of the $[p,q]$ and $[p,q]^*$ classes.
In the {\bf electric} phase the theory corresponds to $SU(N_c)_{k}$ SQCD with $N_f$ 
fundamentals $Q$, $N_a$ antifundamentals $\tilde Q$, 
and $W=0$.
As in the classification of \cite{Benini:2011mf}, the dual {\bf magnetic} model depends instead on the relative value of 
$2k$ with respect to $\Delta F \equiv |N_f-N_a|$.
If  $\Delta F \leq 2k $ the dual model is 
  $SU(\widetilde N_c)_{-k} \times U(1)_{\widetilde N_c-k}$, where 
$\widetilde N_c \equiv \frac{1}{2}(N_f+N_a) +k-N_c$, with $N_f$ 
fundamentals, $N_a$ antifundamentals, the meson $M = Q \tilde Q$, and $W=M q \tilde q$.
If  $\Delta F > 2k $ the dual model is
$SU(\widetilde N_c)_{-k} \times U(1)_{\tilde N_c-k} \times U(1)_0$, where 
$\widetilde N_c \equiv \max (N_f,N_a) -N_c$,  with $N_f$ 
fundamentals, $N_a$ antifundamentals, the meson $M = Q \tilde Q$,
and $W=M q \tilde q$. There is also a mixed CS term, at level $1$,
between the two $U(1)$ gauge groups.

Observe that the chiral class denoted $[p,0]$ in \cite{Benini:2011mf}, 
corresponding to the case $\Delta F =2k$, was not distinguished from the case
$\Delta F < 2k$ in the analysis of \cite{Aharony:2014uya}. More recently \cite{Nii:2020xgd} claimed that actually these two cases are
distinguished. In the following we will support this latter claim.

 As observed in \cite{Aharony:2014uya} the chiral dualities for $SU(N_c)$ 
 SQCD cannot be easily obtained  from the ones for $U(N_c)$ SQCD by gauging the topological $U(1)$ symmetry.
 This is because there are CS interactions that make this step nontrivial.
 Observe also that the $SU(N_c)$ chiral dualities have been obtained 
 starting from the non-chiral duality for $SU(N_c)_k$ with matter.
 Nevertheless another possibility would be to start from the
 $SU(N_c)$ version of Aharony duality (i.e. at zero CS level) obtained in 
  \cite{Park:2013wta}\footnote{
 A different version of the $SU(N_c)$ duality has been obtained in \cite{Aharony:2013dha}.
 The two are related by local mirror symmetry in the abelian sector of the dual 
 model. Here we will use the version of  \cite{Park:2013wta} because it extends
 more easily to the case with adjoint matter.} and then
 flow to the chiral  $SU(N_c)$  cases.
 This construction would be more similar to the one obtained in \cite{Benini:2011mf} for the 
 $U(N_c)$ case. We are going to perform such a construction in the next section, and comment 
 on the relation with \cite{Aharony:2014uya} in appendix \ref{appAF}.

We end this section by compactly presenting the known dualities for $U(N_c)$ and $SU(N_c)$ chiral SQCD in table \ref{tab:USU}. 
In section \ref{SUAF} we will rederive the chiral dualities for $SU(N_c)$ SQCD, finding 
different dual descriptions than the ones presented in table \ref{tab:USU}. Their equivalence is then proven in appendix \ref{appAF}.
In table \ref{tab:USUadj} we anticipate the results for $U(N_c)$ and $SU(N_c)$ chiral SQCD with one adjoint matter field, presented in section \ref{sec:adjUN} and \ref{sec:adjSUN} respectively.

\begin{table}[ht!]
\centering \scriptsize
\begin{tabular}{lcccc}
 \multicolumn{1}{c}{case} & \begin{tabular}{c} $|N_f-N_a|$ \\ vs. $2k$ \end{tabular} & \begin{tabular}{c} electric \\ phase \end{tabular} & \begin{tabular}{c}magnetic \\phase\end{tabular} & ref. \\[10pt]
\hline
\rule{0pt}{25pt}
$[0,0]$ & $0=0$ &  \begin{tabular}{c} $U(N_c)_0$  \\ w/ $(N_f,N_f)$; \\ $W=0$ \end{tabular} &  \begin{tabular}{c} $U(N_f-N_c)_0$ w/ \\ $(N_f,N_f), M=Q\tilde{Q}, T_\pm$; \\ $W=Mq\tilde{q}+T_+ t_- + T_- t_+$ \end{tabular} & \cite{Aharony:1997gp} \\[15pt]
\hline
\rule{0pt}{25pt}
$[p,0]$ & $\Delta F=2k$ &  \begin{tabular}{c} $U(N_c)_k$ \\ w/ $(N_f,N_a), N_a < N_f$; \\ $W=0$ \end{tabular} &  \begin{tabular}{c} $U(N_f-N_c)_{-k}$ w/ \\ $(N_a,N_f), M=Q\tilde{Q}, T_+$; \\ $W=Mq\tilde{q}+T_+ t_- $ \end{tabular} & \cite{Benini:2011mf} \\[15pt]
\hline
\rule{0pt}{25pt}
$[p,p]$ & $0< 2k$ &  \begin{tabular}{c} $U(N_c)_k$  \\ w/ $(N_f,N_f)$; \\$W=0$ \end{tabular} &  \begin{tabular}{c} $U(N_f-N_c +|k|)_{-k}$ w/ \\ $(N_f,N_f), M=Q\tilde{Q}$; \\ $W=Mq\tilde{q}$ \end{tabular} & \cite{Giveon:2008zn} \\[15pt]
\hline
\rule{0pt}{25pt}
$[p,q]$ & $\Delta F < 2k$ & \multirow{5}{*}{ \begin{tabular}{c} $U(N_c)_k$ \\ w/ $(N_f,N_a)$; \\ $W=0$ \end{tabular} }&  \begin{tabular}{c} $U(\tfrac{1}{2}(N_f+N_a)+|k|-N_c)_{-k}$ \\ w/ $(N_a,N_f), M=Q\tilde{Q}$; \\ $W=Mq\tilde{q}$ \end{tabular} & \multirow{5.5}{*}{\cite{Benini:2011mf}} \\[15pt] \cline{1-2} \cline{4-4}
\rule{0pt}{25pt}
$[p,q]^*$ & $\Delta F>2k$ &  &  \begin{tabular}{c} $U(\max(N_f,N_a)-N_c)_{-k}$ \\w/ $(N_a,N_f), M=Q\tilde{Q}$; \\ $W=Mq\tilde{q} $ \end{tabular} &  \\[15pt]
\hline\hline
\rule{0pt}{25pt}
$[0,0]$ & $0=0$ &  \begin{tabular}{c} $SU(N_c)_0$  \\ w/ $(N_f,N_f)$; \\ $W=0$ \end{tabular} &  \begin{tabular}{c} $ U(N_f-N_c)_0$ w/ \\ $(N_f,N_f), (b,\tilde{b}), M=Q\tilde{Q}, Y$; \\ $W=\eqref{WARSW}$ \end{tabular} & \begin{tabular}{c} \cite{Aharony:2013dha} \\ (\cite{Park:2013wta})\end{tabular} \\[15pt]
\hline
\rule{0pt}{30pt}
$[p,q]$ & $\Delta F < 2k$ &  \multirow{12}{*}{\begin{tabular}{c} $SU(N_c)_k$ \\ w/ $(N_f,N_a)$; \\ $W=0$ \end{tabular}} &  \begin{tabular}{c} $SU(\widetilde N_c)_{-k} \times U(1)_{\widetilde N_c -k}$\\ $\widetilde N_c \equiv \tfrac{1}{2}(N_f+N_a)+|k|-N_c$ \\ w/ $(N_a,N_f), M=Q\tilde{Q}$; \\ $W=Mq\tilde{q}$ \end{tabular} & \cite{Aharony:2014uya} \\[20pt] \cline{1-2} \cline{4-5}
\rule{0pt}{30pt}
$[p,0]$ & $\Delta F = 2k$ &  &  \begin{tabular}{c} $SU(\widetilde N_c)_{-k} \times U(1)_{{\widetilde N_c}/{2}-k} $\\ $\widetilde N_c \equiv \tfrac{1}{2}(N_f + N_a)+|k|-N_c$ \\ w/ $(N_a,N_f), M=Q\tilde{Q}, T_-$; \\ $W= $ \cite[Eq. (7.45)]{Nii:2020xgd} \end{tabular} & \cite{Nii:2020xgd} \\[20pt]
\cline{1-2} \cline{4-5}
\rule{0pt}{30pt}
$[p,q]^*$ & $\Delta F>2k$ &  &  \begin{tabular}{c} $SU(\widetilde N_c)_{-k} \times (U(1)_{\widetilde N_c -k} \times U(1)_0)_1$ \\ $\widetilde N_c \equiv \max (N_f,N_a)-N_c$ \\ w/ $(N_a,N_f), M=Q\tilde{Q}$; \\ $W=Mq\tilde{q} $ \end{tabular} & \cite{Aharony:2014uya} 
\end{tabular}
\caption{Summary of chiral dualities for $U(N_c)$ and $SU(N_c)$ SQCD with $(N_f,N_a)\in (\mathbf{N}_c,\overline{\mathbf{N}_c})$ matter fields. Here $U(N_c)_k \equiv U(N_c)_{k,k} = SU(N_c)_k \times U(1)_0$ (neglecting global issues). The notation $(U(1)\times U(1))_1$ means there is a mixed CS term at level $1$ between the two gauge $U(1)$'s. (One of the two factors may be the abelian center of $U(\widetilde N_c)$.) The dictionary between the non-negative integers $p,q$ and the effective CS level $k_\pm$ on the Coulomb branch is as in formulae \eqref{eq:kpm}-\eqref{eq:pq}.}
\label{tab:USU}
\end{table}
\begin{table}[ht!]
\centering \scriptsize
\begin{tabular}{lcccc}
 \multicolumn{1}{c}{case} & \begin{tabular}{c} $|N_f-N_a|$ \\ vs. $2k$ \end{tabular} & \begin{tabular}{c} electric \\ phase \end{tabular} & \begin{tabular}{c}magnetic \\phase\end{tabular}  & ref. \\[10pt]
\hline
\rule{0pt}{27.5pt}
$[0,0]_\text{adj}$ & $0=0$ &  \begin{tabular}{c} $U(N_c)_0$ w/ \\ $(N_f,N_f), X$; \\ $W=\Tr X^{n+1}$ \end{tabular} &  \begin{tabular}{c} $U(nN_f-N_c)_0$ w/ \\ $(N_f,N_f), Y, M_j=QX^j\tilde{Q}$, \\ 
$T_{j,+} =T_+ X^j, T_{j,-} = T_-X^j$,\\ $j=0,\ldots,n-1$; $W=\eqref{wKP}$  \end{tabular} &  \cite{Kim:2013cma} \\[20pt]
\hline
\rule{0pt}{25pt}
$[p,p]_\text{adj}$ & $0< 2k$ &  \begin{tabular}{c} $U(N_c)_k$  w/ \\ $(N_f,N_f), X$; \\$W=\Tr X^{n+1} $ \end{tabular} &  \begin{tabular}{c} $U(n(N_f+|k|)-N_c)_{-k}$ w/ \\ $(N_f,N_f), Y, M_j=QX^j \tilde{Q}$, \\ $j=0,\ldots,n-1$; $W=\eqref{Wniarchos}$ \end{tabular} & \cite{Niarchos:2008jb} \\[15pt] 
\hline
\rule{0pt}{25pt}
$[p,q]_\text{adj}$ & $\Delta F < 2k$ & \multirow{10}{*}{ \begin{tabular}{c} $U(N_c)_k$ w/ \\ $(N_f,N_a), X$; \\ $W=\Tr X^{n+1}$ \end{tabular} }&  \begin{tabular}{c} $U(n(\tfrac{1}{2} (N_f+N_a)+|k|)-N_c)_{-k}$ w/ \\ $(N_a,N_f), Y, M_j=QX^j\tilde{Q}$, \\$j=0,\ldots,n-1$; $W=\eqref{wd}$ \end{tabular} & here \\[15pt] \cline{1-2} \cline{4-4}
\rule{0pt}{25pt}
$[p,0]_\text{adj}$ & $\Delta F = 2k$ & &  \begin{tabular}{c} $U(n(\tfrac{1}{2} (N_f+N_a)+|k|)-N_c)_{-k}$ w/ \\ $(N_a,N_f), Y, M_j=QX^j\tilde{Q}$, \\ $T_{j,-} = T_- X^j,\ j=0,\ldots,n-1$; \\ $W=\eqref{wdd}$ \end{tabular} &  \multirow{5}{*}{\cite{Hwang:2015wna}} \\[15pt] \cline{1-2} \cline{4-4}
\rule{0pt}{25pt}
$[p,q]^*_\text{adj}$ & $\Delta F>2k$ &  &  \begin{tabular}{c} $U(n \max(N_f,N_a) -N_c)_{-k}$ w/ \\ $(N_a,N_f), Y, M_j=QX^j\tilde{Q}$, \\$j=0,\ldots,n-1$; $W=\eqref{Wddd}$ \end{tabular} &  \\[15pt]
\hline\hline
\rule{0pt}{27.5pt}
$[0,0]_\text{adj}$ & $0=0$ &  \begin{tabular}{c} $SU(N_c)_0$ w/ \\ $(N_f,N_f), X$; \\ $W=\Tr X^{n+1}$ \end{tabular} &  \begin{tabular}{c} $(U(nN_f-N_c)_0 \times U(1)_0)_1$ w/ \\ $(N_f,N_f), Y, M_j=QX^j\tilde{Q}$, \\ $(T_{j,+}, T_{j,-}) \in (+,-)$,\\ $j=0,\ldots,n-1$; $W=\eqref{SU00adj}$ \end{tabular} & \cite{Park:2013wta} \\[20pt]
\hline
\rule{0pt}{27.5pt}
$[p,q]_\text{adj}$ & $\Delta F < 2k$ &  \multirow{11.5}{*}{\begin{tabular}{c} $SU(N_c)_k$ w/ \\  $(N_f,N_a), X$; \\ $W=\Tr X^{n+1}$ \end{tabular}} &  \begin{tabular}{c} $(U(n(\tfrac{1}{2} (N_f+N_a)+|k|)-N_c)_{-k} \times U(1)_{-n})_1 $ \\w/ $(N_a,N_f), Y, M_j=QX^j\tilde{Q}$, \\ $j=0,\ldots,n-1$; $W=\eqref{pqsuadj}$ \end{tabular} &\multirow{11.5}{*}{here}  \\[20pt] \cline{1-2} \cline{4-4}
\rule{0pt}{27.5pt}
$[p,0]_\text{adj}$ & $\Delta F = 2k$ &  &  \begin{tabular}{c} $(U(nN_f-N_c)_{-k} \times U(1)_{-n/2})_1$ \\ w/ $(N_a,N_f), Y, M_j=QX^j\tilde{Q}$, \\  $T_{j,+}\in (+)$,\\ $j=0,\ldots,n-1$;  $W=\eqref{Wp0suadj}$ \end{tabular} & \\[20pt]
\cline{1-2} \cline{4-4}
\rule{0pt}{27.5pt}
$[p,q]^*_\text{adj}$ & $\Delta F>2k$ &  &  \begin{tabular}{c} $(U(n \max(N_f,N_a) -N_c)_{-k} \times U(1)_{0})_1$ \\ w/ $(N_a,N_f), Y, M_j=QX^j\tilde{Q}$, \\ $j=0,\ldots,n-1$; $W=\eqref{wpqasdj*}$ \end{tabular} &  
\end{tabular}
\caption{Summary of chiral dualities for $U(N_c)$ and $SU(N_c)$ adjoint SQCD with $(N_f,N_a)\in (\mathbf{N}_c,\overline{\mathbf{N}_c})$ matter fields, and one field $X$ in the $\mathbf{Adj}$ of the electric gauge group. The dual adjoint field in the magnetic phase is $Y$. The notation is as in table \ref{tab:USU}.}
\label{tab:USUadj}
\end{table}

\clearpage

\normalsize
%
%
%
%
%
   \section{\texorpdfstring{Reformulating the dualities for $SU(N_c)$ chiral SQCD}{Reformulating the dualities for SU(Nc) chiral SQCD}}
\label{SUAF}
%
%
%
%
%
In this section we discuss the existence of families of 
3d chiral dualities for $SU(N_c)_k$ CS matter theories 
very similar to the ones derived in \cite{Aharony:2014uya}.
The main differences between the two families of dualities (i.e. ours and theirs) are the following:
\begin{itemize}
\item the case $\Delta F \equiv |N_f - N_a| < 2k$ and $\Delta F= 2k$ are treated separately in our description;
\item there is a different amount of $U(1)$ gauge factors.
\end{itemize}
In appendix \ref{appAF} we will then show that the dualities obtained here and the ones
discussed in \cite{Aharony:2014uya,Nii:2020xgd} are related by some mirror transformations, 
similarly to what is expected in the non-chiral case \cite{Aharony:2013dha}.

We derive the chiral dualities by
flowing from the generalization of Aharony duality  to $SU(N_c)$ non-chiral SQCD with vanishing CS level. 
We perform a real mass flow on the electric side 
considering large vacuum expectation values (VEVs) for some of  the scalars in the background vector multiplets associated to the global symmetries.  
This procedure leads to a chiral $SU(N_c)$ model.
Then we match this limit on the magnetic side, by considering also large 
 VEVs for the scalars in the vector multiplet of the
 gauge symmetry.

On the electric side we have  $SU(N_c)_0$ SQCD with global 
$SU(N_f)_L  \times SU(N_f)_R \times U(1)_B \times U(1)_A \times  U(1)_R$
symmetry and vanishing superpotential.  
The charges are as follows:
\begin{equation}
\begin{array}{c|cccccc}
          &   SU(N_c)  &   SU(N_f)_L & SU(N_f)_R & U(1)_B &U(1)_A & U(1)_R \\
             \hline
Q           &  \mathbf{N}_c   & \mathbf{N}_f& \mathbf{1} & 1& 1&\Delta\\ 
\tilde Q   &  \overline {\mathbf{N}_c}  &\mathbf{1} &\overline {\mathbf{N}_f}  & -1& 1&\Delta\\ 
\end{array}
\end{equation}

The dual model is a $U(\widetilde N_c)_0 \times U(1)_0$ gauge theory, $\widetilde{N}_c \equiv N_f - N_c$, with a mixed CS term
between the center $U(1) \subset U(\widetilde N_c)$ and the other $U(1)$ gauge factor.
The matter content is summarized in the following table:
 \begin{equation}
 \label{tablech}
\begin{array}{c|cccccc}
          &   U(\widetilde N_c) \times U(1)   &   SU(N_f)_L & SU(N_f)_R & U(1)_B &U(1)_A & U(1)_R \\
             \hline
\rule{0pt}{15pt} q           &  \widetilde{\mathbf{N}}_{c,0}    & \overline {\mathbf{N}_f}   &\mathbf{1} & 0& -1&1-\Delta\\ 
\tilde q   &  \overline {\widetilde{\mathbf{N}}_c }_{,0}  & \mathbf{1} &\mathbf{N}_f &0& -1&1-\Delta\\ 
M & \mathbf{1}_0&\mathbf{N}_f&\overline {\mathbf{N}_f} &0&2&2 \Delta\\
T_+       &  \mathbf{1}_{1}    & \mathbf{1}   &\mathbf{1}& N_c& N_f &N_f(1-\Delta)-N_c+1\\ 
T_-   &   \mathbf{1}_{-1}  & \mathbf{1}&\mathbf{1}&-N_c& N_f&N_f(1-\Delta)-N_c+1\\ 
\end{array}
\end{equation}
The superpotential is
\begin{equation}
W = M q \tilde q  
+ T_- t_+  + T_+ t_-
\end{equation}
where $t_\pm $ and are  the monopole and antimonopole of the $U(\widetilde N_c)$  gauge group, respectively.
 The fields $T_\mp$ are charged with charge $-1$ and $+1$ (respectively) with respect to the 
 $U(1)$ gauge group.
Observe that the normalization of the baryonic symmetry is conventional, because it 
can be combined arbitrarily with the gauge symmetry.

A very useful way to understand this freedom consists of studying this duality at the level of the three-sphere partition function.
The identity between the three-sphere partition functions of the electric of the magnetic theory can be obtained starting from the one for the $U(N_c)$ Aharony duality, namely
\begin{align}\label{eq:Uaharony}
Z_\text{ele}^{U(N_c)}(\mu;\nu;\lambda) 
=& \
\prod_{a,b=1}^{N_f} \Gamma_h(\mu_a+\nu_a)
Z_\text{mag}^{U(N_f-N_c)}(\omega-\nu;\omega-\mu;-\lambda) \cdot  \nonumber \\
&\ \cdot \Gamma_h \left(\pm \frac{\lambda}{2} - m_A N_f +\omega (N_f-N_c +1) \right)\ ,
\end{align}
where 
\begin{equation}
\label{defPF}
Z^{U(N)}(\mu;\nu;\lambda) 
\equiv
\int \prod_{i=1}^{N} d \sigma_i\, e^{i \pi \lambda \sum_i\sigma_i}
\prod_{a=1}^{N_f} \left(\Gamma_h (\mu_a + \sigma_i) 
 \Gamma_h (\nu_a - \sigma_i) \right)
\!\!\!\!\!
 \prod_{1\leq i < j \leq N} 
 \!\!\!\!\!
 \Gamma_h^{-1} (\pm (\sigma_i - \sigma_j))\ .
\end{equation}
In this definition $\Gamma_h$ are hyperbolic hypergeometric gamma functions
and they correspond to the one-loop determinants for the matter and vector multiplets obtained
from localization.
We refer the reader to \cite{VanDeBult} for a definition of these functions 
and to \cite{Benini:2011mf} for a more physical approach.
Furthermore in formula (\ref{defPF})  we used the shortcut $\Gamma_{h} ( \pm x)\equiv \Gamma_{h} (x) \Gamma_{h} (-x)$. 
The vectors $\mu_a$ and $\nu_a$, collecting the masses of the fundamental and antifundamental fields respectively, can be further constrained by the symmetries of the problem.
In this case the constraints $\sum_{a=1}^{N_f} \mu_a = \sum_{a=1}^{N_f} \nu_a = N_f m_A$ reproduce the 
presence of the axial mass (in addition to the nonabelian flavor symmetries).
In this notation the R-charge is hidden in the imaginary part of these masses, while the parameter 
$\lambda$ is a Fayet--Iliopoulos (FI) term, corresponding to the real mass of the topological symmetry.
Another useful term often appearing in the partition function, and that will play a prominent role in this paper, 
is the CS contribution at level $k$.  It corresponds to a term $e^{-i \pi k \sum_i \sigma_i^2}$
coming from the classical action in the localization procedure.

We now add a term $\frac{1}{2}e^{-i \pi \lambda N_c m_B}$ to both sides of \eqref{eq:Uaharony}, and gauge the topological $U(1)$ symmetry
by integrating over $\lambda$.
On the electric side we can first shift $\sigma_i \rightarrow \sigma_i+m_B$ 
and then use the identity 
\begin{equation}
\label{delta}
\frac{1}{2} \int d\lambda e^{  i \pi \lambda \sum_{i=1}^{N_c} \sigma_i}= \delta \left(\sum_{i=1}^{N_c} \sigma_i \right)
\end{equation}
such that the electric theory has $SU(N_c)$ gauge group.

On the dual side we cannot use the same trick because the monopoles are charged under the 
topological symmetry. Moreover, to be consistent with the baryonic charges of the dual quarks in  
(\ref{tablech}), we do not shift the gauge symmetry by $m_B$.
Then the magnetic side has partition function $Z_\text{mag}(\omega-\nu;\omega-\mu;-2 \xi)$ reading
\begin{align}
\label{eq:Z_mag-SQCD}
Z_\text{mag} =&
\prod_{a,b=1}^{N_f} \Gamma_h(\mu_a+\nu_b)
\int d\xi \,\Gamma_h \left( \pm \xi  - m_A N_f +\omega (\widetilde N_c+1) \right) \cdot \nonumber \\
&\int \prod_{i=1}^{\widetilde N_c} d\tilde \sigma_i \prod_{i<j} \Gamma_h( \pm  (\tilde \sigma_i -\tilde \sigma_j))^{-1} \, e^{-2 i \pi \xi  \left(\sum_i \tilde \sigma_i+ ({N_c}/{\widetilde N_c} )m_B\right)} \cdot \nonumber \\
&\cdot \prod_{i=1}^{\widetilde N_c}\left( \prod_{a=1}^{N_f} \Gamma_h(\omega-\mu_a - \tilde \sigma_i) \Gamma_h(\omega-\nu_a+\tilde \sigma_i) \right)\ ,
\end{align}
where we have used the substitution $\lambda = 2\xi$ in order to have a proper charge normalization for the 
fields  $T_\pm$  in the $U(1)$ gauge sector.
It is also possible to apply mirror symmetry locally to the $U(1)$ gauge sector
associated to the gauging of the topological symmetry in this magnetic phase.
This sector corresponds to SQED with one flavor and
local mirror symmetry introduces a superpotential
for its monopole and antimonopole, identified with the baryon $b$ and the antibaryon $\tilde b$ of the dual
$U(N_f-N_c)$ SQCD. These fields 
interact through a new singlet $Y$, 
corresponding to the monopole operator of the electric model.
The dual model in this case has superpotential 
\begin{equation}
\label{WARSW}
W = M q \tilde q + Y b \tilde b + t_+ + t_-\  .
\end{equation}
In this way one obtains the construction of \cite{Aharony:2013dha} 
for the $SU(N_c)$ version of Aharony duality.
The dual partition function becomes 
\begin{align}
\label{eq:Z_mag-SQCD2}
Z_\text{mag}= &\ 
\prod_{a,b=1}^{N_f} \Gamma_h(\mu_a+\nu_b)
\Gamma_h \Big(
2\omega (N_f-N_c+1) -\sum(\mu_a+\nu_a) \Big) \cdot \nonumber \\
&\cdot  \int
\frac{\prod_{i=1}^{\widetilde N_c}  d\tilde \sigma_i}{\prod_{i<j} \Gamma_h( \pm  (\tilde \sigma_i -\tilde \sigma_j))} \Gamma_h \left(\pm \sum_{i=1}^{\widetilde N_c}  \tilde \sigma_i + \frac{1}{2} \sum(\mu_a+\nu_a)
-\omega \widetilde N_c \right) \cdot \nonumber \\
&\cdot \prod_{i=1}^{\widetilde N_c} \left( \prod_{a=1}^{N_f} \Gamma_h \left(\omega-\mu_a - \tilde \sigma_i+m_B \frac{N_c}{\widetilde N_c}\right) \Gamma_h \left(\omega-\nu_a+\tilde \sigma_i-m_B \frac{N_c}{\widetilde N_c}\right)\right) \ .
\end{align}
We will mostly focus on the first version of the duality, with the extra $U(1)$ sector.
This is because it is easily generalizable to the case with adjoint matter.\footnote{In order to study the case with adjoint matter without gauging the topological 
$U(1)$ symmetry we should reproduce the relation between (\ref{eq:Z_mag-SQCD})
and (\ref{eq:Z_mag-SQCD2}) by applying an opportune version of mirror symmetry.
In such a case, as we will review later on in the paper,  there are dressed monopoles
in the this $U(1)$ sector. It follows that one has to apply local mirror symmetry to a $U(1)$ sector 
with $n$ flavors, where $n$ corresponds to the power in the superpotential $W = \Tr X^{n+1}$.}
In appendix \ref{appB} we will discuss the flows in the magnetic dual identified by 
(\ref{eq:Z_mag-SQCD2}) in the case without adjoint matter, in order to compare with the results discussed here and in \cite{Aharony:2014uya,Nii:2020xgd}.

The next step consists of performing a real mass flow in order to obtain three families of dualities, that we classify according to the relative value of $\Delta F = |N_f-N_a|$ with respect to $2k$, and dub as follows:
\begin{itemize}
\item $\Delta F  < 2k$: the $SU$ \emph{generalization} of the $[p,q]$ case;
\item $\Delta F  = 2k$: the $SU$ \emph{generalization} of the $[p,0]$ case;
\item $\Delta F  > 2k$: the $SU$ \emph{generalization} of the $[p,q]^*$ case.
\end{itemize}
Let us discuss the three cases separately.
%
%
%
%
%
\subsection{\texorpdfstring{$\Delta F  < 2k$: $SU$ generalization of the $[p,q]$ case}{Delta F  < 2k: SU generalization of the [p,q] case}}
\label{pqPP}
%
%
%
%
%
In this case we assign a positive large real mass to 
$N_f-N_f^{(1)}$ fundamentals and a positive large real mass to 
$N_f-N_f^{(2)}$ antifundamentals.
The electric theory is $SU(N_c)_k$ with $N_f^{(1)}$ fundamentals and 
$N_f^{(2)}$ antifundamentals.
The CS level generated by this real mass flow is $k = N_f-\tfrac{1}{2}(N_f^{(1)}+N_f^{(2)} )$
and $\Delta F = |N_f^{(1)}-N_f^{(2)}|  < 2k =2N_f- N_f^{(1)}-N_f^{(2)} $.

On the magnetic side the situation is more complicated. First of all, because of the normalization of the baryonic symmetry, we need to consider a nonzero vacuum for the scalar in the vector multiplet.
Furthermore we also need to shift the scalar in the vector multiplet of the gauged topological symmetry.
We are left with
 $U(k+\tfrac{1}{2}(N_f^{(1)}+N_f^{(2)})-N_c)_{-k} \times U(1)$ gauge symmetry with a mixed CS level, at level $1$, 
between the $U(1)$ symmetries.
There are $N_f^{(1)}$ dual antifundamentals and $N_f^{(2)}$ dual fundamentals and 
there is a superpotential $W = M q \tilde q$. 
Furthermore in presence of nonzero CS levels  there are also nontrivial contact terms
in the two-point functions of the global symmetry currents \cite{Closset:2012vp,Closset:2012vg}.
Their difference is physical and it appears explicitly in the partition function.
This can be checked in all the dualities studied in this paper.

The real mass flow just described can be reproduced on the three-sphere partition function. 
It corresponds to assigning the following mass parameters:
\begin{equation}
\label{pqrm}
\left\{
\begin{array}{llll}
m_A & \to & m_A+ \frac{2N_f -N_f^{(1)}-N_f^{(2)}}{2 N_f} s;& \\
m_B & \to & m_B-\frac{N_f^{(1)}-N_f^{(2)}}{2N_f}s; & \\
m_a & \to & m_a-\frac{N_f-N_f^{(1)}}{N_f} s,& a=1,\dots N_f^{(1)};  \\
m_a & \to & m_a+\frac{N_f^{(1)}}{N_f}s ,& a= N_f^{(1)}+1,\dots N_f;  \\
n_a & \to & n_a -\frac{N_f-N_f^{(2)}}{N_f} s,& a=1,\dots N_f^{(2)};  \\
n_a & \to & n_a+\frac{N_f^{(2)}}{N_f} s ,& a= N_f^{(2)}+1,\dots N_f ; \\
\tilde{\sigma}_i & \to &\tilde \sigma_i-\frac{N_f^{(1)}-N_f^{(2)}}{2 N_f }s;  & \\
\xi & \to & \xi+\frac{N_f^{(1)}-N_f^{(2)}}{2} s.   \\
\end{array}
\right.
\end{equation}
The real mass flow consists of studying the large 
$s$ limit on both sides of the identity 
between the electric and the magnetic $SU(N)$ 
 non-chiral partition functions.
The divergent contributions cancel between the electric and 
the magnetic phase and we obtain a new identity
$Z_\text{ele} = Z_\text{mag}$ that reproduces the duality 
discussed above.
The electric and the magnetic partition functions read respectively
\begin{align}
 \label{eq:Z_ele-SQCD-SU-pq}
Z_\text{ele} =&\ 
\frac{1}{N_c!}
\int \prod_{i=1}^{N_c} 
d\sigma_i \, \delta \left(\sum_{i=1}^{N_c} \sigma_i \right) \prod_{i<j} \Gamma_h( \pm  (\sigma_i -\sigma_j))^{-1} 
\,e^{-i \pi k \sum_i \sigma_i^2} \cdot   \nonumber \\
 & \cdot \prod_{i=1}^{N_c}\left( \prod_{a=1}^{N_f^{(1)}} \Gamma_h(\mu_a +m_B+ \sigma_i)  
 \cdot
\prod_{b=1}^{N_f^{(2)}} \Gamma_h(\nu_a-m_B-\sigma_i)\right)\ ,
\end{align}
and
\begin{align}
\label{eq:Z_mag-SQCD-SU-pq}
Z_\text{mag} =&\ \frac{e^{i \pi \phi}}{ \widetilde N_c!}
\prod_{a=1}^{N_f^{(1)}} \prod_{b=1}^{N_f^{(2)}} \Gamma_h(\mu_a+\nu_b)
\int d\xi  
e^{ i \pi \xi( \xi+2  m_B N_c)} \int
\prod_{i=1}^{ \widetilde N_c}  d\tilde \sigma_i 
 \, e^{i  \pi  (k  \sum_i \tilde \sigma_i^2-(\eta-2 \xi ) \sum_i\tilde \sigma_i)} 
\cdot 
\nonumber \\
&\cdot
\prod_{i=1}^{ \widetilde N_c} 
\left(
\prod_{a=1}^{N_f^{(1)}} \Gamma_h(\omega-\mu_a - \tilde \sigma_i)
\cdot
\prod_{b=1}^{N_f^{(2)}} \Gamma_h(\omega-\nu_b+\tilde \sigma_i)
\right)
 \prod_{i<j} 
\Gamma_h( \pm  (\tilde \sigma_i -\tilde \sigma_j))^{-1}
\ ,
\end{align}
where
\begin{equation}
\eta \equiv \left(N_f^{(1)}-N_f^{(2)}\right) m_A \ .
\end{equation}
The complex exponent $\phi$ is needed to match the CS contact terms 
as discussed above. It reads
\begin{align} \label{eq:phiSUpq}
\phi =& \ 
m_B^2 N_c (N_f^{(1)}+N_f^{(2)}+4 k)+2  \omega  m_B(N_f^{(1)}-N_f^{(2)}) N_c -2 m_A m_B (N_f^{(1)}-N_f^{(2)})  N_c+
\nonumber \\ & 
-2 \omega  m_A [2 N_f^{(1)} N_f^{(2)}-(N_f^{(1)}+N_f^{(2)}) N_c]-m_A^2 [2 (N_f^{(1)}+N_f^{(2)}) k+(N_f^{(1)}-N_f^{(2)})^2]
+
\nonumber \\ & +
\omega ^2 [2 N_c^2-3 (N_f^{(1)}+N_f^{(2)}) N_c-4 k N_c+(N_f^{(1)}+N_f^{(2)}+2 k)^2+N_f^{(1)} N_f^{(2)}]
+
\nonumber \\ 
&  -(N_f^{(1)}+2 k) \sum _{a=1}^{N_f^{(1)}} m_a^2-(N_f^{(2)}+2 k) \sum _{b=1}^{N_f^{(2)}} n_b^2 \ .
\end{align}
We can also compare this result with the one expected from the duality obtained in \cite{Aharony:2014uya}.
This can be done by computing the gausssian integral over $\xi$ in formula (\ref{eq:Z_mag-SQCD-SU-pq}).
Completing the square with a term $(\sum_{i=1}^{\tilde N_c} \sigma_i+m_B N_c )^2$ 
yields
\begin{align}
\label{eq:Z_mag-SQCD-SU-pq3}
Z_\text{mag} =&\  \frac{e^{i \pi \phi}}{ \widetilde N_c!}
\prod_{a=1}^{N_f^{(1)}} \prod_{b=1}^{N_f^{(2)}} \Gamma_h(\mu_a+\nu_b)
\int  \frac{\prod_{i=1}^{ \widetilde N_c} d\tilde \sigma_i}{\prod_{i<j} \Gamma_h( \pm  (\tilde \sigma_i -\tilde \sigma_j))}
\, e^{i  \pi  (k  \sum_i \tilde \sigma_i^2 - \eta \sum_i \tilde \sigma_i)} \cdot \nonumber \\
&\cdot e^{i \pi \left( \sum_i \tilde \sigma_i+m_B N_c  \right)^2}
\prod_{i=1}^{ \widetilde N_c} \left(
\prod_{a=1}^{N_f^{(1)}} \Gamma_h(\omega-\mu_a - \tilde \sigma_i)
\cdot
\prod_{b=1}^{N_f^{(2)}} \Gamma_h(\omega-\nu_b+\tilde \sigma_i)\right) \ .
\end{align}
Using the results of appendix \ref{appC}, we observe that 
the magnetic gauge group is $SU(\widetilde N_c)_{-k} \times U(1)_{\widetilde N_c - k}$,
consistently with the duality discussed in \cite{Aharony:2014uya}.

%
%
%
%
%
\subsection{\texorpdfstring{$\Delta F=2k$: $SU$ generalization of the $[p,0]$ case}{Delta F = 2k: SU generalization of the [p,0] case}}
\label{p0PP}
%
%
%
%
%
In this case we assign a positive large real mass to 
$N_f-N_f^{(1)}$ antifundamentals.
The electric theory is $SU(N_c)_k$ with $N_f$ fundamentals and 
$N_f^{(1)}$ antifundamentals.
The CS level generated by this real mass flow is $k = \tfrac{1}{2}(N_f-N_f^{(1)} )$
and $\Delta F = |N_f-N_f^{(1)}|  = 2k$.
On the magnetic side we take again  a nonzero vacuum for the scalar in the vector multiplet,
and we shift the scalar in the vector multiplet of the gauged topological symmetry.
We are left with $U\big(\widetilde{N}_c\equiv k+\tfrac{1}{2}(N_f+N_f^{(1)})-N_c\big)_{-k} \times U(1)$ gauge symmetry with a mixed CS level, at level $1$, between the $U(1)$ symmetries.
In the $U(1)$ sector the field with charge $+1$ is massive and is integrated out.
The field with charge $-1$ is massless, because the shift induced by 
real masses is canceled against the one of the real scalar of 
the gauged topological $U(1)$.
In the $U(\widetilde{N}_c)$ sector there are $N_f$ dual antifundamentals and $N_f^{(1)}$ dual fundamentals interacting  through the superpotential $W = M q \tilde q +{T_-} t_+$.

One of the main differences with respect to the analysis of \cite{Aharony:2014uya}
is the fact that, starting from the duality for $SU(N_c)$ SQCD at vanishing CS level, we can distinguish the $\Delta F<2k$ case (in other words, the $[p,q]$ case of \cite{Benini:2011mf}) from the 
$\Delta F = 2k$ case (the $[p,0]$ case). 
Another important difference with \cite{Aharony:2014uya} is that 
here there is  charged matter field in the $U(1)$ gauge sector.

The real mass flow just described can be reproduced on the three-sphere partition function. 
This corresponds to assigning the following mass parameters:
\begin{equation}
\label{p0rm}
\left\{
\begin{array}{llll}
m_A & \to &m_A+\frac{N_f-N_f^{(1)}}{2N_f}s;& \\
m_B & \to &m_B- \frac{N_f-N_f^{(1)}}{2N_f} s; & \\
n_a  & \to &n_a -\frac{N_f-N_f^{(1)}}{N_f}s,& a=1,\dots N_f^{(1)};  \\
n_a & \to &n_a+\frac{N_f^{(1)}}{N_f}s, & a=N_f^{(1)}+1,\dots,N_f-N_f^{(1)};\\
\tilde{\sigma}_i & \to &\tilde \sigma_i- \frac{N_f-N_f^{(1)}}{2N_f} s;  & \\
\xi & \to & \xi+\frac{N_f-N_f^{(1)}}{2}s.& \\
\end{array}
\right.
\end{equation}
The real mass flow consists of studying the large 
$s$ limit on both sides of the identity 
between the electric and the magnetic $SU(N)$ 
 non-chiral partition functions.
The divergent contributions cancel between the electric and 
the magnetic phase, and we obtain a new identity
$Z_\text{ele} = Z_\text{mag}$, where
\begin{align}
 \label{eq:Z_ele-SQCD-SU-p0}
Z_\text{ele} =& \ 
\frac{1}{N_c!}
\int \prod_{i=1}^{N_c} d\sigma_i \,
 \delta \left(\sum_{i=1}^{N_c} \sigma_i\right)
\prod_{i<j} \Gamma_h( \pm  (\sigma_i -\sigma_j))^{-1}\, e^{-i \pi k \sum_i \sigma_i^2} \cdot  \nonumber \\
 &\cdot 
 \prod_{i=1}^{N_c} \left(
 \prod_{a=1}^{N_f} \Gamma_h(\mu_a + \sigma_i+m_B)  
 \cdot
\prod_{b=1}^{N_f^{(1)}} \Gamma_h(\nu_a-\sigma_i-m_B)\right) \ ,
\end{align}
and
\begin{align}
\label{eq:Z_mag-SQCD-SU-p0}
Z_\text{mag} =&\ \frac{e^{\frac{i \pi}{2} \phi}}{ \widetilde N_c!}
\prod_{a=1}^{N_f} \prod_{b=1}^{N_f^{(1)}} \Gamma_h(\mu_a+\nu_b) \cdot 
\int d\xi 
e^{i \pi \xi \rho } \Gamma_h( N_f(\omega-m_A)-\omega(N_c-1) + \xi ) \nonumber \\ 
& \int \frac{\prod_{i=1}^{  \widetilde N_c} d\tilde \sigma_i}{\prod_{i<j} \Gamma_h( \pm  (\tilde \sigma_i -\tilde \sigma_j))} \, 
e^{i \pi k \sum_i\tilde \sigma_i^2} e^{i \pi \frac{\xi^2}{2} - i \pi (\eta-2\xi) \sum_i \tilde \sigma_i}  \cdot
\nonumber \\ &
\cdot 
\prod_{i=1}^{ \widetilde N_c}\left(  
\prod_{a=1}^{N_f} \Gamma_h(\omega-\mu_a- \tilde \sigma_i)
\cdot
\prod_{b=1}^{N_f^{(1)}} \Gamma_h(\omega-\nu_b+\tilde \sigma_i)\right) \ ,
\end{align}
with
\begin{equation}
\eta \equiv  2 k m_A, 
\quad 
\rho \equiv  N_f (m_A-\omega) +N_c (2 m_B+\omega) \ .
\end{equation}
The complex exponent $\phi$ is needed to match the CS contact terms and reads
\begin{align}
\phi = & \
2 N_c \left(k \left(m_B+\omega \right) \left(m_B-2 m_A+\omega \right)+\omega  N_f \left(m_A-\omega \right)\right)
+   \\
&
+N_f \left(\omega -m_A\right) \left(m_A \left(6 k-N_f\right)+\omega  \left(N_f-2 k\right)\right)+\omega ^2 N_c^2
-2 k \sum _{a=1}^{N_f} m_a^2\ .
\nonumber
\end{align}

We conclude the analysis by matching the result obtained here with the one discussed recently in \cite{Nii:2020xgd}.
In order to compare the result we need to get rid of the integral over the $U(1)$ sector identified by $\xi$.
This is done by using the formula \cite[Eqs. (6.10) \& (6.11)]{Benini:2011mf}
\begin{equation}
\label{half}
 \int d \xi
 e^{i \pi  \left( \frac{\xi ^2}{2} +\xi (2 \lambda +\omega -m ) \right)}
\Gamma_h (m+\xi)
 =
 e^{i \pi  \left(m^2-m (\lambda +\omega )-\frac{\lambda ^2}{2}\right)}
  \Gamma_h( \omega + \lambda - m)\ ,
\end{equation}
corresponding to the application of local mirror symmetry on this $U(1)_{-1/2}$ sector with one 
negatively charged chiral superfield (the so-called ``half mirror symmetry'' of \cite{Dorey:1999rb}).\footnote{Notice that in \eqref{half} we have one positively, rather than negatively, charged chiral. This amounts to changing the phase appearing on the RHS of \cite[Eq. (6.11)]{Benini:2011mf}, in such a way that \eqref{half} holds.}
Using (\ref{half}) we obtain
\begin{align}\label{eq:halfmirrorpost}
& \int d\xi 
e^{i \pi \left(\frac{\xi^2}{2} +  \xi
( N_f (\omega -m_A) -N_c (2 m_B+\omega)+ 2 \sum_{i=1}^{\tilde N_c}\tilde \sigma_i) \right)}
\Gamma_h( N_f(\omega-m_A)-\omega(N_c-1) +\xi )\  = \nonumber \\
& e^{i \pi  \left[
N_f \left(m_A-\omega \right) \left(N_c \left(m_B+2 \omega \right)-\omega \right)+N_f^2 \left(\omega -m_A\right){}^2-N_c \left(\frac{1}{2} m_B^2 N_c-\omega  \left(N_c-1\right) \left(m_B+\omega \right)\right)
\right]} \, \cdot
\nonumber\\
&\cdot 
e^{i \pi  \left((N_f (m_A-\omega )-N_c \left(m_B-\omega \right)-\omega) \sum _{i} \tilde \sigma _i -\frac{1}{2} \left(\sum _{i} \tilde \sigma _i\right)^2\right)}
\Gamma_h \left(N_f (m_A\!-\!\omega)\!+\!N_c (m_B\!+\!\omega) \!+\!\sum_{i=1}^{\widetilde N_c} \tilde \sigma_i \right)\ .
\end{align}
We then plug this integral into (\ref{eq:Z_mag-SQCD-SU-p0}), thus reproducing the dual phase discussed in \cite{Nii:2020xgd}
(see appendix \ref{sub:ARSWp0} for details on the flow studied in \cite{Nii:2020xgd} on the partition function).
Indeed using the results of appendix \ref{appC} the magnetic gauge group is found to be $SU(\widetilde N_c)_{-k} \times U(1)_{-k+\widetilde N_c/2}$, with superpotential 
\begin{equation}
W = M q \tilde q + t_+\ .
\end{equation}
The $\Gamma_h$ on the RHS of \eqref{eq:halfmirrorpost} corresponds to the contribution of the charged baryon $b$ in \eqref{WARSW} that remains massless after the real mass flow.

%
%
%
%
%
\subsection{\texorpdfstring{$\Delta F >2k$: $SU$ generalization of the $[p,q]^*$ case}{Delta F > 2k: SU generalization of the [p,q]* case}}
\label{pq*PP}
%
%
%
%

In this case we assign a positive large real mass to 
$N_f^{(1)}$ antifundamentals and a negative large real mass to 
$N_f^{(2)}$ antifundamentals.
The electric theory is $SU(N_c)_k$ with $N_f$ fundamentals and 
$N_a$ antifundamentals with $N_a=N_f-N_f^{(1)}-N_f^{(2)}$.
The CS level generated by this real mass flow is $k = \frac{1}{2}(N_f^{(1)}-N_f^{(2)} )$
and $\Delta F = N_f-N_a = N_f^{(1)}+N_f^{(2)} >2k = N_f^{(1)} - N_f^{(2)}$.
On the magnetic side the situation is more complicated. First of all, because of the normalization of the baryonic symmetry, we need to consider a nonzero vacuum for the scalar in the vector multiplet.
Furthermore we also need to shift the scalar in the vector multiplet of the gauged topological symmetry.
We are left with $U(N_f-N_c)_{-k} \times U(1)$ gauge symmetry with a mixed CS level, at level $1$, between the two $U(1)$ symmetries.
In the $U(1)$ sector obtained by gauging the topological symmetry both the fields
with charge $+1$ and $-1$ are massive.
In the $U(\widetilde{N}_c)$ sector  there are $N_f$ dual antifundamentals and $N_a$ dual fundamentals and  there is a superpotential $W = M q \tilde q$.

By a parity transformation one can also define the case where $N_f<N_a$. In general one has
an electric $U(N_c)_k$ theory with
$N_f$ fundamentals, $N_a$ antifundamentals, and $|N_f-N_a| > k$,  dual to  
$U( \, \max (N_f,N_a) -N_c)_{-k}$ with
$N_f$ dual antifundamentals, $N_a$ dual fundamentals, and the superpotential 
$W = M q \tilde q$.

The real mass flow just described can be reproduced on the three-sphere partition function. 
This corresponds to assigning the following mass parameters:
\begin{equation}
\label{pq*rm}
\left\{
\begin{array}{llll}
m_A & \to & m_A+\frac{N_f^{(1)}-N_f^{(2)}}{2N_f}s;& \\
m_B & \to &m_B-\frac{N_f^{(1)}-N_f^{(2)}}{2N_f}s; & \\
n_a & \to &n_a - \frac{N_f^{(1)}-N_f^{(2)}}{N_f}s,& a=1,\dots, N_f-N_f^{(1)}-N_f^{(2)} =N_a; \\
n_a & \to &n_a+\frac{N_f-N_f^{(1)}+N_f^{(2)} }{N_f} s,& a=N_a+1,\dots,N_f^{(1)}+N_a = N_f-N_f^{(2)};\\
n_a & \to &n_a-\frac{N_f+N_f^{(1)}-N_f^{(2)} }{N_f} s,& a= N_f-N_f^{(2)}+1,\dots,N_f;\\
\tilde{\sigma}_i & \to &\tilde \sigma_i-\frac{N_f^{(1)}-N_f^{(2)}}{2N_f}s;  & \\
\xi & \to & \xi+2(N_f^{(1)}+N_f^{(2)})s.& \\
\end{array}
\right.
\end{equation}
The real mass flow consists of studying the large 
$s$ limit on both sides of the identity 
between the electric and the magnetic $SU(N)$ 
 non-chiral partition functions.
The divergent contributions cancel between the electric and 
the magnetic phase and we obtain a new identity
$Z_\text{ele} = Z_\text{mag}$, where
\begin{align}
\label{eq:Z_ele-SQCD-SU-pq*}
Z_\text{ele} =&\  \frac{1}{N_c!} 
\int \prod_{i=1}^{N_c} d\sigma_i 
\prod_{i<j} \Gamma_h( \pm  (\sigma_i -\sigma_j))^{-1}\, \delta\left( \sum_{i=1}^{N_c}\sigma_i\right) \, e^{ - i \pi k \sum_i \sigma_i^2} \cdot \nonumber \\
&\cdot  \prod_{i=1}^{N_c} \left(
\prod_{a=1}^{N_f} \Gamma_h(\mu_a + \sigma_i + m_B)  
\cdot
\prod_{b=1}^{N_a} \Gamma_h(\nu_b  -  \sigma_i -  m_B)\right) \ ,
\end{align}
and 
\begin{align}
\label{eq:Z_mag-SQCD-SU-pq*}
Z_\text{mag} =&\ \frac{e^{i \pi \phi}}{ \widetilde N_c!}
\prod_{a=1}^{N_f}\prod_{b=1}^{N_a} \Gamma_h(\mu_a+\nu_b)
\int d\xi 
e^{-2i \pi \xi \rho } \int \prod_{i=1}^{ \widetilde N_c} d\tilde \sigma_i 
\prod_{i<j} 
\Gamma_h( \pm  (\tilde \sigma_i -\tilde \sigma_j))^{-1} \cdot
\nonumber \\
&\cdot 
e^{ i \pi \left( k \sum_i \tilde \sigma_i^2+ (\eta+2 \xi )\sum_i \tilde \sigma_i \right)} \prod_{i=1}^{ \widetilde N_c} \left(
\prod_{a=1}^{N_f} \Gamma_h(\omega-\mu_a - \tilde \sigma_i)
\prod_{b=1}^{N_a} \Gamma_h(\omega-\nu_a+\tilde \sigma_i)\right) \ ,
\end{align}
where
\begin{equation}
\eta \equiv 
2 k m_A,
\quad 
\rho = 2 (N_f  (m_A-\omega)+N_c (m_B +\omega))\ .
\end{equation}
The complex exponent $\phi$ is needed to match the CS contact terms, and reads:
\begin{equation}
\phi =
-k
\bigg(
\sum _{a=1}^{N_f} m_a^2
+
N_c \left(m_B+\omega \right) \left(2 m_A\!-\!m_B\!-\!\omega \right)+N_f \left(3 m_A^2-4 \omega  m_A+\omega ^2\right)
\!\!
\bigg)\ .
\end{equation}

\subsection*{Summary of this section} 

In this section we have derived the chiral dualities for 3d $\mathcal{N}=2$  $SU(N_c)$ SQCD 
borrowing the classification of 
\cite{Benini:2011mf} (where similar dualities were obtained for $U(N_c)$ SQCD).
The dualities found here have been obtained by matching the real mass flows between the electric and the magnetic phase 
of the duality of \cite{Park:2013wta}. We have studied this flow on the integral identities corresponding to the matching of the partition functions.
These results can be compared against the ones obtained by \cite{Aharony:2014uya}, where the real mass flow was performed starting from the duality for $SU(N_c)_k$ SQCD, i.e. the $SU(N_c)$ version of Giveon--Kutasov duality obtained in \cite{Aharony:2013dha}.
We have studied this flow on the partition function in appendix \ref{appAF}, matching the results with the ones of this section.
A further check of the flow and of the dualities studied here can be done using a different version of the dual model
of \cite{Park:2013wta}. As discussed above, this dual model was obtained in \cite{Aharony:2013dha}
and its partition function has been reported in formula (\ref{eq:Z_mag-SQCD2}).
We have performed this check in appendix \ref{appB}, discussing the matching for the $[p,q]$ and $[p,0]$ cases, whereas we have not found the flow that reproduces the $[p,q]^*$ case.

\begin{figure}[ht!]
\begin{center}
\includegraphics[scale=.26]{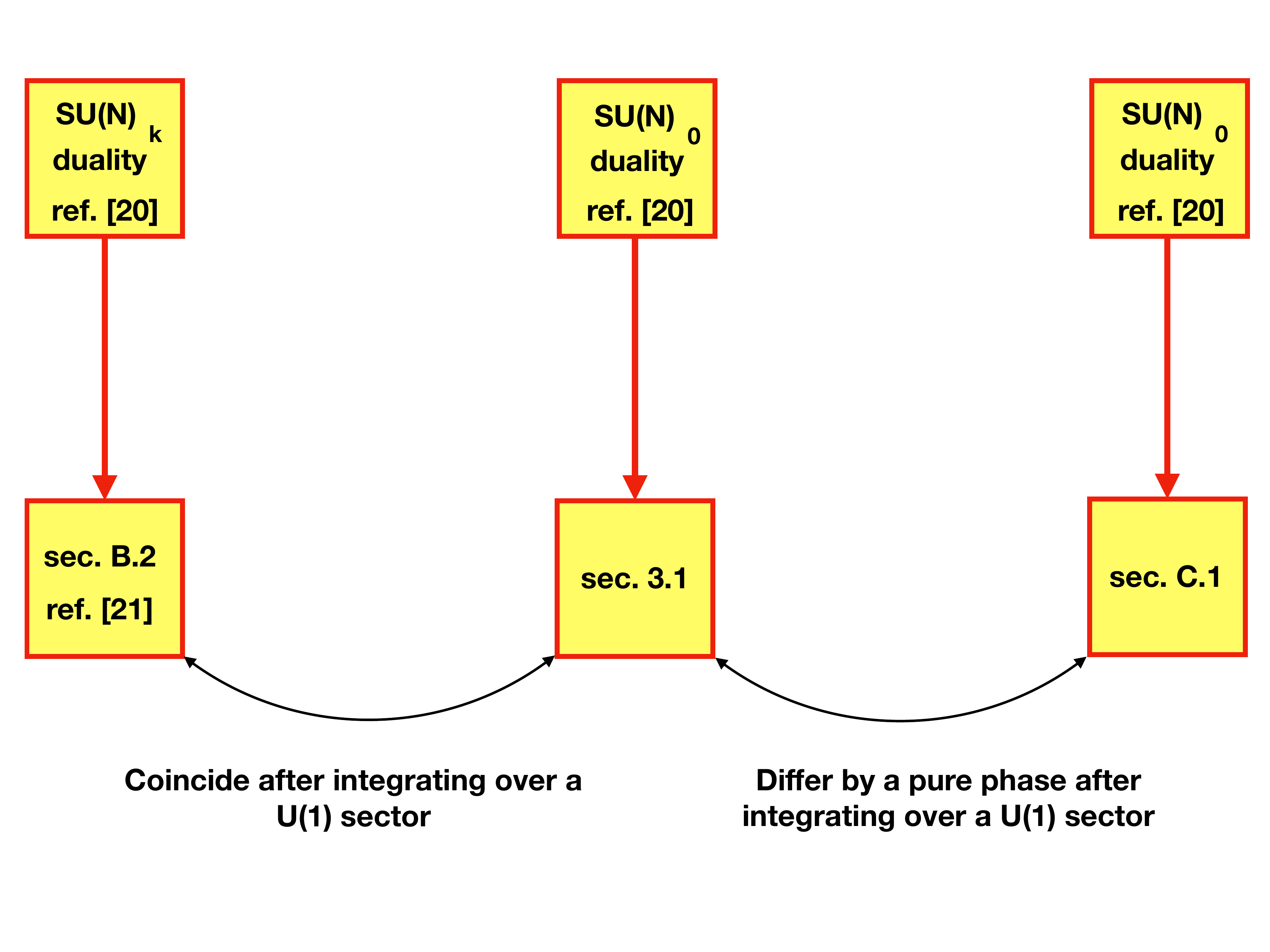}
\caption{The various real mass flows leading to the $[p,q]$ case studied in this paper and their relation.
We report the references corresponding to the UV duality where the RG flows considered here start.
}
\label{pqfig}
\end{center}
\end{figure}

\begin{figure}[ht!]
\begin{center}
\includegraphics[scale=.26]{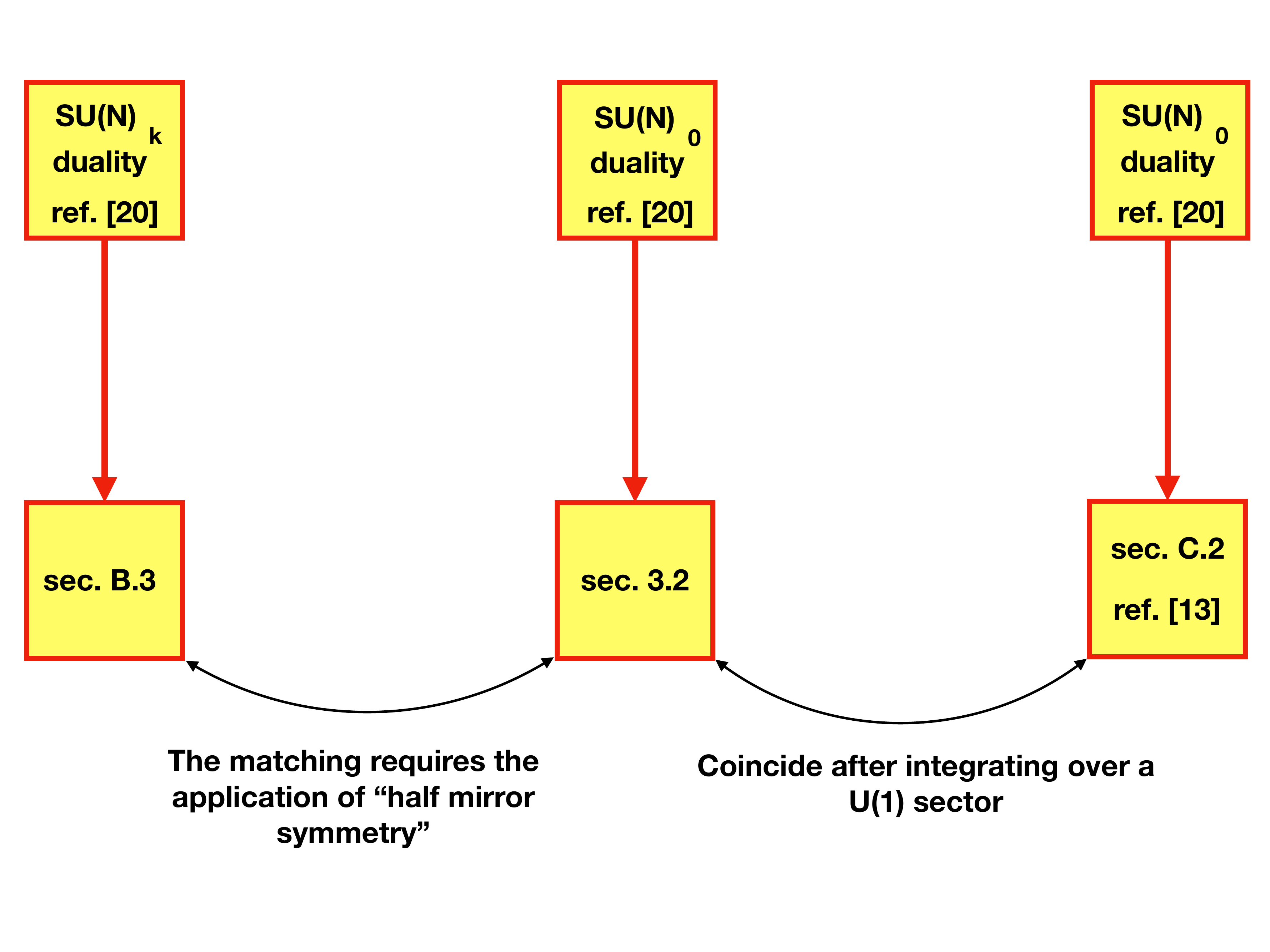}
\caption{The various real mass flows leading to the $[p,0]$ case studied in this paper and their relation.
We report the references corresponding to the UV duality where the RG flows considered here start.
}
\label{p0fig}
\end{center}
\end{figure}

\begin{figure}[ht!]
\begin{center}
\includegraphics[scale=.26]{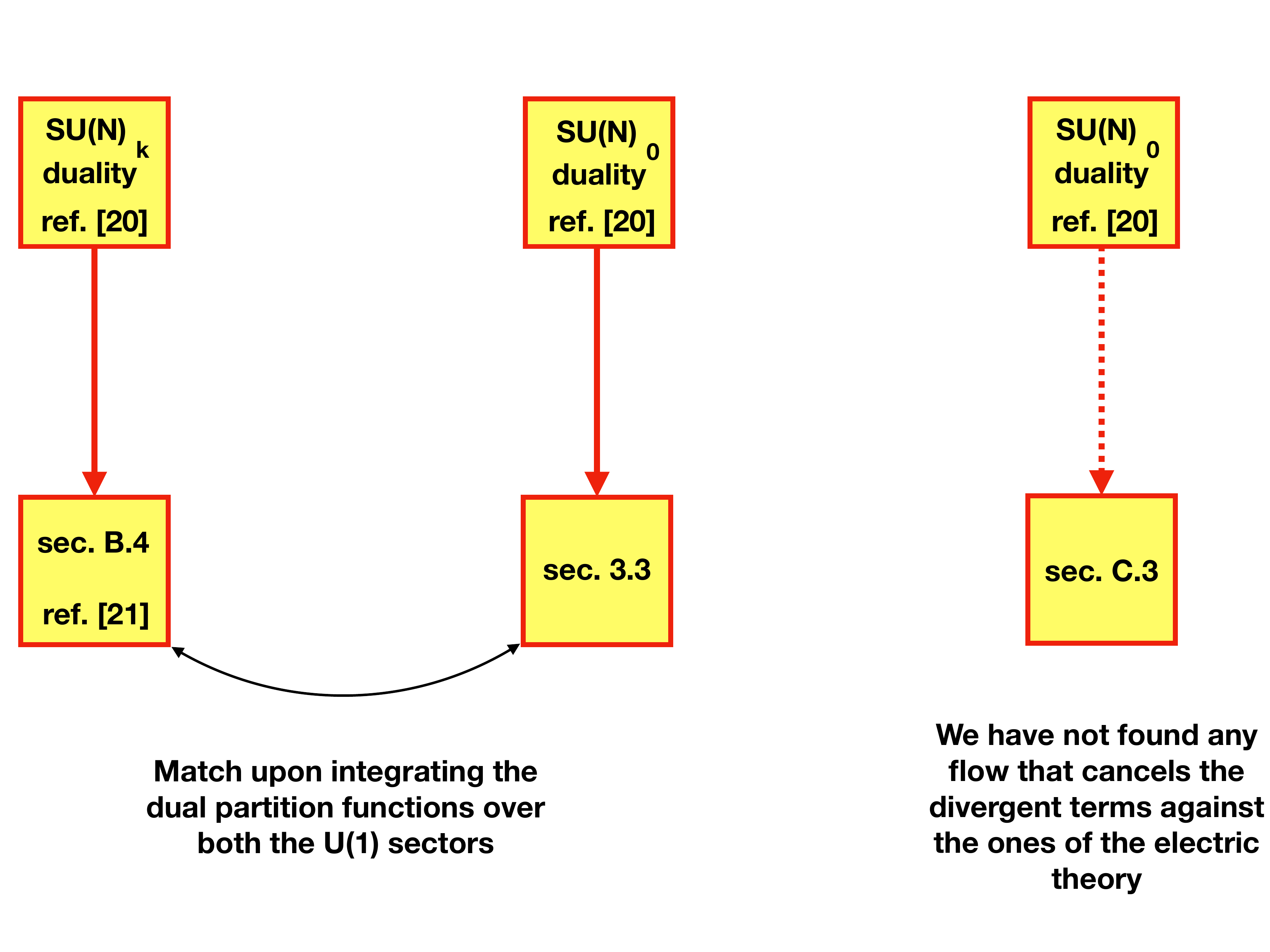}
\caption{The various real mass flows leading to the $[p,q]^*$ case studied in this paper and their relation.
We report the references corresponding to the UV duality where the RG flows considered here start.
}
\label{pq*fig}
\end{center}
\end{figure}

In order to simplify the reading of the appendices and the interrelation with the results obtained here 
we summarize the situation in figures \ref{pqfig}, \ref{p0fig} and \ref{pq*fig}, 
where the various flows and their relation is made explicit for the $[p,q]$, $[p,0]$, and $[p,q]^*$ case respectively.


\newpage
\section{\texorpdfstring{Dualities for $U(N_c)$ chiral adjoint SQCD}{Dualities for U(Nc) chiral adjoint SQCD}}
\label{sec:adjUN}

In this section we extend the chiral dualities of \cite{Benini:2011mf} 
to $U(N_c)$ gauge theories with adjoint matter.
Again we can distinguish three cases, identified by the relative size of
$\Delta F$ with respect to the CS level $2k$.
In order to furnish a uniform picture 
 we once again borrow the notation of \cite{Benini:2011mf},
and distinguish the three classes $[p,q]_\text{adj}$,
$[p,0]_\text{adj}$, and $[p,q]_\text{adj}^*$, where the subscript $\text{adj}$ denotes the 
presence of an adjoint multiplet.
Actually the last two cases analyzed in this section 
have already been studied in \cite{Hwang:2015wna},
while the third case has not appear in the literature yet (to the best of the authors' knowledge).

Let us first state the duality in the non-chiral case \cite{Kim:2013cma} (see also \cite{Nii:2014jsa}).
The electric theory is 3d $\mathcal{N}=2$ $U(N_c)$ 
SQCD, with $N_f$ pairs of fundamentals $Q$ and antifundamentals $\tilde Q$
and one adjoint $X$. There is a superpotential 
\begin{equation}
\label{aleadjW}
W = \Tr X^{n+1}\ .
\end{equation}
The fields are charged under the various symmetries as follows:
\begin{equation}
\begin{array}{c|cccccc}
          &   U(N_c)   &   SU(N_f)_L & SU(N_f)_R & U(1)_T &U(1)_A & U(1)_R \\
             \hline
Q           &  \mathbf{N}_c   & \mathbf{N}_f& \mathbf{1} & 0& 1&\Delta\\ 
\tilde Q   &  \overline {\mathbf{N}_c}  & \mathbf{1}&\overline {\mathbf{N}_f}  & 0& 1&\Delta\\ 
X &  \mathbf{Adj}  & \mathbf{1}&\mathbf{1}  & 0& 0&\frac{2}{n+1}\\ 
\end{array}
\end{equation}
The dual theory is 3d $\mathcal{N}=2$ $U(n N_f-N_c)$ 
SQCD, with $N_f$ pairs of dual fundamentals $q$ and dual antifundamentals $\tilde q$
and one adjoint $Y$. There are also $n N_f^2$ singlets $M_j = Q X^j \tilde Q$ with $j=0,\dots,n-1$,
corresponding to the (dressed) mesons, which are non-vanishing in the chiral ring of the electric theory,  and $2n$ singlets $T_{j,+}=T_+ X^j$ and $ T_{j,-}= T_{-} X^j$ 
with $j=0,\dots,n-1$, corresponding to the (dressed) monopoles and antimonopoles (respectively) of the electric theory acting as singlets in the magnetic phase.
The dual superpotential  is
\begin{equation}
\label{wKP}
W = \Tr Y^{n+1} + \sum_{j=0}^{n-1} (M_j q Y^j \tilde q+ T_{j,+}  t_{n-1-j,-}+ T_{j,-} t_{n-1-j,+})\ .
\end{equation}
The fields are charged under the various symmetries as follows:
\begin{equation} 
\begin{array}{c|cccccc}
             & U(\widetilde N_c)  &SU(N_f)_L & SU(N_f)_R & U(1)_T   & U(1)_A & U(1)_R  \\
\hline
\rule{0pt}{15 pt} q            &\widetilde{\mathbf{N}}_c&   \overline{\mathbf{N}_f}    &   \mathbf{1} &  0   &   -1    & \frac{2}{k+1}-\Delta    \\
\tilde q &\overline{\widetilde{\mathbf{N}}_c} &  \mathbf{1}    & \mathbf{N}_f&  0& - 1    & \frac{2}{k+1}-\Delta      \\
Y&\mathbf{Adj} &\mathbf{1} &\mathbf{1}&0&0&\frac{2}{k+1}\\
M_j&\mathbf{1}&\mathbf{N}_f&\overline{\mathbf{N}_f}&0&2&2 \Delta+\frac{2j}{k+1}\\
T_{j,+} &\mathbf{1} &\mathbf{1}  &\mathbf{1} & 1 &-N_f  &N_f(1-\Delta)-\frac{2}{k+1}(N_c-1-j) \\
 T_{j,-} &\mathbf{1} &\mathbf{1}  &\mathbf{1}  &- 1 &-N_f &N_f(1-\Delta)-\frac{2}{k+1}(N_c-1-j) \\
\end{array}
\end{equation}

The equivalence between the electric and the magnetic partition function was
discussed in \cite{Amariti:2014iza} and corresponds to the identity
\begin{align}
\label{amcl}
Z_\text{ele}^{U(N_c)}  (\tau;\mu;\nu;\lambda) 
=&\ 
\prod_{j=0}^{n-1}
\Gamma_h \left(\pm \frac{\lambda}{2} - \frac{1}{2}  \sum(\mu_a+\nu_a)
+\tau(j-N_c+1) + \omega N_f \right)
\nonumber \\
&\cdot
\prod_{a,b=1}^{N_f} \Gamma_h(\mu_a+\nu_b+j \omega \tau)
\cdot
Z_\text{mag}^{U(n N_f-N_c)}  (\tau;\tau-\nu;\tau-\mu;-\lambda) 
\end{align}
where the partition function of a $U(N_c)$ gauge theory with 
an adjoint and $N_f$ pairs of fundamentals and antifundamentals is 
\begin{align}
Z^{U(N_c)}  (\tau;\mu;\nu;\lambda) =&\  
\frac{\Gamma_h( \tau)^{N_c}}{N_c!}
\int \prod_{i=1}^{N_c} d\sigma_i \prod_{i<j} 
\frac{\Gamma_h(\tau \pm  (\sigma_i -\sigma_j))}{\Gamma_h( \pm  (\sigma_i -\sigma_j))}
e^{i \pi \lambda \sum_i \sigma_i} \cdot \nonumber \\
&\cdot \prod_{i=1}^{N_c} \left(
\prod_{a=1}^{N_f} \Gamma_h(\mu_a + \sigma_i) \cdot
\Gamma_h(\nu_a-\sigma_i) \right)\ .
\end{align}
The mass parameters are 
$\mu_a \equiv m_A +m_a$ and $\nu_a \equiv m_A+n_a$, 
and are constrained by
$\sum_{a=1}^{N_f} m_a = \sum_{a=1}^{N_f} n_a = 0$.
Moreover the superpotential (\ref{aleadjW})
imposes $\tau = \frac{2}{n+1} \omega$.

A comment regarding the identity \eqref{amcl} is in order. That relation was obtained in \cite{Amariti:2014iza} by reducing the 
identity between the electric and the magnetic side of the 4d Kutasov--Schwimmer--Seiberg duality \cite{Kutasov:1995ve,Kutasov:1995np,Kutasov:1995ss}, i.e. by reducing the superconformal index of each side and equating the two. As is customary for many superconformal index identities, its existence was not established rigorously, although several hints towards the existence of such a relation were furnished in \cite{Spiridonov:2009za}.
Furthermore in \cite{Dolan:2008qi} a perturbative proof was discussed at large $N_c$ at fixed $N_f$. Here we will assume \eqref{amcl} holds.

The chiral dualities in the classes $[p,q]_\text{adj}$,
$[p,0]_\text{adj}$, and $[p,q]_\text{adj}^*$ can be obtained by 
studying the real mass flows analogous to the ones  formulated 
in \cite{Benini:2011mf}.
Again we study these three cases  separately.
%
%
%
%
%
\subsection{\texorpdfstring{$\Delta F  < 2k$: the $[p,q]_\mathrm{adj}$ case}{Delta F  < 2k: the [p,q]_adj case}}
\label{sub:adjUNpq}
%
%
%
%
%
In this case we assign a positive large real mass to 
$N_f-N_f^{(1)}$ fundamentals and a positive large real mass to 
$N_f-N_f^{(2)}$ antifundamentals.
The electric theory is $U(N_c)_{k}$ with $N_f^{(1)}$ fundamentals and 
$N_f^{(2)}$ antifundamentals.
The CS level generated by this real mass flow is $k = N_f-\tfrac{1}{2}(N_f^{(1)}+N_f^{(2)})$
and $\Delta F = |N_f^{(1)}-N_f^{(2)}|  < 2k =2N_f- N_f^{(1)}-N_f^{(2)} $.
The dual gauge group is $U( n N_f-N_c)_{-k}$, with 
$N_f^{(1)}$ antifundamentals and $N_f^{(2)}$ fundamentals.
The (dressed) monopoles and antimonopoles  acting as singlets in the dual phase are massive and we integrate them out. We are left with the superpotential
\begin{equation}
\label{wd}
W = \Tr Y^{n+1} + \sum_{j=0}^{n-1} M_j q Y^j \tilde q \ .
\end{equation}

The real mass flow just described can be reproduced on the three-sphere partition function. 
This corresponds to assigning the following mass parameters:
\begin{equation}
\left\{
\begin{array}{l l l l l}
m_A & \to &m_A -s \frac{N_f^{(1)}+N_f^{(2)}-2 N_f}{2N_f};&
\\
m_a & \to & m_a+s \frac{N_f^{(1)}-N_f}{2N_f},& a =1,\dots,N_f^{(1)} ;
\\
m_a & \to & m_a+s \frac{N_f^{(1)}}{2N_f},&a =N_f^{(1)}+1,\dots,N_f-N_f^{(1)} ;
\\
n_a & \to & n_a+s \frac{N_f^{(2)}-N_f}{2N_f},&a =1,\dots,N_f^{(2)} ;
\\
n_a & \to & n_a+s \frac{N_f^{(2)}}{2N_f},&a =N_f^{(2)}+1,\dots,N_f-N_f^{(2)} ;
\\
\sigma_i & \to &\sigma_i  -s \frac{N_f^{(1)}-N_f^{(2)}}{2 N_f},&
i=1,\dots,N_c;
\\
\tilde{\sigma}_i & \to &\tilde{\sigma}_i  -s \frac{N_f^{(1)}-N_f^{(2)}}{2 N_f},&
i=1,\dots,n N_f-N_c;
\\
\lambda & \to &\lambda  +s \big(N_f^{(1)}-N_f^{(2)} \big)&
\end{array}
\right.
\end{equation}
The real mass flow consists of studying the large 
$s$ limit on both sides of the identity 
(\ref{amcl}).
The divergent contributions cancel between the electric and 
the magnetic phase and we obtain a new identity
$Z_\text{ele} = Z_\text{mag}$ that reproduces the duality 
discussed above. The electric partition function is
\begin{align}
 \label{eq:Z_ele-SQCD-U-pqA}
Z_\text{ele} =&\ 
\frac{\Gamma_h( \tau)^{N_c}}{N_c!}
\int \prod_{i=1}^{N_c} d\sigma_i \prod_{i<j} 
\frac{\Gamma_h( \tau\pm  (\sigma_i -\sigma_j))}{\Gamma_h( \pm  (\sigma_i -\sigma_j))}
\,
e^{-i \pi k \sum_i \sigma_i (\sigma_i-\hat \lambda)} \cdot \nonumber \\
&\cdot \prod_{i=1}^{N_c} \left(
  \prod_{a=1}^{N_f^{(1)}} \Gamma_h(\mu_a + \sigma_i)  
 \cdot
\prod_{b=1}^{N_f^{(2)}} \Gamma_h(\nu_a-\sigma_i)\right)\ ,
\end{align}
with
\begin{equation}
\hat \lambda \equiv \lambda +(N_f^{(1)} -N_f^{(2)}) (m_A-\omega) \ .
\end{equation}
The magnetic one is
\begin{align}
 \label{eq:Z_mag-SQCD-U-pqA}
Z_\text{mag} =&\ \frac{e^{i \pi \phi} \Gamma_h( \tau)^{\widetilde N_c}}{\widetilde N_c!}\prod_{a=1}^{N_f^{(1)}} \prod_{b=1}^{N_f^{(2)}} 
\Gamma_h(\mu_a+\nu_b+j \omega \tau) \cdot \nonumber \\
& \int \prod_{i=1}^{\widetilde N_c} d\sigma_i \prod_{i<j} 
\frac{
\Gamma_h( \tau\pm  (\sigma_i -\sigma_j))
}
{
\Gamma_h( \pm  (\sigma_i -\sigma_j))
}
\,
e^{\frac{ i \pi }{4} \lambda ^2 n}
e^{i \pi k \sum_i \sigma_i ( \sigma_i- \tilde{\lambda} ) }
\nonumber
\\
&\cdot \prod_{i=1}^{\widetilde N_c}\left(
 \prod_{a=1}^{N_f^{(1)}} \Gamma_h(\tau-\mu_a - \sigma_i)  
 \cdot
\prod_{b=1}^{N_f^{(2)}} \Gamma_h(\tau-\nu_a + \sigma_i)\right)\ ,
\end{align}
with
\begin{equation}
\tilde{\lambda} \equiv \lambda +(N_f^{(1)} -N_f^{(2)}) (m_A-\tau+\omega)
\end{equation}
The complex exponent $\phi$ reads
\begin{align}
\phi = & 
-\frac{1}{2}  \bigg\{
\frac{1}{2} n \bigg[ (N_f^{(1)}-N_f^{(2)}+2 k) \sum _{a=1}^{N_f^{(1)}} m_a^2+(N_f^{(2)}-N_f^{(1)}+2 k) \sum _{b=1}^{N_f^{(2)}} n_b^2 \bigg] +
\nonumber \\
&+
m_A \bigg[ \frac{1}{2} (N_f^{(2)}+N_f^{(1)}) (N_f^{(2)}+N_f^{(1)}+2 k) (n m_A+2 ((n-2) \omega +\tau )) + 
\nonumber \\
&
-4 N_f^{(2)} N_f^{(1)} (n m_A+\tau -2 \omega )+2 N_c (N_f^{(2)}+N_f^{(1)}) (\tau -2 \omega )\bigg]\bigg\} +
\nonumber \\
&
+  n \tau  \bigg \{ \frac{\tau}{2}   \big[ (\widetilde N_c^2+(N_f^{(1)}-N_c) (N_f^{(2)}-N_c))-\frac{(n-1)^2}{8} (N_f^{(2)}+N_f^{(1)}+2 k) \cdot
\nonumber \\
&
\cdot
 (N_f^{(2)}+N_f^{(1)}) 
\big]
-\frac{n-1}{12}  \omega \big[\frac{1}{4} (N_f^{(2)}+N_f^{(1)}+2 k)^2-N_f^{(1)} N_f^{(2)}-2 \big] \bigg \}\ .
\end{align}

%
%
%
%
%
\subsection{\texorpdfstring{$\Delta F  = 2k$: the $[p,0]_\mathrm{adj}$ case}{Delta F  = 2k: the [p,0]_adj case}}
\label{sub:adjUNp0}
%
%
%
%
%
In this case we assign a positive large real mass to 
$N_f-N_f^{(1)}$ fundamentals.
The electric theory is $U(N_c)_k$ with $N_f^{(1)}$ fundamentals and 
$N_f$ antifundamentals.
The CS level generated by this real mass flow is $k=\tfrac{1}{2}( N_f-N_f^{(1)})$
and $\Delta F =N_f- N_f^{(1)}=2k $.
The dual gauge group is $U( n N_f-N_c)_{-k}$, with 
$N_f^{(1)}$ antifundamentals and 
$N_f$ fundamentals.
The (dressed) monopoles acting as singlets in the dual phase are massive and we integrate them out. 
On the other hand the antimonopoles remain massless because the shift due to the
large mass is compensated by a shift of the FI.
 The dual superpotential is 
\begin{equation}
\label{wdd}
W = \Tr Y^{n+1} + \sum_{j=0}^{n-1} M_j q Y^j \tilde q + {T}_{j,-}  t_{n-1-j,+} \ .
\end{equation}
Equivalently we can assign a positive large real mass to 
$N_f-N_f^{(2)}$ antifundamentals, this latter case being related to the former by a parity transformation.
The difference is that the last term in the dual superpotential 
(\ref{wdd}) becomes  $  {T}_{j,-}  t_{n-1-j,+} $.

The real mass flow just described can be reproduced on the three-sphere partition function. 
This corresponds to assigning the following mass parameters:
\begin{equation}
\left\{
\begin{array}{l l l l }
m_A&\to&m_A +s \frac{N_f-N_f^{(1)}}{2 N_f};&
\\
m_a&\to&m_a-s \frac{N_f-N_f^{(1)}}{N_f},& a =1,\dots,N_f^{(1)} ;
\\
m_a&\to& m_a+s \frac{N_f^{(1)} }{N_f},&a =N_f^{(1)}+1,\dots,N_f ;
\\
\sigma_i&\to &\sigma_i  +s \frac{N_f-N_f^{(1)}}{2 N_f},&
i=1,\dots,N_c;
\\
\tilde \sigma_i&\to &\tilde{\sigma}_i  + s \frac{N_f-N_f^{(1)}}{2 N_f},&
i=1,\dots,n N_f-N_c;
\\
\lambda&\to& \lambda + s (N_f-N_f^{(1)}).&
\end{array}
\right.
\end{equation}
The real mass flow consists of studying the large 
$s$ limit on both sides of the identity 
(\ref{amcl}).
The divergent contributions cancel between the electric and 
the magnetic phase and we obtain a new identity
$Z_\text{ele} = Z_\text{mag}$ that reproduces the duality 
discussed above. The electric partition function is
\begin{align}
 \label{eq:Z_ele-SQCD-U-p0A}
Z_\text{ele} =&\ 
\frac{\Gamma_h( \tau)^{N_c}}{N_c!}
\int \prod_{i=1}^{N_c} d\sigma_i \prod_{i<j} 
\frac{
\Gamma_h( \tau\pm  (\sigma_i -\sigma_j))
}
{
\Gamma_h( \pm  (\sigma_i -\sigma_j))
}
\,
e^{-i \pi k \sum_i \sigma_i (\sigma_i-\hat \lambda)} \cdot \nonumber \\
&\cdot  \prod_{i=1}^{N_c}
\left(
  \prod_{a=1}^{N_f^{(1)}} \Gamma_h(\mu_a + \sigma_i)  
 \cdot
\prod_{b=1}^{N_f} \Gamma_h(\nu_a-\sigma_i)
\right)\ ,
\end{align}
with
\begin{equation}
\hat \lambda \equiv \lambda -\big(N_f- N_f^{(1)} \big)(m_A-\omega)\ .
\end{equation}
The magnetic one is
\begin{align}
 \label{eq:Z_mag-SQCD-U-p0A}
Z_\text{mag} =&\ 
\frac{e^{\frac{i \pi}{2} n  \phi} \Gamma_h( \tau)^{\widetilde N_c}}{\widetilde N_c!} \prod_{a=1}^{N_f^{(1)}} \prod_{b=1}^{N_f} 
\Gamma_h(\mu_a+\nu_b+j \omega \tau) \nonumber \\
&\int \prod_{i=1}^{\widetilde N_c} d\tilde \sigma_i \prod_{i<j} 
\frac{
\Gamma_h( \tau\pm  (\tilde \sigma_i -\tilde \sigma_j))
}
{
\Gamma_h( \pm  (\tilde \sigma_i -\tilde \sigma_j))
}
\,
e^{-i \pi k \sum_i \tilde \sigma_i ( \tilde \sigma_i+ \tilde{ \lambda} ) } \cdot
\nonumber
\\
&\cdot \prod_{i=1}^{\widetilde N_c} \left(
 \prod_{a=1}^{N_f^{(1)}} \Gamma_h(\tau-\mu_a -\tilde \sigma_i)  
 \cdot
\prod_{b=1}^{N_f} \Gamma_h(\tau-\nu_a + \tilde \sigma_i)\right)\ ,
\end{align}
with
\begin{equation}
\tilde{ \lambda} \equiv \
\lambda -\big(N_f-N_f^{(1)} \big)(m_A-\tau+\omega)\ .
\end{equation}
The complex exponent $\phi$ reads
\begin{align}
\phi =& \
N_f (N_f^{(1)}-4 k)m_A^2
-\frac{\omega}{3} \bigg[\tau -\omega +(N_f^{(1)}+2 k)(10 k (\tau -\omega )-3N_f^{(1)}\omega)\bigg]
-2 k \sum _{b=1}^{N_f} n_b^2+
  \nonumber
 \\
&
+2m_A(2 k \tau  N_f+N_f^{(1)}\tau  N_c-\omega  N_f)
+
N_c \tau ((N_f-N_f^{(1)}+N_c)\tau -2 \omega  N_f)\ .
\end{align}
%
%
%
%
%
\subsection{\texorpdfstring{$\Delta F  > 2k$: the $[p,q]_\mathrm{adj}^*$ case}{Delta F  > 2k: the [p,q]*_adj case}}
\label{sub:adjUNpq*}
%
%
%
%
%
This is the case that has not been discussed in \cite{Hwang:2015wna}.
In this case one assigns a positive  mass to $N_f^{(1)}$ antifundamentals and 
a negative  mass to $N_f^{(2)}$ antifundamentals.
The CS level is $k=\tfrac{1}{2}(N_f^{(1)}-N_f^{(2)})$ and 
$\Delta F = N_f^{(1)}+N_f^{(2)}>2k \equiv N_f^{(1)}-N_f^{(2)}$.
The electric theory is $U(N_c)_k$ with $N_f$ fundamentals and 
$N_a$ antifundamentals with $N_a=N_f-N_f^{(1)}-N_f^{(2)}$.
The dual gauge group is $U( n N_f-N_c)_{-k}$
with $N_a$ fundamentals and 
$N_f$ antifundamentals.
The (dressed) monopoles and antimonopoles  acting as singlets in the dual phase are massive and we integrate them out. We are left with the dual superpotential
\begin{equation}
\label{Wddd}
W = \Tr Y^{n+1} + \sum_{j=0}^{n-1} M_j q Y^j \tilde q \ .
\end{equation}
By a parity transformation one can also obtain the case where $N_f<N_a$. In general one has
an electric $U(N_c)_k$ theory with
$N_f$ fundamentals, $N_a$ antifundamentals, and $|N_f-N_a| > 2k$,  dual to  
$U(n \, \max (N_f,N_a) -N_c)_{-k}$ with
$N_f$ dual antifundamentals, $N_a$ dual fundamentals, and the superpotential 
(\ref{Wddd}).

The real mass flow just described can be reproduced on the three-sphere partition function. 
This corresponds to assigning the following mass parameters:
\begin{equation}
\left\{
\begin{array}{l l l l }
m_A &\to &m_A +s \frac{N_f^{(1)}-N_{f}^{(2)}}{N_f};  &
\\
n_a &\to &n_a-s \frac{N_f^{(1)}-N_{f}^{(2)}}{N_f} ,    &  a =1,\dots,N_a ;
\\
n_a &\to & n_a+ s \frac{N_f-N_f^{(1)}+N_{f}^{(2)}}{N_f},     &  a =N_a+1,\dots,N_a+N_f^{(2)} = N_
f-N_f^{(1)};
\\
n_a &\to & n_a-s \frac{N_f+N_f^{(1)}-N_{f}^{(2)}}{N_f} ,   & a =N_a+N_f^{(2)} +1,\dots,N_f ;
\\
\sigma_i &\to &\sigma_i  -s \frac{N_f^{(1)}-N_{f}^{(2)}}{N_f}   ,&
i=1,\dots,N_c;
\\
\tilde{\sigma}_i &\to &\tilde{\sigma}_i  -s \frac{N_f^{(1)}-N_{f}^{(2)}}{N_f} ;&
i=1,\dots,n N_f-N_c,
\\
\lambda &\to & \lambda - s (N_f^{(1)}+N_f^{(2)});&
\end{array}
\right.
\end{equation}
The real mass flow consists of studying the large 
$s$ limit on both sides of the identity 
(\ref{amcl}).
The divergent contributions cancel between the electric and 
the magnetic phase and we obtain a new identity
$Z_\text{ele} = Z_\text{mag}$ that reproduces the duality 
discussed above. The electric partition function is
\begin{align}
 \label{eq:Z_ele-SQCD-U-pqA*}
Z_\text{ele} =&\ 
\frac{\Gamma_h( \tau)^{N_c}}{N_c!}
\int \prod_{i=1}^{N_c} d\sigma_i \prod_{i<j} 
\frac{
\Gamma_h( \tau\pm  (\sigma_i -\sigma_j))
}
{
\Gamma_h( \pm  (\sigma_i -\sigma_j))
}
\,
e^{-i \pi k \sum_i \sigma_i (\sigma_i-\hat \lambda)} \cdot \nonumber \\
&\prod_{i=1}^{N_c}\left(
  \prod_{a=1}^{N_f} \Gamma_h(\mu_a + \sigma_i)  
 \cdot
\prod_{b=1}^{N_a} \Gamma_h(\nu_a-\sigma_i)
\right)\ ,
\end{align}
with
\begin{equation}
\hat \lambda \equiv \lambda - 2k (m_A-\omega)\ .
\end{equation}
The magnetic one is
\begin{align}
 \label{eq:Z_mag-SQCD-U-pq*A}
Z_\text{mag} =&\ \frac{e^{i \pi n k  \phi} \Gamma_h( \tau)^{\widetilde N_c}}{\widetilde N_c!}
e^{i \pi  \lambda  n \left(\tau  N_c-N_f \left(\omega -m_A\right)\right)}
\prod_{a=1}^{N_f} \prod_{b=1}^{N_a} 
\Gamma_h(\mu_a+\nu_b+j \omega \tau) \cdot \nonumber \\
&\int \prod_{i=1}^{\widetilde N_c} d\tilde \sigma_i \prod_{i<j} 
\frac{
\Gamma_h( \tau\pm  (\tilde \sigma_i -\tilde \sigma_j))
}
{
\Gamma_h( \pm  (\tilde \sigma_i -\tilde \sigma_j))
}
\,
e^{i \pi k \sum_i \tilde \sigma_i ( \tilde \sigma_i- \tilde{ \lambda} ) }\cdot 
\nonumber
\\
&\cdot \prod_{i=1}^{\widetilde N_c}\left(
 \prod_{a=1}^{N_f} \Gamma_h(\tau-\mu_a-\tilde \sigma_i)  
 \cdot
\prod_{b=1}^{N_a} \Gamma_h(\tau-\nu_a + \tilde \sigma_i)\right)\ ,
\end{align}
with
\begin{equation}
\tilde{ \lambda} \equiv
\lambda - 2k (m_A-\tau+\omega)
\ .
\end{equation}
The complex exponent $\phi$ reads
\begin{align}
\phi  = 
\tau  N_c \left(\tau -2 m_A\right)-N_f \left(3 m_A^2-2 m_A (\tau +\omega )-\frac{1}{3} (n-4) \tau  \omega \right)
-\sum _{a=1}^{N_f} m_a^2\ .
\end{align}
%
%
%
%
%
%
%
%
\section{\texorpdfstring{Dualities for $SU(N_c)$ chiral adjoint SQCD}{Dualities for SU(Nc) chiral adjoint SQCD}}
\label{sec:adjSUN}
%
%
%
%
%
%
%
%
In this section we generalize the construction to the case of $SU(N_c)$ SQCD with $N_f$ 
flavors $Q$ and $\tilde Q$ and an adjoint $X$, interacting through the superpotential in (\ref{aleadjW}).
The dual model was derived in \cite{Park:2013wta} (see also \cite{Nii:2014jsa}). It is a $U(nN_f-N_c) \times U(1)$ gauge theory with 
$N_f$ flavors $q$ and $\tilde q$ in the nonabelian sector, and $n$ pairs $(T_{j,+},T_{j,-}$ (with $j=0,\dots,n-1$) of opposite 
gauge charge in the $U(1)$ gauge sector. These latter fields interact through the superpotential  
with the dressed monopoles $t_{i,+}$ and antimonopoles  $t_{i,-}$ of the $U(nN_f-N_c)$ sector.
Such monopoles and antimonopoles carry charge $-1$ and $+1$ (respectively) under the $U(1)$ gauge group as well. 
The superpotential of the dual gauge theory is 
\begin{equation}
\label{SU00adj}
W = \Tr Y^{n+1} + \sum_{j=0}^{n-1} M_j q Y^j \tilde q  + T_{j,+}  t_{n-1-j,-}
+ T_{j,-}  t_{n-1-j,+}\ ,
\end{equation}
where $M_j=Q X^j \tilde Q$ are the $n N_f^2$ dressed mesons of the electric theory acting as singlets in the dual phase.
There is also a level-$1$ mixed CS term between the abelian center $U(1) \subset U(n N_f-N_c)$ and the other $U(1)$ 
gauge factor.

The fields are charged under the various symmetries as follows:
\begin{equation}
\begin{array}{c|cccccc}
          &   SU(N_c)   &   SU(N_f)_L & SU(N_f)_R & U(1)_B &U(1)_A & U(1)_R \\
             \hline
Q           &  \mathbf{N}_c   & \mathbf{N}_f & \mathbf{1} & 1& 1&\Delta\\ 
\tilde Q   &  \overline {\mathbf{N}_c}  & \mathbf{1} &\overline {\mathbf{N}_f}  & -1& 1&\Delta\\ 
X & \mathbf{Adj} & \mathbf{1} &\mathbf{1} &0&0&\frac{2}{n+1}  \\
\end{array}
\end{equation}
The dual model is a $U(\widetilde N_c) \times U(1)$ gauge theory, with a mixed CS term
between $U(1) \subset U(\widetilde N_c)$ and the other $U(1)$ gauge symmetry.
The fields are charged under the various symmetries as follows:
 \begin{equation}
\begin{array}{c|cccccc}
          &   U(\widetilde{N}_c) \times U(1)   &   SU(N_f)_L & SU(N_f)_R & U(1)_B &U(1)_A & U(1)_R \\
             \hline
\rule{0pt}{15pt} q           &  \widetilde{\mathbf{N}}_{c,0}    & \overline {\mathbf{N}_f}   & \mathbf{1} & 0& -1&1-\Delta\\ 
\tilde q   &  \overline {\widetilde{\mathbf{N}}_c}_{,0}  & \mathbf{1}&\mathbf{N}_f &0& -1&1-\Delta\\ 
Y &\mathbf{Adj}_0 &\mathbf{1}&\mathbf{1}&0&0&\frac{2}{n+1}  \\
M_j & \mathbf{1}_0 &\mathbf{N}_f&\overline {\mathbf{N}_c} &0&2&2 \left(\Delta+\frac{j}{n+1}\right)\\
T_{j,+}        &  \mathbf{1}_{1}    & \mathbf{1}   & \mathbf{1}& N_c& N_f &N_f(1-\Delta)+\frac{2(j-N_c+1)}{n+1}\\ 
T_{j,-}  &   \mathbf{1}_{-1}  & \mathbf{1}&\mathbf{1}&-N_c& N_f&N_f(1-\Delta)+\frac{2(j-N_c+1)}{n+1}\\ 
\end{array}
\end{equation}
The duality for $SU(N_c)$ adjoint SQCD can be 
represented at the level of the partition function, by
manipulating the  identity (\ref{amcl}).
We first multiply the partition function
by a factor $\frac{1}{2} e^{-i \pi \lambda N_c m_B}$.
Then we gauge the topological symmetry, i.e. we integrate over $\lambda$.
By shifting $\sigma_i \rightarrow \sigma_i + m_B$ we produce 
a factor  $\delta(\sigma_1+ \dots+\sigma_{N_c})$. 
The partition function of the electric theory then becomes 
\begin{align}
\label{eq:Z_ele-SQCD-adj}
Z_\text{ele}
=&\ \frac{\Gamma_h(\tau)^{N_c}}{N_c!}
\int \prod_{i=1}^{N_c} d\sigma_i \, \delta \left(\sum_{i=1}^{N_c} \sigma_i \right) \prod_{i<j} 
\frac{\Gamma_h(\tau \pm  (\sigma_i -\sigma_j))} {\Gamma_h( \pm  (\sigma_i -\sigma_j))} \cdot \nonumber \\ 
&\cdot \prod_{i=1}^{N_c} \left( \prod_{a=1}^{N_f} \Gamma_h(\mu_a + \sigma_i+m_B) \cdot  \Gamma_h(\nu_a-\sigma_i-m_B) \right)\ .
\end{align}
On the dual side the theory is $U(n N_f -N) \times U(1)$, and there is a 
mixed CS term between the two $U(1)$s. The partition function is then
\begin{align}
\label{eq:Z_mag-SQCD-adj}
Z_\text{mag} =&\ 
\frac{\Gamma_h(\tau)^{\widetilde N_c}}{\widetilde N_c!}
\,
\prod_{j=0}^{n-1}
\prod_{a,b=1}^{ N_f } \Gamma_h(\mu_a+\nu_b+j \omega \tau)\cdot \nonumber \\
& \int d\xi \prod_{j=0}^{n-1}
\Gamma_h \left(\pm \xi - \frac{1}{2}\sum(\mu_a+\nu_a)
+\tau(j-N_c+1) + \omega N_f \right) \cdot \nonumber \\
& \int  \prod_{i=1}^{\widetilde N_c} d\tilde \sigma_i  \prod_{i<j} 
\frac{\Gamma_h(\tau \pm  (\tilde \sigma_i -\tilde \sigma_j))}{\Gamma_h( \pm  (\tilde \sigma_i -\tilde \sigma_j))} \,
e^{-2i \pi \xi \sum_i (\tilde \sigma_i+ \frac{N_c}{n N_f -N_c}m_B)} \cdot
\nonumber \\
&\cdot
\prod_{i=1}^{n N_f  - N_c}\left(
\prod_{a=1}^{ N_f } \Gamma_h(\tau-\mu_a - \tilde \sigma_i)
\Gamma_h(\tau-\nu_a+\tilde \sigma_i) \right)\ .
\end{align}
We can shift $\tilde \sigma_i$ in the magnetic partition function and define a more canonical baryonic symmetry.
Again, in order to have a proper charge normalization for the fields in the $U(1)$ sector we have used the substitution
$\lambda = 2\xi$ in \eqref{amcl}.

The equivalence between 
(\ref{eq:Z_ele-SQCD-adj}) and (\ref{eq:Z_mag-SQCD-adj}) 
corresponds to the duality of \cite{Park:2013wta}  
at the level of the partition function.
Observe that in both cases there is an overall factor $\Gamma_h(\tau)^{\rk (G)}$,
where $\rk (G)$ is $N_c$ on the electric  and $\widetilde N_c = n N_f-N_c$ on the dual side.
Actually, in the final identity, because we are dealing
with an $SU(N_c)$ gauge group on the electric side the rank is rather given by  $\rk(G) = N_c-1$.
On the dual side this corresponds to considering a new singlet $\beta$ and a superpotential coupling $W \propto \beta\, \Tr Y$.
By integrating out the massive fields we are left with the overall contribution $\Gamma_h(\tau)^{\widetilde N_c-1}$
on the magnetic side, that corresponds to a traceleness requirement for the adjoint of the dual gauge group. Equivalently on the dual side we consider the adjoint of the $SU(\widetilde N_c) \subset U(\widetilde N_c)$ subgroup.

In the following, starting from the 
identity between (\ref{eq:Z_ele-SQCD-adj}) and (\ref{eq:Z_mag-SQCD-adj}),
we consider the three different
real mass flows that correspond to the three generalizations of $[p,q]_\text{adj}^*$,
$[p,0]_\text{adj}$ and $[p,q]_\text{adj}$.
%
%
%
%
\subsection{\texorpdfstring{$\Delta F  < 2k$: $SU$ generalization of the $[p,q]_\mathrm{adj}$ case }{Delta F  < 2k: SU generalization of the [p,q]_adj case}}
\label{sub:adjSUpq}
%
%
%
%
In this case the real mass flow generalizes the one studied in section \ref{pqPP}.
We  assign a positive large real mass to 
$N_f-N_f^{(1)}$ fundamentals and a positive large real mass to 
$N_f-N_f^{(2)}$ antifundamentals.
The electric theory is $SU(N_c)_k$ adjoint SQCD with $N_f^{(1)}$ fundamentals and 
$N_f^{(2)}$ antifundamentals.
The CS level generated by this real mass flow is $k = N_f-\tfrac{1}{2}(N_f^{(1)}+N_f^{(2)} )$
and $\Delta F = |N_f^{(1)}-N_f^{(2)}|  < 2k =2N_f- N_f^{(1)}-N_f^{(2)} $.

On the dual we  must consider a nonzero vacuum for the scalar in the vector multiplet and 
we must shift the scalar in the vector multiplet of the gauged topological symmetry.
We are left with
 $U\big(n(\tfrac{1}{2}(N_f^{(1)}+N_f^{(2)})+k)-N_c \big)_{-k} \times U(1)_{-n}$ gauge symmetry with a mixed CS level, at level $1$,  between the two $U(1)$ symmetries.
There are $N_f^{(1)}$ dual antifundamentals and $N_f^{(2)}$ dual fundamentals and 
there is a superpotential
\begin{equation}
\label{pqsuadj}
W = \Tr Y^{n+1} + \sum_{j=0}^{n-1} M_j q Y^j \tilde q \  .
\end{equation}

The real mass flow just described corresponds to the  assignment (\ref{pqrm}) followed by a large $s$ limit.
Defining $\mu_a  \equiv m_a + m_A +m_B$ and $\nu_a \equiv n_a+m_A-m_B$ we have
\begin{align}
 \label{eq:Z_ele-SQCD-adj-pq}
Z_\text{ele} =&\ 
\frac{1}{N_c!}
\int \prod_{i=1}^{N_c} d\sigma_i \, 
e^{- i \pi k   \sum_i \sigma_i^2}
\,
 \delta \left(\sum_{i=1}^{N_c} \sigma_i\right)
\prod_{i<j} 
\frac{\Gamma_h(\tau \pm  (\sigma_i -\sigma_j))}{\Gamma_h( \pm  (\sigma_i -\sigma_j))} \,  \cdot \nonumber \\
 &\cdot \prod_{i=1}^{N_c} \left(
 \prod_{a=1}^{N_f^{(1)}} \Gamma_h(\mu_a + \sigma_i)  
 \cdot
\prod_{b=1}^{N_f^{(2)}} \Gamma_h(\nu_a-\sigma_i) \right)\ .
\end{align}
The magnetic partition function is
\begin{align}
\label{eq:Z_mag-SQCD-adj-pq}
Z_\text{mag} =&\ \frac{e^{i \pi \phi}}{\widetilde N_c !}
\prod_{j=0}^{n-1}
\prod_{a=1}^{N_f^{(1)}} \prod_{b=1}^{N_f^{(2)}} \Gamma_h(\mu_a+\nu_b+j \omega \tau)
\int d\xi 
e^{i \pi (n \xi^2+2 \xi m_B N_c)} \cdot
\nonumber \\
&\int 
\prod_{i=1}^{\widetilde N_c} d\sigma_i 
\, e^{i \pi (k \sum_i \tilde \sigma_i^2
+ 2(\xi-\eta) \sum_i \tilde \sigma_i )}
\,\prod_{i<j} 
\frac{\Gamma_h(\tau \pm  (\tilde \sigma_i -\tilde\sigma_j))}{\Gamma_h( \pm  (\tilde\sigma_i -\tilde\sigma_j))}\,
\cdot \nonumber \\
&\cdot \prod_{i=1}^{\widetilde N_c}
\left(
\prod_{a=1}^{N_f^{(1)}} \Gamma_h(\tau-\mu_a - \tilde\sigma_i)
\cdot
\prod_{b=1}^{N_f^{(2)}} \Gamma_h(\tau-\nu_b+\tilde \sigma_i)
\right)\ ,
\end{align}
where
\begin{equation}
\eta\equiv \frac{N_f^{(1)}-N_f^{(2)}}{2} (m_A-\tau +\omega )\ .
\end{equation}
The complex exponent $\phi$ reads
\begin{align}
\phi =&\ \frac{n}{4}  \bigg[(2 k+N_f^{(1)}-N_f^{(2)}) \sum _{a=1}^{N_f^{(1)}} m_a^2+(2 k-N_f^{(1)}+N_f^{(2)}) \sum _{b=1}^{N_f^{(2)}} n_b^2 \bigg]-k N_c m_B^2+
\nonumber \\
&
+
N_c m_A m_B (N_f^{(1)}-N_f^{(2)})+\frac{n m_A}{4}  \bigg\{(n-1) \tau  \big[N_f^{(1)} (N_f^{(1)}+2 k)+N_f^{(2)} (N_f^{(2)}+2 k)\big]
+
\nonumber \\
&
-4 N_c \tau  (N_f^{(1)}+N_f^{(2)})+4 (\tau +\omega ) N_f^{(1)} N_f^{(2)} \bigg\}
+\frac{n m_A^2}{4}  \bigg\{2 k (N_f^{(1)}+N_f^{(2)})-6 N_f^{(1)} N_f^{(2)}
+
\nonumber \\
&
+{N_f^{(1)}}^2+ {N_f^{(2)}}^2 \bigg\}
-N_c \omega  m_B (N_f^{(1)}-N_f^{(2)})-\frac{n \tau}{24}   \bigg\{ 3 \tau  \bigg[ n^2 (N_f^{(1)}+N_f^{(2)}+2 k)^2
+
\nonumber \\
&
-\frac{1}{2} (N_f^{(1)}+N_f^{(2)}+2 k) ((n-1)^2 (N_f^{(1)}+N_f^{(2)})+8 n N_c)-4 N_c (N_f^{(1)}+N_f^{(2)})
+
\nonumber \\
&
+4 N_f^{(1)} N_f^{(2)}+8 N_c^2 \bigg]+2 \omega (n-1) \bigg[N_f^{(1)} N_f^{(2)}+2-\frac{1}{4} (N_f^{(1)}+N_f^{(2)}+2 k)^2 \bigg]\bigg\}\ .
\end{align}


\subsection{\texorpdfstring{$\Delta F  = 2k$: $SU$ generalization of  the $[p,0]_\mathrm{adj}$ case}{Delta F = 2k: SU generalization of the [p,0]_adj case}}
\label{sub:adjSUp0}

In this case the real mass flow generalizes the one studied in section \ref{p0PP}.
We assign a positive large real mass to 
$N_f-N_f^{(1)}$ antifundamentals.
The electric theory is $SU(N_c)_k$ adjoint SQCD with $N_f$ fundamentals and 
$N_f^{(1)}$ antifundamentals and superpotential $W = \Tr X^{n+1}$
The CS level generated by this real mass flow is $k = \tfrac{1}{2}(N_f-N_f^{(1)} )$
and $\Delta F = |N_f-N_f^{(1)}|  = 2k$.
On the magnetic side we take again  a nonzero vacuum for the scalar in the vector multiplet,
and we shift the scalar in the vector multiplet of the gauged topological symmetry.
We are left with $U\big(\widetilde{N}_c\equiv n \big(k+\tfrac{1}{2}(N_f+N_f^{(1)})\big)-N_c\big)_{-k} \times U(1)_{-\frac{n}{2}}$ gauge symmetry with a mixed CS level, at level $1$, between the $U(1)$ symmetries.
In the $U(1)$ sector the fields $T_{j,-}$  are massive and are integrated out.
The fields $T_{j,+}$ are massless, because the shift induced by 
real masses is canceled against the one of the real scalar of 
the gauged topological $U(1)$.
In the $U(\widetilde{N}_c)$ sector there are also $N_f$ dual antifundamentals and $N_f^{(1)}$ dual fundamentals.
The dual superpotential is
\begin{equation}
\label{Wp0suadj}
W = \Tr Y^{n+1} + \sum_{j=0}^{n-1} M_j q Y^j \tilde q  +T_{j,+} \, t_{n-1-j,-}\ .
\end{equation} 
The real mass flow corresponds to the  assignment (\ref{p0rm}), followed by a large $s$ limit.
The electric partition function is
\begin{align}
 \label{eq:Z_ele-SQCD-adj-p0}
Z_\text{ele} =&\ 
\frac{1}{N_c!}
\int \prod_{i=1}^{N_c} d\sigma_i \,
e^{ - i \pi k \sum_i\sigma_i^2}\,
 \delta \left(\sum_{i=1}^{N_c} \sigma_i\right)
\prod_{i<j} 
\frac{\Gamma_h(\tau \pm  (\sigma_i -\sigma_j))}{\Gamma_h( \pm  (\sigma_i -\sigma_j))}\, 
 \cdot
 \nonumber \\
 &\cdot
 \prod_{i=1}^{N_c} \left(
 \prod_{a=1}^{N_f} \Gamma_h(\mu_a + \sigma_i+m_B)  
 \cdot
\prod_{b=1}^{N_f^{(1)}} \Gamma_h(\nu_a-\sigma_i-m_B) \right)\ .
\end{align}
The magnetic one is
\begin{align}
\label{eq:Z_mag-SQCD-adj-p0}
Z_\text{mag} =&\ \frac{e^{i \pi \phi}}{\widetilde N_c!}
\prod_{j=0}^{n-1}
\prod_{a=1}^{N_f} \prod_{b=1}^{N_f^{(1)}} \Gamma_h(\mu_a+\nu_b+j \omega \tau)\cdot \nonumber \\
& \int d\xi 
e^{\frac{i\pi}{2}n\xi^2 +i\pi \xi 
(n N_f (\omega -m_A) +N_c (2 m_B+n \tau)) } \prod_{j=0}^{n-1}
\Gamma_h((\omega-m_A) N_f - \xi 
+\tau(j-N_c+1) )\cdot
\nonumber \\
&\int \prod_{i=1}^{\widetilde N_c}  d\tilde \sigma_i 
\, e^{ i \pi (k \sum_i\tilde \sigma_i^2  +2 (\xi - \eta )\sum_i \tilde \sigma_i)}
\,\prod_{i<j} 
\frac{\Gamma_h(\tau \pm  (\tilde \sigma_i -\tilde \sigma_j))}{\Gamma_h( \pm  (\tilde \sigma_i -\tilde \sigma_j))}\, 
 \cdot \nonumber \\
&\cdot \prod_{i=1}^{\widetilde N_c} \left(
\prod_{a=1}^{N_f} \Gamma_h(\tau-\mu_a - \tilde \sigma_i)
\cdot
\prod_{b=1}^{N_f^{(1)}} \Gamma_h(\tau-\nu_b+\tilde \sigma_i)
\right)\ ,
\end{align}
with
\begin{equation}
\eta \equiv   k  \left(m_A-\tau +\omega \right)\ .
\end{equation}
The complex exponent $\phi$ reads
\begin{align}
\phi =&\ 
n(N_f^{(1)}-N_f)  \sum _{a=1}^{N_f} m_a^2-2 \eta  (n N_f (m_A-\omega )+N_c (m_B+n \tau ))\ +
\nonumber \\
&+
N_f^{(1)} (n N_f (m_A-\omega )^2-N_c (m_B+n\tau )^2)
+
N_f N_c (2 n \tau  (m_A+m_B)+m_B^2)\ +
\nonumber \\
&+
n \tau  (\tau  N_c (N_c-N_f)-\frac{1}{6} (N_f (N_f-N_f^{(1)})-1) (n-1) \omega )\ .
\end{align}
%
%
%
\subsection{\texorpdfstring{$\Delta F  > 2k$: $SU$ generalization of the $[p,q]_\mathrm{adj}^*$ case}{Delta F > 2k: SU generalization of the [p,q]_adj case}}
\label{sub:adjSUpq*}
%
%
%
In this case the real mass flow generalizes the one studied in section \ref{pq*PP}.
We assign a positive large real mass to 
$N_f^{(1)}$ antifundamentals and a negative large real mass to 
$N_f^{(2)}$ antifundamentals.
The electric theory is $SU(N_c)_k$ with $N_f$ fundamentals and 
$N_a$ antifundamentals with $N_a=N_f-N_f^{(1)}-N_f^{(2)}$.
The CS level generated by this real mass flow is $k = \frac{1}{2}(N_f^{(1)}-N_f^{(2)} )$
and $\Delta F = N_f-N_a = N_f^{(1)}+N_f^{(2)} >2k = N_f^{(1)} - N_f^{(2)}$.
On the magnetic side  we need to consider a nonzero vacuum for the scalar in the vector multiplet
and we also need to shift the scalar in the vector multiplet of the gauged topological symmetry.
We are left with $U(n N_f-N_c)_{-k} \times U(1)_{0}$ gauge symmetry with a mixed CS level, at level $1$, between the two $U(1)$ symmetries.
In the $U(1)$ sector obtained by gauging the topological symmetry all the fields $T_{j,\pm}$ 
are massive.
In the $U(\widetilde{N}_c)$ sector  there are $N_f$ dual antifundamentals and $N_a$ dual fundamentals and  there is a superpotential 
\begin{equation}
\label{wpqasdj*}
W = \Tr Y^{n+1} + \sum_{j=0}^{n-1} M_j q Y^j \tilde q \ .
\end{equation}
By a parity transformation one can also define the case where $N_f<N_a$. In general one has
an electric $SU(N_c)_k$ theory with
$N_f$ fundamentals, $N_a$ antifundamentals, and $|N_f-N_a| > 2k$,  dual to  
$(U(n \max (N_f,N_a) -N_c)_{-k} \times U(1)_0)_1$ with
$N_f$ dual antifundamentals, $N_a$ dual fundamentals, and the superpotential \eqref{wpqasdj*}.

The real mass flow corresponds to the  assignment (\ref{pq*rm})
, followed by a large $s$ limit:
The electric partition function is
\begin{align}
\label{eq:Z_ele-SQCD-adj-pq*}
Z_\text{ele} =&\ \frac{1}{N_c!}
\int \prod_{i=1}^{N_c} d\sigma_i \,
 e^{ - i \pi k  \sum_i\sigma_i^2}
 \,
 \delta \left(\sum_{i=1}^{N_c} \sigma_i\right)
\prod_{i<j} 
\frac{\Gamma_h(\tau \pm  (\sigma_i -\sigma_j))}{\Gamma_h( \pm  (\sigma_i -\sigma_j))} \,
\cdot
\nonumber \\
&\cdot
\prod_{i=1}^{N_c} \left(
\prod_{a=1}^{N_f} \Gamma_h(\mu_a + \sigma_i + m_B)  
\prod_{b=1}^{N_f-N_f^{(1)}-N_f^{(2)}} \Gamma_h(\nu_b  -  \sigma_i -  m_B) \right)\ .
\end{align}
The magnetic one is
\begin{align}
\label{eq:Z_mag-SQCD-adj-pq*}
Z_\text{mag} =&\ \frac{e^{i \pi \phi}}{\widetilde N_c!}
\prod_{j=0}^{n-1}
\prod_{a=1}^{N_f}\prod_{b=1}^{N_f-N_f^{(1)}-N_f^{(2)}} \Gamma_h(\mu_a+\nu_b+j \omega \tau)
\int d \xi \,
e^{2 i \pi  \xi (nN_f (m_A-\omega)+N_c (m_B +n\tau))}
\nonumber \\
&\cdot 
 \prod_{i=1}^{\widetilde N_c} d\tilde \sigma_i \,
 e^{i \pi (k  \sum_i \tilde \sigma_i^2+ 2i\pi (\xi- \eta)  \sum_i\tilde \sigma_i)}\,
\prod_{i<j}
\frac{\Gamma_h(\tau \pm  (\tilde \sigma_i -\tilde \sigma_j))}{\Gamma_h( \pm  (\tilde \sigma_i -\tilde \sigma_j))}\, \cdot
\nonumber \\
& 
\cdot \prod_{i=1}^{\widetilde N_c} \left(
\prod_{a=1}^{N_f} \Gamma_h(\tau-\mu_a - \tilde \sigma_i)
\cdot
\prod_{b=1}^{N_f-N_f^{(1)}-N_f^{(2)}} \Gamma_h(\tau-\nu_a+\tilde \sigma_i)\right) \ ,
\end{align}
with
\begin{equation}
 \eta \equiv
k (m_A+ \omega-\tau) \ .
\end{equation}
The complex exponent $\phi$ reads
\begin{align}
\phi =& \ k  \bigg\{
 N_c (\tau-2 m_A+m_B ) (n \tau+m_B )-n \sum _{a=1}^{N_f} m_a^2 \ + \nonumber  \\ 
& +  n N_f \bigg[ m_A ((n+3) \tau - m_A)+\frac{(n-4)}{3} \tau  \omega \bigg] \bigg\} \ .
\end{align}
%
%
%
\section{Conclusions}
\label{sec:conc}
%
%
%
In this paper we have provided a full classification of 3d $\mathcal{N}=2$ dualities for adjoint SQCD
with unitary and special unitary gauge group and a different amount of
fundamentals $N_f$ and antifundamentals $N_a$.
The classification generalizes the construction of \cite{Benini:2011mf} for $U(N_c)$ SQCD
and the one of \cite{Aharony:2014uya,Nii:2020xgd} for $SU(N_c)$ SQCD.
It distinguishes three cases, depending on the 
relative value of the difference $\Delta F = |N_f-N_a|$ with respect to the CS level $k$.
We have kept the notation of \cite{Benini:2011mf} where the $\Delta F<2k$ case is denoted $[p,q]$,
 $\Delta F= 2k$ is denoted $[p,0]$, and $\Delta F > 2k$ is denoted $[p,q]^*$.
 The analysis of \cite{Benini:2011mf} was performed starting from the original 
case of Aharony duality, where $N_f=N_a$ and the CS level is vanishing, and subsequently applying a real mass flow to produce chiral dualities.
The dualities were corroborated by an analysis of the flow at the level of three-sphere partition functions (i.e. by matching them across dual phases). Here we have derived chiral dualities 
for $SU(N_c)$ SQCD starting from the $SU(N_c)$ version of Aharony duality (i.e. at zero CS level).
This is different from the analysis of \cite{Aharony:2014uya} where the starting point was the 
generalization of Giveon--Kutasov duality for $SU(N_c)_k$ SQCD.
We have commented in appendix \ref{appAF} on the relation between the two approaches.
In this way we have also studied the flow of \cite{Aharony:2014uya} on the three-sphere partition function.
Then we have extended  our analysis of chiral dualities to the case of adjoint SQCD. We have considered the
non-chiral dualities for $U(N_c)$ and $SU(N_c)$ adjoint SQCD of \cite{Nii:2014jsa,Amariti:2014iza} and performed the real 
mass flow on the corresponding integral identities between three-sphere partition functions.
In this way we have obtained a complete classification. Notice that one can further simplify the integral identities of section \ref{sec:adjSUN} by performing the $\xi$ integral (corresponding to the $U(1)$ gauge sector). The analysis in the $[p,q]_\text{adj}$ and $[p,q]^*_\text{adj}$ is straightforward (one performs a gaussian integral in the former and uses identity \eqref{delta} in the latter), analogously to what done in section \ref{SUAF}. The $[p,0]_\text{adj}$ case is more complex because the $\xi$ integral corresponds to $U(1)_{-n/2}$ with $n$ negatively charged fields, which is not (known to be) dual to a set of singlets.

As a bonus, we have obtained integral identities for $U(N_c)$ CS
matter theories where the CS level of the $SU(N_c)$ factor differs from the one of the $U(1)$
factor. These types of integrals have not been thoroughly investigated so far, but they can 
be interesting for checking e.g. the dualities recently studied in \cite{Nii:2020ikd}.

Further checks and generalizations of our analysis are possible.
For instance it would be desirable to match the moduli space between the  
dual phases, extending the analysis of \cite{Nii:2014jsa} to the case
of adjoint SQCD.
Another independent check of the dualities that we have constructed here consists of computing 
the superconformal indices of the dual phases, and matching them at least for small rank $N_c$.

It could also be possible to obtain  the chiral dualities for adjoint SQCD studied in section \ref{sec:adjUN}  starting from the 
non-chiral dualities at nonzero CS level discussed in \cite{Niarchos:2008jb,Niarchos:2009aa,Kapustin:2011vz}. 
In such a case the electric theory corresponds to $U(N_c)_k$ SQCD with $N_f$ pairs of fundamentals and antifundamentals $Q$ and $\tilde Q$ and an adjoint $X$ with superpotential $W=\Tr X^{n+1}$, while the dual model is 
$U( n (N_f +|k|)- N_c )_{-k}$ SQCD with $N_f$ pairs of fundamentals and antifundamentals, an adjoint $Y$ and $n N_f$ singlets $M_j = Q X^j \tilde Q$, with superpotential 
\begin{equation}
\label{Wniarchos}
W = \Tr Y^{n+1} + \sum_{j=0}^{n-1} M_j q Y^j \tilde q \ .
\end{equation}
The relevant integral identities to match the partition functions were given in \cite{Amariti:2014iza}. 
The $SU(N_c)$ version of this duality can be obtained by gauging the topological symmetry and it can be used to 
also reproduce  the dualities studied in section \ref{sec:adjSUN}.

Other chiral dualities can be derived  from the non-chiral ones of \cite{Giacomelli:2017vgk,Giacomelli:2019blm} and of \cite{Pasquetti:2019tix}: the latter case is interesting because the integral identities needed to match the three-sphere partition 
functions are known. Another possibility would be constructing chiral dualities for SQCD with two adjoints (which is studied in \cite{Hwang:2018uyj}).

\section*{Acknowledgments}

We wish to thank O.~Aharony and S.~Benvenuti for useful correspondence and discussions. This work has been supported in part by the Italian Ministero dell’Istruzione, Universit\`a e Ricerca (MIUR), in part by Istituto Nazionale di Fisica Nucleare (INFN) through the ``Gauge Theories, Strings, Supergravity'' (GSS) research project and in part by MIUR-PRIN contract 2017CC72MK-003.
The work of M.F. is supported in part by the European Union's Horizon 2020 research and innovation
programme under the Marie Sk\l odowska-Curie grant agreement No. 754496 - FELLINI. 
\appendix

%
%
%
\section{\texorpdfstring{$SU(N_c)_{k_1} \times U(1)_{k_2} $ and the partition function}{SU(Nc)_{k1} x U(1)_{k2} and the the partition function}} 
\label{appC}
%
%
%

In order to compare the results obtained here with the ones discussed in \cite{Aharony:2014uya,Nii:2020xgd},
it is necessary to distinguish the values of the CS levels of the $SU(N_c)$  and the $U(1)$ factor 
of $U(N_c) =(SU(N_c) \times U(1))/\mathbb{Z}_{N_c}$.
In general we have obtained here CS contributions to the partition function of the form
\begin{equation}
\label{CSUvsSU}
\exp \left[-i \pi k_1 \sum_{i=1}^{N_c} \sigma_i^2 -i \pi k_2 \left(\sum_{i=1}^{N_c} \sigma_i\right)^2 \right]\ .
\end{equation}
We want to isolate the $U(1)$ factor from the first term in the exponent.
This can be done by redefining $\sigma_i =  \frac{x_0}{\sqrt {N_c}}- x_i$ with $x_1 + \dots + x_{N_c} = 0$.
In this way (\ref{CSUvsSU}) becomes 
\begin{equation}
\label{CSUvsSU2}
\delta \left(\sum_{i=1}^{N_c} x_i \right)
\exp \left[-i \pi k_1 \sum_{i=1}^{N_c} x_i^2 - i \pi \left(k_1+ N_c k_2\right) x_0 ^2 \right]\ ,
\end{equation}
i.e. we have shown that (\ref{CSUvsSU}) corresponds to the CS contribution to the partition function 
of the gauge group $U(N_c)_{k_1,k_1 + N_c k_2}$, where 
$k_1$ is the CS level of the $SU(N_c)$ factor (with eigenvalues $\{ x_i\}_{i=1}^{N_c}$) and $k_1 + N_c k_2$ is the CS level of 
 the $U(1)$ factor (with eigenvalue $x_0$).
\section{The flow of \cite{Aharony:2014uya} on the partition function}
\label{appAF}

In this section we discuss the real mass flow from the $SU(N)_k$ Giveon--Kutasov duality presented in 
\cite[Eq. (3.23)]{Aharony:2013dha} to the chiral dualities. This is the flow originally considered in \cite{Aharony:2014uya} and here we study this flow as an infinite limit on the real masses on the partition function.
The starting point is the identity between the electric $SU(N)_k$ (for $k>0$) theory with $N_f$ pairs of fundamentals and antifundamentals
and its $U(N_f -N_c+k)_{-k, N_f-N_c}$ dual with $N_f$ pairs of dual fundamentals and antifundamentals and superpotential $W =  M q \tilde q$.
Up to our knowledge the integral identity relating these two phases has not appeared in the literature so far: for 
this reason we start our analysis by deriving such an identity by gauging the topological symmetry of Giveon--Kutasov duality.
Then we study the $[p,q]$, $[p,0]$, and $[p,q]^*$ cases separately, comparing the relations obtained here with the ones 
discussed in the section \ref{SUAF}.

\subsection{Integral identity for $SU(N_c)_k$  duality}
\label{sub:intGK}

Our starting point is the identity 
\begin{eqnarray}
Z_{U(N_c)_k}(\mu;\nu;2 \xi)
=
\prod_{a,b=1}^{N_f} \Gamma_h(\mu_a + \nu_b) Z_{U(N_f-N_c+k)_{-k}}(\omega-\nu;\omega-\mu;-2 \xi)\ .
\end{eqnarray}
We then add a term $e^{2 i \pi m_B \xi}$ on both sides and gauge the topological symmetry by integrating over $\xi$. We arrive at the identity $Z_\text{ele} = Z_\text{mag}$ with
\begin{align}
\label{aaa}
Z_\text{ele} =&\ \frac{1}{N_c!} \int  \prod_{i=1}^{N_c} d\sigma_i\, \delta \left (\sum_{i=1}^{N_c} \sigma_i \right) \prod_{i<j} 
\Gamma_h( \pm  (\sigma_i -\sigma_j))^{-1}\, 
e^{-i \pi k  \left(m_B^2 N_c+ \sum_{i} \sigma_i^2  \right)} \cdot \nonumber \\
&\cdot \prod_{i=1}^{N_c}\left(\prod_{a=1}^{N_f} 
  \Gamma_h(\mu_a + m_B + \sigma_i)  
\Gamma_h(\nu_a-m_B-\sigma_i)\right) \ , \\
\label{bbb}
Z_\text{mag}  = & \ \frac{e^{-\frac{i \pi}{2} \phi}}{\widetilde N_c!}
\prod_{a,b=1}^{N_f} \Gamma_h(\mu_a+\nu_b)
\int  \prod_{i=1}^{\widetilde N_c} d\tilde \sigma_i 
e^{i \pi \left( k \sum_i \tilde \sigma_i^2  -\left( N_c m_B- \sum_i \tilde \sigma_i  \right)^2 \right)}
\nonumber \\
&\cdot \prod_{i=1}^{\widetilde N_c}\left(
\prod_{a=1}^{N_f} \Gamma_h(\omega-\mu_a -  \tilde \sigma_i)
\Gamma_h(\omega-\nu_a + \tilde \sigma_i)\right)
\prod_{i<j}  
{\Gamma_h( \pm  (\tilde \sigma_i -\tilde \sigma_j))}^{-1}\ .
\end{align}
The complex exponent $\phi$ reads
\begin{align}
\phi=&\ k \sum _{a=1}^{N_f} \left(m_a^2+n_a^2\right)+2 \omega \widetilde N_c \left(k \omega -2 m_A N_f\right)\ +
\nonumber \\
&-4 k \omega  m_A N_f+2 m_A^2 N_f (k-N_f)-2 \omega ^2 \widetilde N_c^2- k^2 \omega ^2\ .
\end{align}
The dual partition function in (\ref{bbb}) coincides with the one obtained in (\ref{eq:Z_mag-SQCD-SU-pq3}).

\subsection{\texorpdfstring{$\Delta F  < 2k$}{Delta F  < 2k}}
\label{sub:GKpq}

Here we study the flow of \cite{Aharony:2014uya} corresponding the the $\Delta F <2k$, i.e. $[p,q]$ case.
In this case the real mass flow corresponds to shifting the mass parameters appearing in the partition 
function  as follows:
\begin{equation}
\left\{
\begin{array}{llll}
m_A & \to &m_A-s\frac{ N_f^{(m)}}{2 N_f}; & \\
m_B & \to &m_B-s\frac{ N_f^{(m)}}{2 N_f}; & \\
m_a & \to &m_a-s\frac{ \left(N_f-N_f^{(m)}\right)}{N_f}, & a=1,\dots, N_f^{(m)};\\
m_a & \to &m_a+s\frac{ N_f^{(m)}}{N_f}, & a=N_f^{(m)}+1,\dots,N_f-N_f^{(m)}
\end{array}
\right.
\end{equation}
with $s>0$. On the magnetic side we also shift $\sigma$ by a term $- s\frac{N_f^{(m)}}{2 N_f}$.
After canceling the infinite contributions, that coincide on the electric and magnetic side, we 
obtain the identity  $Z_\text{ele} = Z_\text{mag}$, where
\begin{align}
Z_\text{ele} =&\ 
\frac{1}{N_c!}
\int 
\prod_{i=1}^{N_c} 
d\sigma_i \, \delta \left (\sum_{i=1}^{N_c} \sigma_i \right)
\prod_{i<j} 
\Gamma_h( \pm  (\sigma_i -\sigma_j))^{-1} \, 
e^{-i \pi \left( k + \tfrac{1}{2}N_f^{(m)}  \right) \sum_{i} \sigma_i^2  }\cdot \nonumber \\
&\cdot \prod_{i=1}^{N_c} \left(
\prod_{a=1}^{N_f-N_f^{(m)}} 
  \Gamma_h(\mu_a + m_B + \sigma_i)  
  \prod_{a=1}^{N_f} 
\Gamma_h(\nu_a-m_B-\sigma_i)\right)\ ,
 \end{align}
and
\begin{align}
Z_\text{mag} = & \ \frac{e^{-\frac{i \pi}{2} \phi}}{\widetilde N_c!}
\prod_{a=1}^{N_f-N_f^{(m)}}\prod_{b=1}^{N_f} \Gamma_h(\mu_a+\nu_b)
\int  \prod_{i=1}^{\widetilde N_c} d\tilde \sigma_i \,
e^{i \pi \left( \eta \sum_{i} \tilde \sigma_i  + \left(k+ \tfrac{1}{2}N_f^{(m)} \right) \sum_{i} \tilde \sigma_i^2  -\left(\sum_{i} \tilde \sigma_i  \right)^2 \right)}\cdot
\nonumber \\
&\cdot 
\prod_{a=1}^{N_f-N_f^{(m)}} \Gamma_h(\omega-\mu_a -  \tilde \sigma_i)
\prod_{a=1}^{N_f} \Gamma_h(\omega-\nu_a + \tilde \sigma_i)
\prod_{i<j}  
\frac{1}{\Gamma_h( \pm  (\tilde \sigma_i -\tilde \sigma_j))}\ ,
\end{align}
with $\widetilde N_c \equiv N_f-N_c+k$ and $\eta \equiv 2 N_c m_B - N_f m_A$. The complex exponent $\phi$ reads
\begin{align}
 \phi 
=&
-k \sum _{a=1}^{N_f-N_f^{(m)}} m_a^2
-
(N_f^{(m)}+k) \sum _{b=1}^{N_f} n_b^2 
+ 
N_c \left[ 2 m_B (m_A-\omega ) N_f^{(m)} + \right. \nonumber \\ &+ \left. 
\omega  ((\omega -2 m_A) N_f^{(m)}- 
4 N_f (\omega -m_A)-2 k \omega )
+m_B^2 (N_f^{(m)}+2 k) \right]\ + \nonumber \\
&+ N_f (\omega -m_A) \left[ m_A (3 N_f^{(m)}+2 k ) +k m_A^2 N_f^{(m)}+2 N_f^2 (\omega -m_A)^2 + \right. \nonumber \\
&+\left. 2 N_c^2 (\omega ^2-m_B^2)+k^2 \omega ^2 +\omega  (2 k-N_f^{(m)})  \right] \ .
\end{align}
It is straightforward to prove that this result coincides with the one obtained in section \ref{pqPP} upon using the dictionary
\begin{equation}
\{N_f,N_f^{(m)},k \}_{\text{here}} =
\{N_f^{(2)},N_f^{(2)}-N_f^{(1)},N_f-N_f^{(2)} \}_{\text{sec.} \, \ref{pqPP}}\ .
\end{equation}

\subsection{\texorpdfstring{$\Delta F  = 2k$}{Delta F  = 2k}}
\label{sub:GKp0}

The $\Delta F=2k$, $[p,0]$ case is obtained by studying the above flow with $s<0$ and $N_f^{(m)}=k$. Actually the analysis of \cite{Aharony:2014uya} for $s<0$ was performed in the whole range $N_f^{(m)} \leq k$. 
In this case one considers the real masses
\begin{equation}
\left\{
\begin{array}{llll}
m_A &\to &m_A-s\frac{ N_f^{(m)}}{2 N_f}; & \\
m_B &\to &m_B-s\frac{ N_f^{(m)}}{2 N_f}; & \\
m_a &\to &m_a-s\frac{ \left(N_f-N_f^{(m)}\right)}{N_f}, & a=1,\dots, N_f^{(m)};\\
m_a &\to &m_a+s\frac{ N_f^{(m)}}{N_f}, & a=N_f^{(m)}+1,\dots,N_f-N_f^{(m)}.
\end{array}
\right.
\end{equation}
In the dual phase the shift on the scalar in the vector multiplet breaks the gauge symmetry as well. Explicitly, the shift is given by 
\begin{equation}
\left\langle \tilde \sigma \right\rangle \to \begin{bmatrix}  \tilde \sigma_{(N_f-N_c)\times(N_f-N_c)} & 0\\  0 & \theta_{k \times k} \end{bmatrix} 
- s \begin{bmatrix}  \left(\frac{N_f^{(m)}}{2N_f}\right) \mathbf{1}_{N_f-N_c}& 0 \\
0 &    \left(\frac{N_f^{(m)}}{2N_f}-1\right)\mathbf{1}_{k} \end{bmatrix} \ ,
\end{equation}
where $\mathbf{1}_n$ denotes the $n\times n$ identity matrix. In this case the divergent phase does not cancel in general between the electric and the magnetic side.
In effect, we are left with a term $e^{-2 i \pi s (N_f^{(m)} -k) \varphi}$ where  
\begin{equation}
\label{vph}
\varphi = N_f^{(m)} (m_A-\omega) + \sum_{j=0}^{k} \theta_j\ . 
\end{equation}
For $N_f^{(m)} <k$ we expect that the divergent term is balanced by a shift in the FI due to the VEV acquired by  the dual fundamentals needed to solve the D-term equation (3.9) of  \cite{Aharony:2014uya}.
Such a shift can be expected to arise from the Higgs branch localization \cite{Fujitsuka:2013fga,Benini:2013yva}.
Following the discussion of \cite{Fujitsuka:2013fga} one can see that the Higgsing does not affect
the one-loop determinants of the chiral fields while it affects the contribution from the classical action. This naive argument supports the fact that the dual model is equivalent to the $[p,q]$ one discussed above.\footnote{It would be interesting to arrange such a computation directly from the Higgs branch localization. We are grateful to O.~Aharony for very valuable discussions on this point.}

On the other hand the term (\ref{vph}) cancels out when $N_f^{(m)} = k$. In such a case we arrive at the identity
$Z_\text{ele} = Z_\text{mag}$, where
 \begin{align}
Z_\text{ele} =&\ \frac{1}{N_c!}\int  \prod_{i=1}^{N_c} 
d\sigma_i \, \delta \left (\sum_{i=1}^{N_c} \sigma_i \right)
\prod_{i<j} 
\Gamma_h( \pm  (\sigma_i -\sigma_j))^{-1}\, 
e^{-i \pi \frac{k}{2} \sum_{i} \sigma_i^2 } \cdot \nonumber \\
&\cdot  \prod_{i=1}^{N_c} \left( \prod_{a=1}^{N_f-k} 
  \Gamma_h(\mu_a + m_B + \sigma_i)    \prod_{a=1}^{N_f} 
\Gamma_h(\nu_a-m_B-\sigma_i)\right)\ ,
\end{align}
and
\begin{align}
\label{b14}
Z_\text{mag} = & \ 
\frac{e^{\frac{i \pi}{2} \phi}}{(N_f-N_c)! k!} 
\prod_{a=1}^{N_f-k}\prod_{b=1}^{N_f} \Gamma_h(\mu_a+\nu_b)
\int  \prod_{i=1}^{N_f- N_c} d\tilde \sigma_i \prod_{j=1}^{k} d\theta_j 
\nonumber \\
&\cdot 
e^{i \pi \left( \frac{k}{2} \sum_{i=1}^{N_f- N_c} \tilde \sigma_i^2  -( \sum_{i=1}^{N_f- N_c} \tilde \sigma_i )^2 \right)} \cdot
\nonumber \\
&\cdot
e^{i \pi \left( \frac{k}{2} \sum_{j=1}^{k} \theta_j^2  -( \sum_{j=1}^{k} \theta_j )^2 \right)
+i \pi  \sum _{j=1}^k \theta _j \left(m_A (2 N_f-k)-2 m_B N_c-2 \omega  \left(N_f-N_c\right)\right)}\cdot
\nonumber \\
&\cdot
e^{-i \pi  \sum _{i=1}^{N_f-N_c} \tilde \sigma _i \left(k \left(m_A-2 \omega \right)+2 m_B N_c\right)-2 i \pi  \sum _{i=1}^{N_f-N_c} \tilde \sigma _i \sum _{j=1}^k \theta _i}\cdot
\nonumber \\
&\cdot \prod_{i=1}^{N_f- N_c}\left(
\prod_{a=1}^{N_f-k} \Gamma_h(\omega-\mu_a -  \tilde \sigma_i)
\prod_{a=1}^{N_f}\right)
\Gamma_h(\omega-\nu_a + \tilde \sigma_i)
\prod_{i<j}  
{\Gamma_h( \pm  (\tilde \sigma_i -\tilde \sigma_j))}^{-1}
\nonumber \\
&\cdot
\prod_{j=1}^{k} \prod_{a=1}^{k} \Gamma_h(\omega- \mu_a -  \theta_j)
\prod_{i<j}  
{\Gamma_h( \pm  (\theta_i -\theta_j))}^{-1}\ ,
\end{align}
where the $\theta_j$ sector corresponds to $U(k)_{-\frac{k}{2},\frac{k}{2}}$ with $k$ fundamentals. 
The complex exponent $\phi$ reads
\begin{align}
\phi = & -k \left(\sum _{a=1}^k m_a^2+\sum _{b=1}^{N_f} n_b^2\right)-k N_c \left(m_B-\omega \right) \left(2 m_A-m_B-3 \omega \right)
+2 N_c^2 \left(\omega ^2-m_B^2\right)+
\nonumber \\
 &
+N_f \left(m_A-\omega \right) \left(k \left(m_A-3 \omega \right)+4 \omega  N_c\right)+2 N_f^2 \left(m_A-\omega \right){}^2+k^2 \left(\omega ^2-m_A^2\right)\ .
\end{align}
In order to reproduce the $[p,0]$ dualities obtained in section \ref{p0PP}  
we should show that this $\theta$ sector can be dualized into a singlet. This is easily achieved for $k=1$, since $U(1)_{1/2}$ with one negatively charged field is dual to a singlet. (This is the half mirror symmetry of \cite{Dorey:1999rb} we already encountered in \eqref{half}.) For higher $k$ one then proceeds by dualizing one massive chiral at a time.

\subsection{\texorpdfstring{$\Delta F  > 2k$}{Delta F  > 2k}}
\label{sub:GKpq*}

We conclude our analysis by studying the $[p,q]^*$ case.  In this case we consider the real masses
\begin{equation}
\left\{
\begin{array}{llll}
m_A &\to & m_A-s\frac{ N_f^{(m)}}{2 N_f}; & \\
m_B &\to & m_B-s\frac{ N_f^{(m)}}{2 N_f}; & \\
m_a &\to & m_a-s\frac{ \left(N_f-N_f^{(m)}\right)}{N_f}, & a=1,\dots,N_f^{(m)};\\
m_a &\to & m_{a}+s\frac{ N_f^{(m)}}{N_f}, & a=N_f^{(m)}+1,\dots, N_f,
\end{array}
\right.
\end{equation}
with $s<0$ and $N_f^{(m)}<k$.
In the dual phase the shift on the scalar in the vector multiplet
breaks the gauge symmetry as well. Explicitly the shift is 
\begin{equation}
\left\langle \tilde \sigma \right\rangle \to \begin{bmatrix}  \tilde \sigma_{(N_f-N_c)\times(N_f-N_c)} & 0\\  0 & \theta_{k \times k} \end{bmatrix} 
+ s \begin{bmatrix}  \left(\frac{N_f^{(m)}}{2N_f}\right) \mathbf{1}_{N_f-N_c}& 0 \\
0 &  \left(\frac{N_f^{(m)}}{2}\left(\frac{1}{N_f}-\frac{2}{k} \right)\right) \mathbf{1}_{k} \end{bmatrix} \ ,
\end{equation}
In this case the divergent phase  cancel  between the electric and the magnetic side
and we are left with the identity $Z_\text{ele} = Z_\text{mag}$, where
 \begin{align}
Z_\text{ele} =&\ 
\frac{1}{N_c!} \int  \prod_{i=1}^{N_c}  d\sigma_i \,  \delta \left (\sum_{i=1}^{N_c} \sigma_i \right)
\prod_{i<j} 
\Gamma_h( \pm  (\sigma_i -\sigma_j))^{-1}\, 
e^{-i \pi \left(k-\tfrac{1}{2}N_f^{(m)}\right) \sum_{i} \sigma_i^2 }\cdot \nonumber \\
&\cdot \prod_{i=1}^{N_c}\left(
\prod_{a=1}^{N_f-N_f^{(m)}} 
  \Gamma_h(\mu_a + m_B + \sigma_i)  
  \prod_{a=1}^{N_f} 
\Gamma_h(\nu_a-m_B-\sigma_i)\right)\ ,
\end{align}
and
\begin{align}
\label{zmagappb}
Z_\text{mag} = & \ 
\frac{e^{\frac{i \pi}{2} \phi}}{(N_f-N_c)! k!}
\prod_{a=1}^{N_f-N_f^{(m)}}\prod_{b=1}^{N_f} \Gamma_h(\mu_a+\nu_b)
\int  \prod_{i=1}^{\widetilde N_c} d\tilde \sigma_i \prod_{j=1}^{k} d\theta_j \prod_{1\leq i<j\leq k}  
\frac{1}{\Gamma_h( \pm  (\theta_i -\theta_j))}
\nonumber \\
&\cdot \prod_{1\leq i<j\leq \widetilde N_c} \frac{1}{\Gamma_h( \pm  (\tilde \sigma_i -\tilde \sigma_j))}\, 
e^{i \pi \left( \left(k-\tfrac{1}{2}N_f^{(m)}\right) \sum_{i=1}^{N_f- N_c} \tilde \sigma_i^2  -( \sum_{i=1}^{N_f- N_c} \tilde \sigma_i )^2 \right)}\cdot
\nonumber \\
&\cdot
e^{i \pi \left( k \sum_{j=1}^{k} \theta_j^2  -( \sum_{j=1}^{k} \theta_j )^2 \right)
+2i \pi  \sum _{j=1}^k \theta _i 
\left(N_f (m_A -\omega)-N_c( m_B - \omega )\right)}\cdot
\nonumber \\
&\cdot
e^{-i \pi  \sum _{i=1}^{N_f-N_c} \tilde \sigma _i  \left( N_f^{(m)} m_A-2 k  \omega +2 m_B N_c\right)
-2 i \pi  \sum _{i=1}^{N_f-N_c} \tilde \sigma _i \sum _{j=1}^k \theta _i}\cdot
\nonumber \\
&\cdot \prod_{i=1}^{\widetilde N_c} \left(
\prod_{a=1}^{N_f-N_f^{(m)}} \Gamma_h(\omega-\mu_a -  \tilde \sigma_i)
\prod_{a=1}^{N_f}
\Gamma_h(\omega-\nu_a + \tilde \sigma_i)\right)\ ,
\end{align}
The complex exponent $\phi$ reads
\begin{align}
\phi =&\ (N_f^{(m)}-k) \sum _{b=1}^{N_f} n_b^2 -k \sum _{a=1}^{N_f} m_a^2 -N_c N_f^{(m)} \left(m_B-\omega \right)  \left(2 m_A+m_B-\omega \right)\ +
\nonumber \\
&+
N_f \left(m_A-\omega \right) \left[ \left(3 m_A-\omega \right) N_f^{(m)}-2 k \left(m_A+\omega \right)+4 \omega  N_c\right]+2 N_f^2 \left(m_A-\omega \right)^2+
\nonumber \\
&+2 m_B^2 N_c \left(k-N_c\right)\ +
\omega ^2 (k^2-2 k N_c+2 N_c^2)\ .
\end{align}
The relevant part of the partition function (\ref{zmagappb}) that we are going to focus on is 
\begin{align}
\label{zb19}
 & \ \frac{1}{\widetilde N_c! k!}  
\int  \prod_{i=1}^{\widetilde N_c} d\tilde \sigma_i
\frac{e^{i \pi \left( k_\text{eff}  \sum_i \tilde \sigma_i^2  -\left(  \sum_{i} \tilde \sigma_i \right)^2 \right)}}{\prod_{1\leq i<j\leq \widetilde N_c} \Gamma_h( \pm  (\tilde \sigma_i -\tilde \sigma_j))}  \prod_{j=1}^{k} d\theta_j \prod_{1\leq i<j\leq k}   
\frac{1}{\Gamma_h( \pm  (\theta_i -\theta_j))}\nonumber \\
& \cdot
e^{i \pi \left( k \sum_{j=1}^{k} \theta_j^2  - \left( \sum_{j=1}^{k} \theta_j \right)^2 
+2  \rho \sum _{j=1}^k \theta _i - \eta \sum _{i=1}^{\widetilde N_c} \tilde \sigma _i  
-2   \sum _{i=1}^{\widetilde N_c} \tilde \sigma _i \sum _{j=1}^k \theta _i
\right)}\cdot
\nonumber \\
&\cdot \prod_{i=1}^{\widetilde N_c} \left(
\prod_{a=1}^{N_f^{(1)}} \Gamma_h(\omega-\mu_a -  \tilde \sigma_i)
\prod_{a=1}^{N_f^{(2)}}
\Gamma_h(\omega-\nu_a + \tilde \sigma_i)\right)\ ,
\end{align}
with 
\begin{align}
&k_\text{eff} \equiv k-\tfrac{1}{2} N_f^{(m)}\ , \\
&\rho \equiv \left(N_f (m_A -\omega)-N_c( m_B - \omega )\right)\ , \\
\quad
&\eta \equiv  \left( N_f^{(m)} m_A-2 k  \omega +2 m_B N_c\right)\ .
\end{align}
We can further substitute  $\theta_i \rightarrow \xi_i + \frac{\xi}{k}$, with 
$\sum_{i=1}^{k} \xi_i = 0$. The integral  (\ref{zb19}) then becomes 
\begin{align}
\label{zb20}
& \
\frac{1}{\widetilde N_c ! k!}
\int  \prod_{i=1}^{\widetilde N_c} d\tilde \sigma_i \prod_{j=1}^{k} \, d\xi_i \,  \delta \left(\sum_{i=1}^{k} \xi_i \right) \, d\xi
\nonumber \\
&e^{i \pi \left( k_\text{eff}  \sum_{i=1}^{\widetilde N_c} \tilde \sigma_i^2  -\left(  \sum_{i=1}^{\widetilde N_c} \tilde \sigma_i \right)^2
+ k \sum_{j=1}^{k} \xi_i^2  +2  \rho  \xi  - (\eta +2    \xi )\sum _{i=1}^{N} \tilde \sigma _i \right)}\cdot 
\nonumber \\
&\cdot \prod_{i=1}^{\widetilde N_c}\left(
\prod_{a=1}^{N_f^{(1)}} \Gamma_h(\omega-\mu_a -  \tilde \sigma_i)
\prod_{a=1}^{N_f^{(2)}}
\Gamma_h(\omega-\nu_a + \tilde \sigma_i)\right) \cdot
\nonumber \\
&\cdot
\prod_{1\leq i < j \leq \widetilde N_c}  
\frac{1}{\Gamma_h( \pm  (\tilde \sigma_i -\tilde \sigma_j))}
\prod_{1\leq i < j \leq k}  
\frac{1}{\Gamma_h( \pm  (\xi_i -\xi_j))}\ .
\end{align}
The $\xi$ integral yields $\delta \left( \rho -\sum _{i=1}^{\widetilde N_c} \tilde \sigma _i  \right)$,
while the integral over $\xi_i$ corresponds to an $SU(k)_{- k}$ pure CS theory.
Performing the latter yields $e^{\frac{i \pi}{2}k^2 \omega^2}$.
All in all (\ref{zb20}) becomes
\begin{align}\label{eq:magSUpq*manipul}
& \ 
\frac{e^{\frac{i \pi }{2}k^2 \omega^2 +\frac{i\pi}{2} \phi}}{\widetilde N_c!}
\int  \frac{\prod_{i=1}^{\widetilde N_c} d\tilde \sigma_i}{\prod_{i<j}\Gamma_h( \pm  (\tilde \sigma_i -\tilde \sigma_j))}\,  \delta \left(\sum_{i=1}^{\widetilde N_c} \tilde \sigma_i \right) \, e^{i \pi \left( k_\text{eff}  \sum_{i=1}^{\widetilde N_c} \tilde \sigma_i^2  +(k_\text{eff} -1)\rho^2-\rho \eta \right)}\cdot \nonumber \\
&\cdot \prod_{i=1}^{\widetilde N_c}\left(
\prod_{a=1}^{N_f^{(1)}} \Gamma_h \left(\omega-\mu_a -  \tilde \sigma_i + \frac{\rho}{\widetilde N_c} \right)
\prod_{a=1}^{N_f^{(2)}} \Gamma_h \left(\omega-\nu_a   +  \tilde \sigma_i  - \frac{\rho}{\widetilde N_c} \right)\right)\ .
\end{align}
Notice that the above result coincides with \eqref{eq:Z_mag-SQCD-SU-pq*} once we perform the $\xi$ integral in that formula (which yields a $\delta(\rho+\sum_i \sigma_i$)) and shift $\sigma_i \rightarrow \sigma_i -\frac{\rho}{\widetilde N_c}$:
\begin{align}
\label{3e24}
Z_\text{mag} =&\  \frac{-e^{i \pi (\frac{\phi}{2} +\frac{k \rho^2}{ \widetilde{N}_c} +{\eta\rho}) }}{\widetilde N_c!} \prod_{a=1}^{N_f}\prod_{b=1}^{N_a} \Gamma_h(\mu_a+\nu_b) \int \frac{\prod_{i=1}^{ \widetilde N_c} d \tilde \sigma_i}{\prod_{i<j} \Gamma_h( \pm  ({\tilde \sigma_i} -{\tilde \sigma_j}))} \delta\left(\sum_{i=1}^{\tilde N_c} \tilde \sigma_i \right) e^{i\pi {k} \sum_i  \tilde \sigma_i^2} \cdot 
\nonumber \\
& \cdot \prod_{i=1}^{\widetilde N_c} \left( \prod_{a=1}^{N_f} \Gamma_h \left(\omega-\mu_a - { \tilde \sigma_i}+\frac{\rho}{\widetilde N_c} \right) \prod_{b=1}^{N_a} \Gamma_h \left(\omega-\nu_a+{\tilde \sigma_i}-\frac{\rho}{\widetilde N_c} \right)\right)\ .
\end{align}
We have checked that (\ref{eq:magSUpq*manipul}) and (\ref{3e24}) are equivalent up to an irrelevant pure phase, confirming the equivalence of the $[p,q]^*$ case obtained in \cite{Aharony:2014uya} and the one obtained here.

%
%
%
%
\section{$SU(N_c)$ chiral dualities flowing from the duality of \cite{Aharony:2013dha}}
\label{appB}
%
%
%
%

In this appendix we study the real mass flow in the dual models discussed in section \ref{SUAF} starting from the duality for $SU(N_c)$ SQCD derived in \cite{Aharony:2013dha}. In this case  the dual partition function is given by formula (\ref{eq:Z_mag-SQCD2}).
Actually here we shift the gauge symmetry by the baryonic one considering an equivalent version of the
the partition function in (\ref{eq:Z_mag-SQCD2}):
\begin{align}
\label{eq:Z_mag-SQCD33}
Z_{\text{mag}} = &\ 
\prod_{a,b=1}^{N_f} \Gamma_h(\mu_a+\nu_b)
\Gamma_h \left(
2\omega (\widetilde N_c+1) -\sum_a (\mu_a+\nu_a) \right) \cdot
\nonumber \\
& \int
\prod_{i=1}^{\widetilde  N_c}  d\tilde \sigma_i \prod_{i<j} \frac{1}{\Gamma_h( \pm  (\tilde \sigma_i -\tilde \sigma_j))} \prod_{i=1}^{\widetilde  N_c}\left(
\prod_{a=1}^{N_f} \Gamma_h(\omega-\mu_a - \tilde \sigma_i)
\cdot 
\Gamma_h(\omega-\nu_a+\tilde \sigma_i) \right)
\nonumber \\ 
&\cdot
\Gamma_h \left(\pm \left(\sum_{i=1}^{\widetilde N_c}  \tilde \sigma_i + m_B N_c\right) + \frac{1}{2} \sum_a (\mu_a+\nu_a)
-\omega (N_f-N_c ) \right)\ .
\end{align}
Using this formula the analysis is more straightforward, because we can use the real mass flows discussed in
section \ref{SUAF}.
These flows correspond to the shifts on the real masses in the partition function summarized in formulae
(\ref{pqrm}), (\ref{p0rm}), and (\ref{pq*rm}).

\subsection{\texorpdfstring{$\Delta F  < 2k$}{Delta F <2k}}
\label{sub:ARSWpq}

In this case, by applying  the  flow (\ref{pqrm}) to (\ref{eq:Z_mag-SQCD33}) we 
cancel the divergent terms against the one obtained on the electric side and we are left with the 
finite dual magnetic partition function
\begin{align}
\label{eq:Z_mag-pq-U}
Z_\text{mag} = & \ e^{\frac{i \pi}{2} \phi}
\prod_{a=1}^{N_f^{(1)}}\prod_{b=1}^{N_f^{(2)}} \Gamma_h(\mu_a+\nu_b)
\int
\prod_{i=1}^{\widetilde N_c} 
d\tilde \sigma_i \, \frac{e^{-i \pi  k \sum _{i} \tilde \sigma _i^2+i\pi \left(\sum _{i=1}^{\widetilde N_c} \tilde \sigma _i\right)^2-i \pi  \eta_2  \sum_i \tilde \sigma _i}}{\prod_{i<j} \Gamma_h( \pm  (\tilde \sigma_i -\tilde \sigma_j))} \cdot \nonumber \\
&\cdot \prod_{i=1}^{\widetilde N_c} \left(
\prod_{a=1}^{N_f^{(1)}} \Gamma_h(\omega-\mu_a - \tilde \sigma_i)
\cdot 
\prod_{b=1}^{N_f^{(2)}} 
\Gamma_h(\omega-\nu_b+\tilde \sigma_i) \right)\ ,
\end{align}
where 
\begin{equation}
\eta_2 \equiv \left(N_f^{(1)} - N_f^{(2)}\right)m_A +2 m_B N_c\ .
\end{equation}
The complex exponent $\phi$ reads
\begin{align}
\phi =& \
m_A^2(N_f (N_f^{(1)} + N_f^{(2)}) \!-\!4 N_f^{(1)} N_f^{(2)} ) 
+
m_B^2 (2(N_f\!-\!N_c)\!-\!(N_f^{(1)} + N_f^{(2)} ))\ +
\nonumber 
\\
&+
2 \left[ ( (N_f^{(1)} + N_f^{(2)} ) N_c \!-\!2 N_f^{(1)} N_f^{(2)}) m_A +(N_f^{(1)} \!-\! N_f^{(2)}) m_B \right]\omega\ + \nonumber \\
&-2 (N_f^{(1)} \!-\! N_f^{(2)}) m_A m_B N_c + \big[N_f^2 \!-\!1+N_f^{(1)} N_f^{(2)}  -2(2N_f+N_f^{(1)} + N_f^{(2)} ) N_c \ +  \nonumber \\
&+ 2N_c^2\big]\omega^2- (N_f\!-\!N_f^{(2)}) \sum_{a=1}^{N_f^{(1)}} m_a^2
- (N_f-N_f^{(1)})  \sum_{b=1}^{N_f^{(1)}} n_b^2 \ .
\end{align}
 Observe that (\ref{eq:Z_mag-pq-U}) and (\ref{eq:Z_mag-SQCD-SU-pq3}) differ by $e^{i \pi \omega^2}$,
 that is just a pure phase and does not affect the free energy.
 
\subsection{\texorpdfstring{$\Delta F  = 2k$}{Delta F  = 2k}}
\label{sub:ARSWp0}

In this case, by applying  the  flow (\ref{p0rm}) to (\ref{eq:Z_mag-SQCD33}) we 
cancel the divergent terms against the one obtained on the electric side and we are left with the 
finite dual magnetic partition function
\begin{align}
\label{eq:Z_mag-SQCD3}
Z_\text{mag} = &\ {e^{\frac{i \pi}{2} \phi}}
\prod_{a=1}^{N_f}  \prod_{b=1}^{N_f^{(1)}} \Gamma_h(\mu_a+\nu_b)
\int \prod_{i=1}^{\widetilde N_c}  d\tilde \sigma_i \, \frac{e^{i \pi k \sum_i \tilde \sigma _i^2 -\frac{i\pi}{2} \big(\sum _i \tilde \sigma _i \big)^2+  i\pi\eta  \sum _i \tilde \sigma _i }}{\prod_{i<j} \Gamma_h( \pm  (\tilde \sigma_i -\tilde \sigma_j))} \cdot \nonumber \\
 &\cdot \prod_{i=1}^{\widetilde N_c} \left( \Gamma_h \left( \sum_{i=1}^{N_f-N_c}  \tilde \sigma_i + N_f (m_A-\omega)+N_c (m_B+\omega)  \right)\right. \cdot \nonumber \\ 
&\cdot
\left.
\prod_{a=1}^{N_f}
 \Gamma_h(\omega-\mu_a - \tilde \sigma_i)
 \cdot 
  \prod_{a=1}^{N_f^{(1)}}
\Gamma_h(\omega-\nu_a+\tilde \sigma_i) 
\right)\ ,
\end{align}
with 
\begin{equation}
\eta \equiv m_A N_f^{(1)}-m_B N_c+\omega  (N_c-N_f-1)\ .
\end{equation}
The complex exponent $\phi$ reads
\begin{align}
\phi =& \ (N_f^{(1)}-N_f) \sum _{a=1}^{N_f} m_a^2
+
N_f \big[ 2 m_A (m_B N_c+\omega  N_c+\eta)
+N_f^{(1)}(m_A-\omega)^2 \ + \nonumber\\
&+ 
(m_B^2-\omega^2) N_c-2 \eta  \omega \big]
+
N_c (m_B+\omega) 
\big[ m_B (N_c-N_f^{(1)})+
\omega  (N_c+N_f^{(1)})+2 \eta \big]\ .
\end{align}
Consistently this result coincides with the one obtained by plugging the identity 
(\ref{half}) into the integral (\ref{eq:Z_mag-SQCD-SU-p0}).

\subsection{\texorpdfstring{$\Delta F  > 2k$}{Delta F  > 2k}}
\label{sub:ARSWpq*}

In this case, by applying  the  flow (\ref{pq*rm}) to (\ref{eq:Z_mag-SQCD33}) we 
cannot cancel the divergent terms against the one obtained on the electric side.
The problem can be confined to the difference in the divergent contributions of
the identity 
\begin{multline}
\int d \xi e^{2 \pi i \xi \sum_{i} \left( \tilde \sigma_i + m_B {N_c}/{\widetilde N_c} \right)} \, \Gamma_h(\pm \xi -N_f m_A +\omega(\tilde N_c+1))
= \\ =  
\Gamma_h\left(2\omega(\widetilde N_c+1)-2 N_f m_A\right)
\Gamma_h\left(\pm \sum_{i=1}^{\widetilde N_c} \left(\tilde \sigma_i +  \frac{N_c}{\widetilde N_c}m_B \right) +N_f m_A -\omega \widetilde N_c \right)\ .
\end{multline}
This identity is the
one that allows us to transform the integral  
(\ref{eq:Z_mag-SQCD}) into (\ref{eq:Z_mag-SQCD2}), proving the equivalence between the duality of \cite{Park:2013wta} and the one of \cite{Aharony:2013dha}.
However if we plug the flow (\ref{pq*rm}) into this identity we obtain a mismatch in the divergent terms.
Furthermore on the LHS side we obtain a purely topological theory while on the RHS one massless singlet
is left over.
This mismatch may be due to a problem when commuting the integral in $d \xi$ and the infinite shift on 
$\xi$. We have not found the correct flow on (\ref{eq:Z_mag-SQCD33}) that leads to the expected duality discussed in section (\ref{pq*PP}). We leave this problem as an open question for future analysis.

\section{Brane engineering of the dualities for $U(N_c)$ chiral SQCD}
\label{sec:BCCbranes}
In this appendix we discuss the  D-brane engineering  of the dualities of
\cite{Benini:2011mf}.
This is done by leveraging the description of Aharony duality \cite{Aharony:1997gp}
in terms of D-branes derived in \cite{Amariti:2015yea,Amariti:2015mva,Amariti:2016kat}.
The latter description contains an aspect that plays a nontrivial role here; namely, the 
brane setup requires the presence of a circle along which the 4d system has been reduced.
This is the reason why this picture differs from the other popular one used to engineer 3d Seiberg like duality, 
obtained by Giveon and Kutasov in \cite{Giveon:2008zn}.

In terms of the brane setup the 4d to 3d reduction consists of performing a T-duality along the compact directions, 
and such a T-duality generates the KK monopole superpotential. 
A further real mass flow, the transition through infinite coupling  and local S-duality 
are the other necessary steps that must be used to complete the brane derivation of
Aharony duality.
Here we use this rather involved picture to derive the dualities of \cite{Benini:2011mf}
for SQCD with a number of fundamentals different from the number of antifundamentals
and with non-vanishing CS term.

Using the notations of \cite{Benini:2011mf} we have four models
\begin{itemize}
\item The $[0,0]$ case corresponds to  the electric theory of Aharony duality: $U(N_c)$ with $N_f$ pairs of
fundamentals and antifundamentals and $W=0$.
\item The $[p,0]$ case is obtained from $[0,0]$ by integrating out $s_1-s_2$ (with $s_1 = N_f)$ fundamentals with negative real mass.
The model is $U(N_c)_k$ with $k = -\frac{1}{2}(s_1-s_2)<0$ and the dual model is $U(s_1- N_c)_{-k}$ with superpotential
$W = M q \tilde q  +  t T$.
\item The $[p,q]$ case is obtained from $[0,0]$ by integrating out 
 $N_f-s_1$  antifundamentals and
 $N_f-s_2$ fundamentals, both with negative real mass.
The models is $U(N_c)_k$ with $k = \frac{1}{2}(s_1+s_2)-N_f<0$ and the dual model is $U(N_f- N_c)_{-k}$ with superpotential
$W = M q \tilde q $.
\item The $[p,q]^*$ case is obtained from $[0,0]$ by integrating out 
 $s_1-\tilde s$ (with $N_f = s_1$)  antifundamentals with positive real mass and
 $\tilde s -s_2$ fundamentals with negative real mass.
The model is $U(N_c)_k$ with $k = \tilde s - \frac{1}{2}(s_1+s_2)$ and the dual model is $U(s_1- N_c)_{-k}$ with superpotential
$W = M q \tilde q $.
\end{itemize}
Our goal consists of finding the brane description of these RG flows from the brane engineering of Aharony duality. The latter is obtained as follows.
\begin{itemize}
\item On the electric side there is a stack of $N_c$ D3-branes extended along $x_{012}$ and 
on a segment along $x_6$. This stack is bounded by an NS5-brane and an NS5'-brane, the first along
$x_{012345}$ and the second along $x_{012389}$, with $x_3$ compact.
There are also $N_f$ D5-branes placed at $x_6=0$ (on the NS5 brane) extended along $x_{012457}$. We must also consider at the position $x_3=\pi$ one extra D5-brane at $x_6=0$. In the large radius limit this sector does not bring any 
new massless mode and we will ignore it, but it is crucial to correctly reproduce the duality.
This corresponds to the real mass flow that is performed on the field theory side in order to 
recover the duality of Aharony starting from the effective duality on the circle.
This last duality was first obtained by \cite{Aharony:2013dha} by reducing  4d Seiberg duality on $S^1$.
\item
The magnetic side is obtained by the Hanany--Witten transition \cite{Hanany:1996ie}, where the NS5 and NS5'-brane are exchanged. During this process, each time a D5-brane crosses the NS5, a D3 is created as to 
preserve the linking number. This is the origin of the different amount of D3-branes in the dual theory, corresponding to 
the different rank for the dual gauge group. 
\end{itemize}

Given this configuration we can keep a finite radius for the circle along $x_3$ and perform a further real mass flow.
In order to understand such a flow we have to study the intersecting brane setup between the NS5 and 
the D5-branes in the electric theory.
This is done in the figure below, where the vertical line is the NS5-brane and the horizontal one is the stack of D5-branes.
Observe that the D5-branes can break on the NS5 and this implies that 
they can move separately on the left and on the right of the NS5 brane. 
Moving such left and right stacks separately along $x_3$ generates the real mass
for the fundamentals and the antifundamentals respectively.
Observe that a similar brane setup was discussed in \cite{Cremonesi:2010ae} for the flavoring of the ABJM model 
in a type IIB string theory setup.\footnote{We are grateful to the referee for pointing out this relation to us.}
Let us discuss the various flows (summarized in figure \ref{figu}):
\begin{figure}
\begin{center}
\includegraphics[width=15cm]{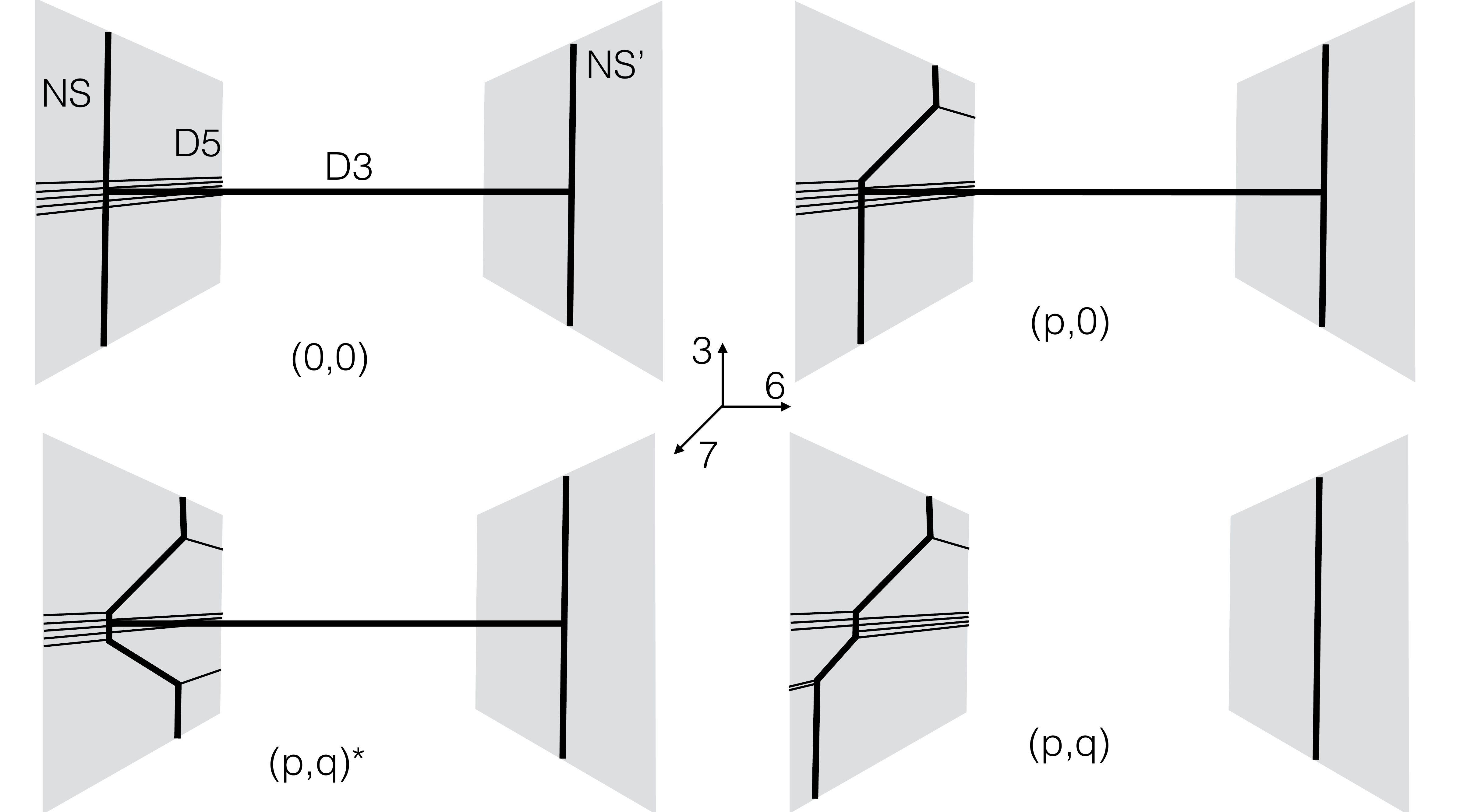}
\caption{The D3-D5-NS5 brane picture summarizing the dualities of \cite{Benini:2011mf}. $x_3$ is a compact direction.}
\label{figu}
\end{center}
\end{figure}
\begin{itemize}
\item Case I: this is Aharony duality. In this case we have the same amount of fundamentals and antifundamentals
and giving a real mass is done by moving one semi-infinite D5-branes on the left and on the right
of the NS5 brane.
In the dual setup the presence of one flavor at $x_3=\pi$ is accompanied by the presence of an abelian gauge factor.
This is because int he Hanany-Witten transition one D3 brane is created at $x_3=\pi$.
By locally mirroring such a sector (corresponding to SQED with one flavor) we obtain the electric monopoles acting as singlets in the dual phase, as expected in the Aharony dual theory.
\item Case II: this corresponds to assigning a positive (or negative) real mass to some fundamentals (or antifundamentals).
Different choices are related by parity transformations.
Observe that on the brane setup this corresponds to moving some of the semi-infinite D5-branes along $x_3$.
In order to preserve supersymmetry this operation requires a motion in the $x_{37}$ plane and the creation of
 a $(1,k)$-fivebrane.
This generates the CS level in the gauge sector. Observe that this motion is not compatible with the effective description 
on the circle, because the NS5  does not close on itself anymore.
Anyway looking at the dual picture we can observe that one of the D1-branes engineering the monopole superpotential survives, and this tells us that we are in presence of half of the original monopole superpotential.
\item Case III: the D5's that have to move are on the same side of the NS5 and two stacks are moved 
along opposite directions. The CS is the difference in this case and the KK monopole is generically broken, as in case III.
This result requires some more care if the number of D5's going up and down is the same.
In this case $k=0$, and in the dual this signals that monopole and antimonopole are integrated
out with opposite real mass. The D1 does not survive in the dual case (following the  picture this becomes evident).
\item Case IV: in this case the flow is generated by moving the left and the right D5-branes in opposite directions.
The CS level is additive and both the monopole and the antimonopole are massive in this case.
\end{itemize}

\bibliographystyle{JHEP}
\bibliography{ref}

\providecommand{\href}[2]{#2}\begingroup\raggedright\begin{thebibliography}{10}

\bibitem{Aharony:2014uya}
O.~Aharony and D.~Fleischer, \emph{{IR Dualities in General 3d Supersymmetric
  SU(N) QCD Theories}},
  \href{https://doi.org/10.1007/JHEP02(2015)162}{\emph{JHEP} {\bfseries 02}
  (2015) 162} [\href{https://arxiv.org/abs/1411.5475}{{\ttfamily 1411.5475}}].

\bibitem{Aharony:2013dha}
O.~Aharony, S.~S. Razamat, N.~Seiberg and B.~Willett, \emph{{3d dualities from
  4d dualities}}, \href{https://doi.org/10.1007/JHEP07(2013)149}{\emph{JHEP}
  {\bfseries 07} (2013) 149} [\href{https://arxiv.org/abs/1305.3924}{{\ttfamily
  1305.3924}}].

\bibitem{Kapustin:2009kz}
A.~Kapustin, B.~Willett and I.~Yaakov, \emph{{Exact Results for Wilson Loops in
  Superconformal Chern-Simons Theories with Matter}},
  \href{https://doi.org/10.1007/JHEP03(2010)089}{\emph{JHEP} {\bfseries 03}
  (2010) 089} [\href{https://arxiv.org/abs/0909.4559}{{\ttfamily 0909.4559}}].

\bibitem{Jafferis:2010un}
D.~L. Jafferis, \emph{{The Exact Superconformal R-Symmetry Extremizes Z}},
  \href{https://doi.org/10.1007/JHEP05(2012)159}{\emph{JHEP} {\bfseries 05}
  (2012) 159} [\href{https://arxiv.org/abs/1012.3210}{{\ttfamily 1012.3210}}].

\bibitem{Hama:2010av}
N.~Hama, K.~Hosomichi and S.~Lee, \emph{{Notes on SUSY Gauge Theories on
  Three-Sphere}}, \href{https://doi.org/10.1007/JHEP03(2011)127}{\emph{JHEP}
  {\bfseries 03} (2011) 127} [\href{https://arxiv.org/abs/1012.3512}{{\ttfamily
  1012.3512}}].

\bibitem{Hama:2011ea}
N.~Hama, K.~Hosomichi and S.~Lee, \emph{{SUSY Gauge Theories on Squashed
  Three-Spheres}}, \href{https://doi.org/10.1007/JHEP05(2011)014}{\emph{JHEP}
  {\bfseries 05} (2011) 014} [\href{https://arxiv.org/abs/1102.4716}{{\ttfamily
  1102.4716}}].

\bibitem{Willett:2011gp}
B.~Willett and I.~Yaakov, \emph{{N=2 Dualities and Z Extremization in Three
  Dimensions}},  \href{https://arxiv.org/abs/1104.0487}{{\ttfamily 1104.0487}}.

\bibitem{Benini:2011mf}
F.~Benini, C.~Closset and S.~Cremonesi, \emph{{Comments on 3d Seiberg-like
  dualities}}, \href{https://doi.org/10.1007/JHEP10(2011)075}{\emph{JHEP}
  {\bfseries 10} (2011) 075} [\href{https://arxiv.org/abs/1108.5373}{{\ttfamily
  1108.5373}}].

\bibitem{Aharony:1997gp}
O.~Aharony, \emph{{IR duality in d = 3 N=2 supersymmetric USp(2N(c)) and
  U(N(c)) gauge theories}},
  \href{https://doi.org/10.1016/S0370-2693(97)00530-3}{\emph{Phys. Lett.}
  {\bfseries B404} (1997) 71}
  [\href{https://arxiv.org/abs/hep-th/9703215}{{\ttfamily hep-th/9703215}}].

\bibitem{Giveon:2008zn}
A.~Giveon and D.~Kutasov, \emph{{Seiberg Duality in Chern-Simons Theory}},
  \href{https://doi.org/10.1016/j.nuclphysb.2008.09.045}{\emph{Nucl. Phys.}
  {\bfseries B812} (2009) 1} [\href{https://arxiv.org/abs/0808.0360}{{\ttfamily
  0808.0360}}].

\bibitem{VanDeBult}
F.~van~de Bult, \emph{{Hyperbolic Hypergeometric Functions,
  http://www.its.caltech.edu/~vdbult/Thesis.pdf}}, {\emph{Thesis} (2008) }.

\bibitem{Hwang:2015wna}
C.~Hwang and J.~Park, \emph{{Factorization of the 3d superconformal index with
  an adjoint matter}},
  \href{https://doi.org/10.1007/JHEP11(2015)028}{\emph{JHEP} {\bfseries 11}
  (2015) 028} [\href{https://arxiv.org/abs/1506.03951}{{\ttfamily
  1506.03951}}].

\bibitem{Nii:2019qdx}
K.~Nii, \emph{{3d "chiral" Kutasov-Schwimmer duality}},
  \href{https://doi.org/10.1016/j.nuclphysb.2020.114920}{\emph{Nucl. Phys.}
  {\bfseries B952} (2020) 114920}
  [\href{https://arxiv.org/abs/1901.08642}{{\ttfamily 1901.08642}}].

\bibitem{Nii:2020xgd}
K.~Nii, \emph{{Coulomb branch in 3d $\mathcal{N}=2$ $SU(N)_k$ Chern-Simons
  gauge theories with chiral matter content}},
  \href{https://arxiv.org/abs/2005.02761}{{\ttfamily 2005.02761}}.

\bibitem{Fazzi:2018rkr}
M.~Fazzi, A.~Lanir, S.~S. Razamat and O.~Sela, \emph{{Chiral 3d SU(3) SQCD and
  $ \mathcal{N}=2 $ mirror duality}},
  \href{https://doi.org/10.1007/JHEP11(2018)025}{\emph{JHEP} {\bfseries 11}
  (2018) 025} [\href{https://arxiv.org/abs/1808.04173}{{\ttfamily
  1808.04173}}].

\bibitem{Amariti:2018wht}
A.~Amariti and L.~Cassia, \emph{{USp(2N$_{c}$) SQCD$_{3}$ with antisymmetric:
  dualities and symmetry enhancements}},
  \href{https://doi.org/10.1007/JHEP02(2019)013}{\emph{JHEP} {\bfseries 02}
  (2019) 013} [\href{https://arxiv.org/abs/1809.03796}{{\ttfamily
  1809.03796}}].

\bibitem{Benvenuti:2018bav}
S.~Benvenuti, \emph{{A tale of exceptional $3d$ dualities}},
  \href{https://doi.org/10.1007/JHEP03(2019)125}{\emph{JHEP} {\bfseries 03}
  (2019) 125} [\href{https://arxiv.org/abs/1809.03925}{{\ttfamily
  1809.03925}}].

\bibitem{Benini:2017dud}
F.~Benini, S.~Benvenuti and S.~Pasquetti, \emph{{SUSY monopole potentials in
  2+1 dimensions}}, \href{https://doi.org/10.1007/JHEP08(2017)086}{\emph{JHEP}
  {\bfseries 08} (2017) 086}
  [\href{https://arxiv.org/abs/1703.08460}{{\ttfamily 1703.08460}}].

\bibitem{Amariti:2018gdc}
A.~Amariti, I.~Garozzo and N.~Mekareeya, \emph{{New 3d $ \mathcal{N} $ = 2
  dualities from quadratic monopoles}},
  \href{https://doi.org/10.1007/JHEP11(2018)135}{\emph{JHEP} {\bfseries 11}
  (2018) 135} [\href{https://arxiv.org/abs/1806.01356}{{\ttfamily
  1806.01356}}].

\bibitem{Park:2013wta}
J.~Park and K.-J. Park, \emph{{Seiberg-like Dualities for 3d N=2 Theories with
  SU(N) gauge group}},
  \href{https://doi.org/10.1007/JHEP10(2013)198}{\emph{JHEP} {\bfseries 10}
  (2013) 198} [\href{https://arxiv.org/abs/1305.6280}{{\ttfamily 1305.6280}}].

\bibitem{Kim:2013cma}
H.~Kim and J.~Park, \emph{{Aharony Dualities for 3d Theories with Adjoint
  Matter}}, \href{https://doi.org/10.1007/JHEP06(2013)106}{\emph{JHEP}
  {\bfseries 06} (2013) 106} [\href{https://arxiv.org/abs/1302.3645}{{\ttfamily
  1302.3645}}].

\bibitem{Niarchos:2008jb}
V.~Niarchos, \emph{{Seiberg Duality in Chern-Simons Theories with Fundamental
  and Adjoint Matter}},
  \href{https://doi.org/10.1088/1126-6708/2008/11/001}{\emph{JHEP} {\bfseries
  11} (2008) 001} [\href{https://arxiv.org/abs/0808.2771}{{\ttfamily
  0808.2771}}].

\bibitem{Closset:2012vp}
C.~Closset, T.~T. Dumitrescu, G.~Festuccia, Z.~Komargodski and N.~Seiberg,
  \emph{{Comments on Chern-Simons Contact Terms in Three Dimensions}},
  \href{https://doi.org/10.1007/JHEP09(2012)091}{\emph{JHEP} {\bfseries 09}
  (2012) 091} [\href{https://arxiv.org/abs/1206.5218}{{\ttfamily 1206.5218}}].

\bibitem{Closset:2012vg}
C.~Closset, T.~T. Dumitrescu, G.~Festuccia, Z.~Komargodski and N.~Seiberg,
  \emph{{Contact Terms, Unitarity, and F-Maximization in Three-Dimensional
  Superconformal Theories}},
  \href{https://doi.org/10.1007/JHEP10(2012)053}{\emph{JHEP} {\bfseries 10}
  (2012) 053} [\href{https://arxiv.org/abs/1205.4142}{{\ttfamily 1205.4142}}].

\bibitem{Dorey:1999rb}
N.~Dorey and D.~Tong, \emph{{Mirror symmetry and toric geometry in
  three-dimensional gauge theories}},
  \href{https://doi.org/10.1088/1126-6708/2000/05/018}{\emph{JHEP} {\bfseries
  05} (2000) 018} [\href{https://arxiv.org/abs/hep-th/9911094}{{\ttfamily
  hep-th/9911094}}].

\bibitem{Nii:2014jsa}
K.~Nii, \emph{{3d duality with adjoint matter from 4d duality}},
  \href{https://doi.org/10.1007/JHEP02(2015)024}{\emph{JHEP} {\bfseries 02}
  (2015) 024} [\href{https://arxiv.org/abs/1409.3230}{{\ttfamily 1409.3230}}].

\bibitem{Amariti:2014iza}
A.~Amariti and C.~Klare, \emph{{A journey to 3d: exact relations for adjoint
  SQCD from dimensional reduction}},
  \href{https://doi.org/10.1007/JHEP05(2015)148}{\emph{JHEP} {\bfseries 05}
  (2015) 148} [\href{https://arxiv.org/abs/1409.8623}{{\ttfamily 1409.8623}}].

\bibitem{Kutasov:1995ve}
D.~Kutasov, \emph{{A Comment on duality in N=1 supersymmetric nonAbelian gauge
  theories}}, \href{https://doi.org/10.1016/0370-2693(95)00392-X}{\emph{Phys.
  Lett.} {\bfseries B351} (1995) 230}
  [\href{https://arxiv.org/abs/hep-th/9503086}{{\ttfamily hep-th/9503086}}].

\bibitem{Kutasov:1995np}
D.~Kutasov and A.~Schwimmer, \emph{{On duality in supersymmetric Yang-Mills
  theory}}, \href{https://doi.org/10.1016/0370-2693(95)00676-C}{\emph{Phys.
  Lett.} {\bfseries B354} (1995) 315}
  [\href{https://arxiv.org/abs/hep-th/9505004}{{\ttfamily hep-th/9505004}}].

\bibitem{Kutasov:1995ss}
D.~Kutasov, A.~Schwimmer and N.~Seiberg, \emph{{Chiral rings, singularity
  theory and electric - magnetic duality}},
  \href{https://doi.org/10.1016/0550-3213(95)00599-4}{\emph{Nucl. Phys.}
  {\bfseries B459} (1996) 455}
  [\href{https://arxiv.org/abs/hep-th/9510222}{{\ttfamily hep-th/9510222}}].

\bibitem{Spiridonov:2009za}
V.~P. Spiridonov and G.~S. Vartanov, \emph{{Elliptic Hypergeometry of
  Supersymmetric Dualities}},
  \href{https://doi.org/10.1007/s00220-011-1218-9}{\emph{Commun. Math. Phys.}
  {\bfseries 304} (2011) 797}
  [\href{https://arxiv.org/abs/0910.5944}{{\ttfamily 0910.5944}}].

\bibitem{Dolan:2008qi}
F.~A. Dolan and H.~Osborn, \emph{{Applications of the Superconformal Index for
  Protected Operators and q-Hypergeometric Identities to N=1 Dual Theories}},
  \href{https://doi.org/10.1016/j.nuclphysb.2009.01.028}{\emph{Nucl. Phys.}
  {\bfseries B818} (2009) 137}
  [\href{https://arxiv.org/abs/0801.4947}{{\ttfamily 0801.4947}}].

\bibitem{Nii:2020ikd}
K.~Nii, \emph{{Generalized Giveon-Kutasov duality}},
  \href{https://arxiv.org/abs/2005.04858}{{\ttfamily 2005.04858}}.

\bibitem{Niarchos:2009aa}
V.~Niarchos, \emph{{R-charges, Chiral Rings and RG Flows in Supersymmetric
  Chern-Simons-Matter Theories}},
  \href{https://doi.org/10.1088/1126-6708/2009/05/054}{\emph{JHEP} {\bfseries
  05} (2009) 054} [\href{https://arxiv.org/abs/0903.0435}{{\ttfamily
  0903.0435}}].

\bibitem{Kapustin:2011vz}
A.~Kapustin, H.~Kim and J.~Park, \emph{{Dualities for 3d Theories with Tensor
  Matter}}, \href{https://doi.org/10.1007/JHEP12(2011)087}{\emph{JHEP}
  {\bfseries 12} (2011) 087} [\href{https://arxiv.org/abs/1110.2547}{{\ttfamily
  1110.2547}}].

\bibitem{Giacomelli:2017vgk}
S.~Giacomelli and N.~Mekareeya, \emph{{Mirror theories of 3d $ \mathcal{N} $ =
  2 SQCD}}, \href{https://doi.org/10.1007/JHEP03(2018)126}{\emph{JHEP}
  {\bfseries 03} (2018) 126}
  [\href{https://arxiv.org/abs/1711.11525}{{\ttfamily 1711.11525}}].

\bibitem{Giacomelli:2019blm}
S.~Giacomelli, \emph{{Dualities for adjoint SQCD in three dimensions and
  emergent symmetries}},
  \href{https://doi.org/10.1007/JHEP03(2019)144}{\emph{JHEP} {\bfseries 03}
  (2019) 144} [\href{https://arxiv.org/abs/1901.09947}{{\ttfamily
  1901.09947}}].

\bibitem{Pasquetti:2019tix}
S.~Pasquetti and M.~Sacchi, \emph{{3d dualities from 2d free field correlators:
  recombination and rank stabilization}},
  \href{https://doi.org/10.1007/JHEP01(2020)061}{\emph{JHEP} {\bfseries 01}
  (2020) 061} [\href{https://arxiv.org/abs/1905.05807}{{\ttfamily
  1905.05807}}].

\bibitem{Hwang:2018uyj}
C.~Hwang, H.~Kim and J.~Park, \emph{{On 3d Seiberg‐Like Dualities with Two
  Adjoints}}, \href{https://doi.org/10.1002/prop.201800064}{\emph{Fortsch.
  Phys.} {\bfseries 66} (2018) 1800064}
  [\href{https://arxiv.org/abs/1807.06198}{{\ttfamily 1807.06198}}].

\bibitem{Fujitsuka:2013fga}
M.~Fujitsuka, M.~Honda and Y.~Yoshida, \emph{{Higgs branch localization of 3d $
  \mathcal{N} =2$ theories}},
  \href{https://doi.org/10.1093/ptep/ptu158}{\emph{PTEP} {\bfseries 2014}
  (2014) 123B02} [\href{https://arxiv.org/abs/1312.3627}{{\ttfamily
  1312.3627}}].

\bibitem{Benini:2013yva}
F.~Benini and W.~Peelaers, \emph{{Higgs branch localization in three
  dimensions}}, \href{https://doi.org/10.1007/JHEP05(2014)030}{\emph{JHEP}
  {\bfseries 05} (2014) 030} [\href{https://arxiv.org/abs/1312.6078}{{\ttfamily
  1312.6078}}].

\bibitem{Amariti:2015yea}
A.~Amariti, D.~Forcella, C.~Klare, D.~Orlando and S.~Reffert, \emph{{The
  braneology of 3D dualities}},
  \href{https://doi.org/10.1088/1751-8113/48/26/265401}{\emph{J. Phys.}
  {\bfseries A48} (2015) 265401}
  [\href{https://arxiv.org/abs/1501.06571}{{\ttfamily 1501.06571}}].

\bibitem{Amariti:2015mva}
A.~Amariti, D.~Forcella, C.~Klare, D.~Orlando and S.~Reffert, \emph{{4D/3D
  reduction of dualities: mirrors on the circle}},
  \href{https://doi.org/10.1007/JHEP10(2015)048}{\emph{JHEP} {\bfseries 10}
  (2015) 048} [\href{https://arxiv.org/abs/1504.02783}{{\ttfamily
  1504.02783}}].

\bibitem{Amariti:2016kat}
A.~Amariti, D.~Orlando and S.~Reffert, \emph{{String theory and the 4D/3D
  reduction of Seiberg duality. A review}},
  \href{https://doi.org/10.1016/j.physrep.2017.08.002}{\emph{Phys. Rept.}
  {\bfseries 705-706} (2017) 1}
  [\href{https://arxiv.org/abs/1611.04883}{{\ttfamily 1611.04883}}].

\bibitem{Hanany:1996ie}
A.~Hanany and E.~Witten, \emph{{Type IIB superstrings, BPS monopoles, and
  three-dimensional gauge dynamics}},
  \href{https://doi.org/10.1016/S0550-3213(97)00157-0,
  10.1016/S0550-3213(97)80030-2}{\emph{Nucl. Phys.} {\bfseries B492} (1997)
  152} [\href{https://arxiv.org/abs/hep-th/9611230}{{\ttfamily
  hep-th/9611230}}].

\bibitem{Cremonesi:2010ae}
S.~Cremonesi, \emph{{Type IIB construction of flavoured ABJ(M) and fractional
  M2 branes}}, \href{https://doi.org/10.1007/JHEP01(2011)076}{\emph{JHEP}
  {\bfseries 01} (2011) 076} [\href{https://arxiv.org/abs/1007.4562}{{\ttfamily
  1007.4562}}].

\end{thebibliography}\endgroup

\end{document}